\documentclass[secnumarabic,amssymb, nobibnotes, aps, prd,onecolumn]{revtex4-2}
\usepackage{graphicx}

\usepackage{xcolor}
\usepackage{bbold}
\usepackage{hyperref}
\usepackage{geometry}
\usepackage{amsmath,amssymb,bm}
\usepackage{booktabs}
\usepackage{amsfonts}
\expandafter\let\csname equation*\endcsname\relax 
\expandafter\let\csname endequation*\endcsname\relax 

\usepackage{lmodern}
\usepackage{moresize}
\usepackage{aas_macros}

\usepackage{natbib}
\usepackage{tikz}
\usetikzlibrary{shapes.symbols,decorations.pathmorphing}
\usepackage{cabin}
\usepackage{pgfplots}
\pgfplotsset{compat = newest}

\usepackage{array}
\usepackage{longtable}

\def\q {\bm{q}}
\def\x {\bm{x}}
\def\r {\bm{r}}
\def\d {{\rm{d}}}

\def\k {\bm{k}}
\def\x {\bm{x}}

\newcommand{\fnl}{{f}_\text{NL}}

\newcommand{\D} {{{\rm{D}}}}
\newcommand{\p}{_{{\text{\tiny$\|$}}}}

\newcommand{\HH}{\mathcal{H}\,}

\newcommand{\<}{\langle}
\renewcommand{\>}{\rangle}

\def\aj{\ref@jnl{AJ}}                   
\def\actaa{\ref@jnl{Acta Astron.}}      
\def\araa{\ref@jnl{ARA\&A}}             
\def\apj{\ref@jnl{ApJ}}                 
\def\apjl{\ref@jnl{ApJ}}                
\def\apjs{\ref@jnl{ApJS}}               
\def\ao{\ref@jnl{Appl.~Opt.}}           
\def\apss{\ref@jnl{Ap\&SS}}             
\def\aap{\ref@jnl{A\&A}}                
\def\aapr{\ref@jnl{A\&A~Rev.}}          
\def\aaps{\ref@jnl{A\&AS}}              
\def\azh{\ref@jnl{AZh}}                 
\def\baas{\ref@jnl{BAAS}}               

\begin{document}

          \title{\boldmath Cosmological zoom-in perturbation theory as a consistent beyond  point-particle approximation framework }

\author{Obinna Umeh }
\affiliation{${}^{1}$Institute of Cosmology \& Gravitation,  University of Portsmouth, Portsmouth PO1 3FX, United Kingdom}
\affiliation{${}^{2}$Department of Physics, University of the Western Cape, Cape Town 7535, South Africa}
\affiliation{${}^{3}$Hierarchical Intelligence Lab, Divine Estate, Charlesco Phase 2, Warri , Delta State, Nigeria}
\email{obinna.umeh@port.ac.uk}
\date{\today}

\begin{abstract}

Modelling structure formation across the full dynamical range of the Universe remains a major challenge in cosmology.
This difficulty originates from a fundamental limitation of geodesics in general relativity: a one-parameter family of geodesics can cease to be geodesic at a finite time.
This implies that the conventional point-particle approximation is not the primary issue; rather, the breakdown of geodesic flow restricts a consistent description across scales. We develop a covariant multi-scale framework that resolves this problem by decomposing spacetime into hierarchical regions separated by matter horizons. We show how to match shared boundary consistently at the level of the action, leading to a covariant backreaction contribution.
The resulting construction provides a first-principles theoretical foundation for cosmological zoom-in simulations and yields an effective energy-momentum tensor capturing the impact of the geometric backreaction effect. As an application, we demonstrate that this backreaction naturally produces flat galaxy rotation curves without invoking an additional dark matter component.
Our results establish a new perspective on nonlinear structure formation, in which long dynamical range is resolved through a hierarchy of discrete geodesic domains.

\end{abstract}

\maketitle
\DeclareGraphicsRule{.wmf}{bmp}{jpg}{}{}
\maketitle

\tableofcontents
\maketitle

\section{Introduction}\label{sec:intro}

The modelling of clustering of matter on small or non-linear scales remains a fundamental open problem in cosmology~\cite{Umeh:2021xqm,Ivanov:2022mrd,Paul:2023yrr,Padmanabhan:2023hfr}.
The key challenge is the simultaneous modelling of physical processes spanning orders of magnitude in both space and time; i.e. from megaparsec scale to sub-galactic scales and from the early universe to today. This has led to restricting consistent relativistic description of matter in the universe to linear scales with a fundamental assumption on the nature of matter content~\cite{Bernardeau:2001qr,Umeh:2015gza,Umeh:2016nuh,Jolicoeur:2017eyi,Jolicoeur:2017nyt,Jolicoeur:2018blf,Koyama:2018ttg,Clarkson:2018dwn,Umeh:2019qyd,Umeh:2019jqg,Maartens:2020jzf}.
This is not a coincidence; it is a consequence of the nature of geodesics in curved spacetime~\cite{Penrose:1969pc,Umeh:2023lbc,Umeh:2026ajv}. The standard Point Particle Approximation (PPA) of the matter distribution in the universe on which the standard model of cosmology is based, neglects the effect of spacetime backreaction on the propagation of geodesic~\cite{Poisson:2011nh,Hahn:2015sia,Ondaro-Mallea:2023qat}.
 The standard or traditional cosmological N-body simulation, which is considered the gold standard for modelling the motion and distribution of matter in the universe is based on PPA~\cite{Springel2006,Adamek:2016zes}.

 The PPA holds in the limit where the characteristic mass and size of the gravitational bound system are much smaller than all external length scales~\cite{Poisson:2011nh}.  
Current efforts to move beyond this limit in cosmology, such as Vlasov perturbation theory via special algebra for higher order moment~\cite{gratus2013colombeau, Garny:2022tlk} and the Effective Field Theory of Large Scale Structure (EFTofLSS)~\cite{Carrasco:2012cv}, remain constrained.
Under the Vlasov perturbation theory, the intention is to capture the impact of higher-order cumulants whose impact renormalises the background spacetime and introduce a new physical scale ~\cite{Garny:2022tlk,Garny:2022kbk,Garny:2025ovs,Garny:2025zlq}. It is computationally intensive and, at the moment, it is based on Newtonian gravity; hence, the full mathematical structure of general relativity is yet to be captured. 
The  EFTofLSS adopt PPA and also assumes separation of scales, which allows it to treat finite-size effects as effective fluids~\cite{Harte:2017hfp,Carrasco:2012cv}. However,  its validity is limited to about the quasi-linear scales, typically up to wavenumbers of $ k \sim  0.4h / \rm{Mpc}$ \cite{Umeh:2015gza, Umeh:2016thy,Ivanov:2022mrd}.  It is generally computationally efficient, but the momentum integrals require regularisation and counterterm renormalisation, which could explode in redshift space~\cite{Pajer:2013jj,Perko:2016puo}.

A more robust and mathematically rigorous extension of the PPA  is the Matched Asymptotic Expansions (MAE)~\cite{O’Malley2014}. It constructs two distinct expansions (Inner and Outer) and then matches them in a shared region~\cite{Poisson:2003nc}. In cosmology or large-scale structures of the universe, this is studied under two-parameter cosmological perturbation theory~\cite{Bruni:2002sma,Goldberg:2016lcq}. The MAE method replaces the single global coordinate system with two different limits to capture the physics at different scales. 

As established in \cite{Umeh:2026ajv}, prior attempts to formalise the MAE method within the context of large-scale structures have faced significant hurdles, primarily due to the omission of the matter horizon  \cite{Ellis:2010fr}. A second critical oversight in earlier applications was the attempt to apply scale decomposition directly to the equations of motion. In contrast, our framework implements scale decomposition at the level of the action \cite{Umeh:2023lbc}. This shift is not merely formal; it empowers us to leverage variational calculus to handle boundary conditions consistently and, crucially, bypasses the mathematical inconsistencies inherent in the product of distributions.

Most significantly, we have demonstrated that MAE is the theoretical equivalent of the cosmological zoom-in N-body simulation approach; a technique traditionally motivated by the challenge brought about by the computational complexity associated with calculating gravitational interactions across millions of light-years while simultaneously resolving the physics of individual stars or galaxies \cite{2005AA...441...55S,Wetzel:2022man}. By utilising the matter horizon as a physically motivated scale separation to define specific spacetime regions, MAE provides the rigorous mathematical foundation for these simulations. This separation of scales creates a robust framework for capturing fast time-scale dynamics by effectively isolating small-scale physics as it evolves against the background of large-scale modes.

Another crucial feature of the MAE in cosmology is its covariant treatment of the back-reaction of background spacetime on the particle trajectory. 
By the backreaction of particle propagation, we refer to the impact of the induced energy-momentum tensor due to dynamical spacetime geometry on the particle propagation~\cite{Clarkson:2011zq}. According to Einstein's field equations, the distribution of energy and momentum determines the curvature of spacetime, so particles propagating through spacetime can, in principle, modify its geometry. 
Here, we describe how the spacetime influences particle propagation, leading to a form of energy-momentum tensor that manifests as dark matter on small scales. As a way of example, we show from first principles how the induced backreaction effect leads naturally to flat rotation curves without invoking any exotic form of matter.  

We describe point particle approximation in section \ref{sec:point_particle_approx}, and the manifold surgery is discussed in section \ref{sec:FLRW_discrete_symmetry}. The derivation of backreaction effects and its interpretation as the effective energy-momentum tensor is discussed in section \ref{sec:backreaction}. The conclusion is presented in section \ref{sec:conc}.

{\bf{Cosmology and Notations}}: For quantitative estimates in this paper, we made use of the  Planck CMB constraint on cosmological parameters. Thats we use:  $h = 0.674$ for the dimensionless Hubble parameter, $\Omega_{\rm b} = 0.0493$ for baryon density parameter,  $\Omega_{\rm{cdm}} = 0.264$ for the dark matter density parameter,  $\Omega_{\rm m} = \Omega_{\rm{cdm}} + \Omega_{\rm b}$ for the matter density parameter,  $n_{\rm s} = 0.9608$ for spectral index,  and  $A_{\rm s} = 2.198 \times 10^{-9}$ for the amplitude of the primordial curvature perturbation~\cite{Aghanim:2018eyx}. 
On the choice of notation, we use the small English alphabet from $a-e$ to denote the full spacetime indices, while $i$ and $j$ denote the spatial indices.  The capital English alphabet from $A-E$ denotes tetrad indices on the screen space.  Plus and Minus ${}^{\pm}$ as a subscript or superscript denote two sheets of spacetime. 


\section{Geodesic breakdown and the origin of multiscale structure}\label{sec:point_particle_approx}

Geodesics appear in three flavours in GR:  timelike geodesics for massive particles, null geodesics for light,  and spacelike geodesics for spatial intervals. The energy-momentum tensor, $T_{ab}$, for the matter distribution in the universe is constructed by approximating the matter distribution in the universe with the action of a set of point particles(a sum of the single-particle relativistic actions: $S_{\ell} = -m_{\ell} \int \mathrm{d}\tau_{\ell}$)
\begin{eqnarray}\label{eq:matter_action}
S_{\text{matter}} =\sum_{\ell} S_{\ell} 
= -\sum_{\ell} m_{\ell} \int \sqrt{-g_{ab} \frac{\d x^a_{\ell} }{\d\tau_{\ell}}\frac{\d x^b_{\ell}}{\d\tau_{\ell}} } \d\tau_{\ell} \,,
\end{eqnarray}
where 
$x_{\ell}^a(\tau_{\ell})$ is the spacetime trajectory of the massive $\ell$-th particle and $m_{\ell }$ is the rest mass of the ${\ell}$-th particle and $\tau_{\ell}$ is the proper time for the ${\ell}$-th particle.  
Our focus is on how an ensemble of massive particles impacts the spacetime, therefore we compute the variation $\delta S_{\text{matter}}$ with respect to $\delta g_{ab}$.
 \begin{eqnarray}
\delta S_{\ell} = \frac{m_{\ell}}{2} \int \mathrm{d}\tau_{\ell}\, \left[\int \d^4 x \delta^4(x^{a} - x^{a}(\tau_{\ell})) \right]\delta g_{ab}(x^{a}(\tau_{\ell})) u^a_{\ell} u^b_{\ell}
= \frac{1}{2}\int \d^4 x \sqrt{-g(t,x^i)} T^{ab}_{\ell} \delta g_{ab}\,,
\end{eqnarray}
where ${u^a_{\ell}}' = {\d x^a_{\ell}}/{\d\tau_{\ell}}$ and $\delta^4(x^a-x_{\ell} (\tau_{\ell}))$ is a 4D Dirac delta function.   We have introduced the energy-mementum tensor  defined as a variation of the action with respect to the metric tensor $g_{ab}$: $T_{\ell }^{ab} \equiv -({2}/{\sqrt{-g}})
  ({\delta\left(\sqrt{-g} {S}_{\ell}\right)})/{\delta g_{ab}} $. The particle is located at $x^a_{\ell} (\tau_{\ell})$, hence,  it only picks up contributions from the particle's trajectory. 
\begin{eqnarray}\label{eq:shifting_property}
\int \d^4 x {\delta^4(x ^{a}- x^{a}(\tau_{\ell})) }=1 \qquad {\rm{and}}\qquad  \int   f(x)  \delta^4(x^{a} - x^{a}_{\ell}(\tau_{\ell}))\, \mathrm{d}^4x \propto f(x^{a}_\ell) \,.
\end{eqnarray}
The  total energy-momentum tensor constructed from the action of a massive particle: 
 \begin{eqnarray}\label{eq:EMT_particles}
 T^{ab} = \sum^N_{\ell}  T^{ab}_{\ell}
 = \sum^N_{\ell}  \frac{m_{\ell}}{\sqrt{-g}} \int {\d}\tau\, u^{a}_{\ell} u^{b}_{\ell} \delta^4(x - x_{\ell}(\tau_{\ell})) \,.
\end{eqnarray}
 Note $x^a$ is a point in the spacetime and the delta function is non-zero only when $x^a$ coincides with  $x^a_{\ell}(\tau_{\ell})$: Again describing the matter distribution in the universe with the singularities as described in equation \eqref{eq:EMT_particles} is valid only in the limit $L \gg L_{\rm{phy}}$ \cite{Adamek:2016zes}. On scales the size of a gravitationally bound system in the universe, it breaks down.  Attempts have been made to replace $ \delta^4(x ^{a}- x^{a}_{\ell}(\tau_{\ell}))$  with a localised, smeared distribution, i.e a Gaussian regulator $G_{\varepsilon} (x^a,{x}^a_{\ell})$ which has a similar mathematical sturcture as described in equation \eqref{eq:shifting_property}~\cite{Poisson:2011nh}.

\subsection{Geodesic propagation in the standard model of cosmology}

Focusing on timelike separation for the moment, we start with the action of a massive particle given in equation \eqref{eq:matter_action}. The first infinitesimal variation of action a massive particle (${\d S_{\ell}}/{\d s}\big|_{s=0}  $) given in equation \eqref{eq:matter_action} gives a geodesc equation~\cite{Umeh:2023lbc}.
\begin{eqnarray}\label{eq:geodeisc_eqn}
u^d\nabla_d u^c = 0= \frac{\d^2 x^a}{\d\tau^2} + \Gamma^{a}_{bc} \frac{\d x^b}{\d \tau}\frac{\d x^c}{\d \tau}\,,
\end{eqnarray}
where we dropped the subscripts $\ell$ on $u^a$ to reduce cluter. We imposed proper variation at the endpoints $[\tau_{\rm{ini}},\tau_f]$.  The massive particle trajectory is geodesic within $[\tau_{\rm{ini}},\tau_f]$.
 Equation \eqref{eq:geodeisc_eqn}  is unchanged under the reparameterisation of the proper time:  $\tau = W\tilde{ \tau} + Z\,$. This induces a re-scaling of the velocity vector $ u^a =  \tilde{u}^a /W$,
 where $W$ and $Z$ are non-zero free parameters.   The freedom to fix $W$ is usually interpreted as the freedom to choose the unit of measurement of the flow of time.  Furthermore, equation \eqref{eq:geodeisc_eqn} is a second-order differential equation; hence, in addition to the freedom to fix $W$ and $Z$,  there is a freedom to choose the initial position, $x^a( \tau_{\rm{ini}})$, initial velocity, $ u^a_{\rm{ini}}$. The $x^a( \tau_{\rm{ini}})$ is fixed at early times when the universe is described by the FLRW spacetime, and $ u^a_{\rm{ini}}$  is determined by the Hubble rate. The current position of the particle $x^i$ can be given in terms of the initial position, $q^i $
\begin{eqnarray}\label{eq:displacement_field}
x^i(t,q^i) = q^i + \Psi^i(t,q^i)\,,
\end{eqnarray}
where  $\Psi^i$ is known as the displacement vector field. It depends on the gradient of $\Phi$. In the  continuum limit, $N\to \infty$ (dust approximation), we can define the scalar rest-frame mass (energy) density $\rho(x)$ by
$\rho(x^i) =   \sum^N_{\ell}  \frac{m_{\ell}}{\sqrt{-g}} \int {\d}\tau\,  \delta^4(x^{a} - x^{a}_{\ell}(\tau_{\ell})) $. 
Since mass is a non-dynamical quantity in Newtonian gravity, one can impose that the mass in $q^i $ coordinate system is the same as the mass at the current time or in the $x^i$ coordinate system: $\bar{\rho}(\tau) \d^3 q = \rho(\tau, {x^i})\d^3 x$, hence, 
 \begin{eqnarray}\label{eq:mass_conservation}
 \frac{\rho(t,{\x})}{\bar{\rho}} = 1+ \delta(t,{\x}) = \frac{1}{{{\rm{det}}[\mathcal{J}](t,{q}) }}\,,
\end{eqnarray}
where $\delta \equiv \delta \rho/ \bar{\rho}$ is the  matter density contrast. The density blows up at  ${{\rm{det}}[\mathcal{J}](t,{q}) } \to 0$. This is interpreted as a position in x-space, where the multiple streams of particles overlap at the same spatial location. 
The caustics or the blow-up are indications that the conservation of mass on which equation \eqref{eq:mass_conservation} is based breaks down at a finite time.

\subsection{Impact of local curvature, matter horizon and separation of scales}\label{sec:matterhorizon}

Working with equation \eqref{eq:geodeisc_eqn} does not capture the impact of the local curvature on particle propagation~\cite{Umeh:2026ajv}. 
One way to capture the impact of local curvature on the geodesics is by taking the second variation of the massive particle action (equation \eqref{eq:matter_action})~\cite{Umeh:2023lbc,Eschenburg1994ComparisonTI}
\begin{eqnarray}\label{eq:second_variation1}
- \frac{\d^2 S}{\d s^2}\bigg|_{s=0} &=&-m\int_{\tau_{\rm{ini}}}^{\tau_i}  \xi_{c} \left[ \frac{\d^2 \xi^{c}}{\d\tau^2} + R^{c}{}_{def} \xi^{d} u^{e} u^{f} \right]\d\tau \,,
\end{eqnarray}
where $ R^{a}{}_{def}$ is the Riemann tensor and we have also introduced a deviation vector, $\xi^a= {\partial x^a(\tau,s)}/{\partial s}$.
The critical point of the second variation (equation \eqref{eq:second_variation1}) (i.e $ {\d^2 S}/{\d s^2} = 0)$ gives the geodesic deviation equation 
\begin{eqnarray}\label{eq:geodesic_deviation}
 \frac{\d^2 \xi^{c}}{\d\tau^2} + R^{c}{}_{def} \xi^{d} u^{e} u^{f}  = 0 \,.
\end{eqnarray} 
Using the fact that the deviation vector $\xi^a$ is Lie dragged along the integral curves of $u^a$, we can decompose equation \eqref{eq:geodesic_deviation} in terms of physical quantities that an observer at rest in the matter frame can measure~\cite{HawkingandEllis:1973lsss.book}.
$
\mathcal{L}_{u} \xi^a = 0 \rightarrow  \frac{{\d} \xi^{a}}{{\d} \tau}  = \nabla_{b} u^a \xi^b  
$, 
where we have decomposed $ \nabla_{b} u_{a} $  into irreducible coordinate independent physical quantities ~\cite{Ellis:1998ct,Ellis:1990gi}
\begin{eqnarray}\label{eq:decomposeCDU}
\nabla_{b} u_{a} = -u^{b}A^{a}+\frac{1}{3}\Theta {h}^{ab }+\sigma^{ab} + \omega^{ab}\,,
\end{eqnarray}
where ${h}_{ab}={g}_{ab}+{u}_a {u}_b$ is the metric on the hypersurface orthogonal to $u^a$,  $A^a$ is the acceleration $A_a=u^d\nabla_{d} u_a$, it vanishes by equation \eqref{eq:geodeisc_eqn}, $\Theta ={\D}_{a}u^a$  describes the expansion/contraction of the nearby family of geodesics. It could be positive  $\Theta >0$ or negative $\Theta<0$.  The actual physical description of $\Theta >0$  or  $\Theta <0$ depends on the coordinate time orientation of the spacetime. 
  $\sigma_{ab} =h_{a}{}^{c}h_{b}^d \nabla_{\<c}u_{d\>}$ is the shear deformation tensor, which describes the rate of change of the deformation of nearby geodesics when compared to flat spacetime.  $\omega_{ab}=h_{a}{}^{c}h_{b}^d \nabla_{[c}u_{d]}$ is the vorticity tensor. It is an anti-symmetric tensor. It describes the twisting of nearby geodesics.  
   It vanishes for orthogonal geodesics
by Frobenius' theorem.

Therefore, putting these in equation \eqref{eq:geodesic_deviation} leads to propagation equation for the expansion $\Theta$,  the shear tensor $\sigma_{ab}$  equatioon and vorticity ${\omega}_{ab}$ equation ~\cite{2012reco.book.....E,Ellis1971grc..conf..104E,Ellis2009,Ellis:1998ct}:
\begin{eqnarray}
\frac{ {\rm{D}} {{\Theta}} }{{\rm{D}} \tau} &=& - \frac{1}{3}{\Theta}^2 - {\sigma}_{ab}{\sigma}^{ab} 
 - {R}_{ab} {u}^a {u}^b\,,
\label{eq:expansion}
\\
\frac{ {\rm{D} } {{\sigma}}_{ab}}{{\rm{D}} \tau} &=& - \frac{2}{3} {\Theta} {\sigma}_{ab} - {\sigma}^{c}{}_{\<a}{\sigma}_{b\>c} 
 - {C}_{acbd}{u}^c {u}^d +\frac{1}{2}S_{\<ab\>}\,,
\label{eq:shear}
\\
\frac{ {\rm{D} } {{\omega}}_{ab}}{{\rm{D}} \tau} &=& -\frac{2}{3}{\Theta}{\omega}_{ab} + {\sigma}^c{}_{[a}{\omega}_{b]c}
\label{eq:vorticity}\,,
\end{eqnarray} 
where  ${ {\rm{D}} {{\cdots}} }/{{\rm{D}} \tau}  = u^a \nabla_{a} \cdots$ is the directional derivative,  ${C}_{acbd}$ is the Weyl tensor and $S_{\<ab\>}=  R_{ab} - h_{ab}R_{cd}h^{cd}/3$ is the anisotropic part of curvature intrinsic to the hypersurface, $R_{ab}$ is the Ricci tensor. 
We plan to solve equations~\eqref{eq:expansion} and \eqref{eq:shear}  in a universe described by a  perturbed metric tensor in Poisson(conformal Newtonian) gauge ~\cite{Umeh:2010pr}
\begin{eqnarray}\label{eq:metric_tensor}
\d  s^2 &=&a^2\left[-(1 + 2\Phi)\d \eta^2 + \omega_{i} \d\eta \d x^i + \left((1-2 \Psi)\delta_{i j} + \chi_{ij} \right)\d x^{i}\d x^{j}\right]\,.
\end{eqnarray}
where $\delta_{ij}$ is the spatial metric of the flat background spacetime, with covariant derivative $\bar{\nabla}_i$,  $\Phi$ and $\Psi$ are scalar potentials, $\omega_{i}$ and $\chi_{ij}$ are vector and tensor perturbations.  We have introduced the conformal time, $\eta$, which is related to the coordinate time, $t$, according to $ \d t= a\d \eta $, where $a$ is the scale factor of the universe, its evolution is governed by the Friedmann equation~\cite{2012reco.book.....E}.  Without loss of generality, we will neglect both tensor and vector perturbations for simplicity. 
The four-velocity in this gauge is given by $[u^0, u^i] = [1 - \Phi ,{ \partial}^{i}v]/a$, where $v$ is the velocity potential.

In a universe described by equation \eqref{eq:metric_tensor}, $\Theta$ can be decomposed into two parts, the global and local parts $\Theta = \Theta_{H}  + \Theta_{L}$., where the global part $\Theta_{H}  = 3 H$ is given by the Hubble rate, 
 The local part, $\Theta_{L}$ is proportional to the divergence of the relative velocity, $v^i$, between the background and the physical spacetime: $ \Theta_{L} = {\D}_{i} v^i$. 
Substituting $\Theta = \Theta_{H}  + \Theta_{L}$ in equation \eqref{eq:expansion} leads to a coupled sysytem of equations for $ \Theta_{H}$  and  $ \Theta_{L}$. $ \Theta_{H}$ obey the following propagation equation
    \begin{eqnarray}\label{eq:Global_part}
\frac{ {\rm{D}} {{\Theta_{H}}} }{{\rm{D}} \tau} &=& - \frac{1}{3}{\Theta}^2_{H} 
-  \frac{1}{2}\kappa\bar{\rho} - \Lambda \,,
\end{eqnarray}
where the shear tensor vanishes on an FLRW background spacetime ${\sigma}_{ab}{\sigma}^{ab} =0$. The symmetry of the background spacetime does not allow coupling to the local expansion.  Furthermore, we made use of the time-time component of the  Einstein field equation, to express $  u^b u^d R_{bd}$ in terms of the matter density field, $  u^b u^d R_{bd} \approx\kappa \rho/2 + \Lambda  = \kappa\left[\bar{\rho} +  \delta \rho \right]/2 + \Lambda, $ where we have decomposed $ \rho$ into global and local parts $ \rho  = \bar{\rho} + \delta \rho$, where $\bar{\rho} $ is the mean density(global component), and $\delta \rho$ is the perturbations or the corresponding local component. 
It is possible to solve equation \eqref{eq:Global_part} as an ordinary differential inequality with the  assumption that the weak energy conditions hold $ \kappa^2\bar{\rho}/2 + \Lambda>0 $
\begin{eqnarray}\label{eq:focussing_theorem}
\frac{ {\rm{D}}  \Theta^{-1}_{H}}{ {\rm{D}} \tau}  \geq \frac{1}{3} \qquad {\rm{with~a~solution }} \qquad 
 \frac{1}{\Theta_{H} } \geq \frac{1}{\Theta _{H\rm{ini}}}+ \frac{\tau - \tau_{\rm{ini}} }{3}\,,
\end{eqnarray}
where $\tau_{\rm{ini}}$ is the proper time at the initial hypersurface and ${\Theta _{H\rm{ini}}} $ is the initial expansion.   $1/\Theta_{H}$ vanishes at the finite time $\tau = \tau_{\rm{ini}}  - 3 /\Theta_{H\rm{ini}}  $ only if the initial expansion is negative ${\Theta _{H\rm{ini}}} <0$.  Hence, the 
 initial global expansion must be strickly positive ${\Theta _{H\rm{ini}}} >0$.   This is consistent with the standard model of cosmology built on expanding FLRW spacetime~\cite{Lemaitre:1927zz,Hubble:1929ig,Planck:2018vyg}.

However, the local expansion ${\Theta_L} $ satisfies the following propagation equation, which is coupled to the global expansion
\begin{eqnarray}\label{eq:local-expansion}
\frac{ {\rm{D}} {{\Theta_L}} }{{\rm{D}} \tau} 
& =   &- \frac{1}{3}{\Theta_L}^2-\frac{2}{3} \Theta_{H}\Theta_{L} - {\sigma}_{ab}{\sigma}^{ab} 
-\frac{1}{2}\kappa\left[  \delta \rho \right] \,,
\end{eqnarray}
Since  ${\sigma}_{ab}{\sigma}^{ab} >0$ is positive definite, equation \eqref{eq:local-expansion} can be solved for over-dense regions $\delta \rho >0$ as  partial differential inequality 
\begin{eqnarray}\label{eq:local-expansion3}
\frac{ {\rm{D}} {{\Theta_L}} }{{\rm{D}} \tau} 
+ \frac{1}{3}{\Theta^2_L}+\frac{2}{3} \Theta_{H}\Theta_{L}& \le   & 0 \,,
\end{eqnarray}
 Integrating equation~\eqref{eq:local-expansion3} leads to 
\begin{eqnarray}
\frac{1}{\Theta_{L}(\tau)} \ge - \frac{1}{\exp\left[\mathcal{I}_{1} (\tau)\right]}  \bigg[ \frac{-1}{\Theta_{L\rm{ini}} }
+\mathcal{I}_{2} (\tau) \bigg]\,,
\label{eq:focusing_theorem2}
\end{eqnarray}
where $\mathcal{I}_{1} (\tau) $ is a function of the background expansion
\begin{eqnarray}
\mathcal{I}_{1} (\tau) &=& -\frac{2}{3}\int_{0}^{\tau} \Theta_{H}(\tau') \d\tau'
 \approx   -2 \ln (1+z)
 \,,
\\
\mathcal{I}_{2} (\tau) &=&-\frac{1}{3}\int_{0}^{\tau} \exp\left[\mathcal{I}_{1} (\tau')  \right] \d \tau' 
\approx
-\frac{1}{3}\int_{z}^{\infty}\frac{ \d z}{(1+z)^3 H(z)}    \qquad{\rm{for}}\qquad \tau \in \left[0,\tau_{\rm{max}}\right] \,
\label{eq:I2}.
\end{eqnarray}
In the second approximation, we made use of $\Theta_{H} = 3 H$ and converted proper time $\tau$ to redshift  using  $\d\tau  = -  d z/(1+z)/H(z)$.
The exponential of any real number gives a positive real number; hence, the common factor in equation \eqref{eq:focusing_theorem2} is positive.   
Most importantly, the terms in the square brackets vanish at finite time $\mathcal{I}_{2} (\tau) = 1/ {\Theta_{L\rm{ini}} }$ in the limit where  $\Theta_{L\rm{ini}} <0$ since $\mathcal{I}_{2} (\tau) <0$.

\begin{figure}[h]
\centering
\includegraphics[width=70mm,height=50mm] {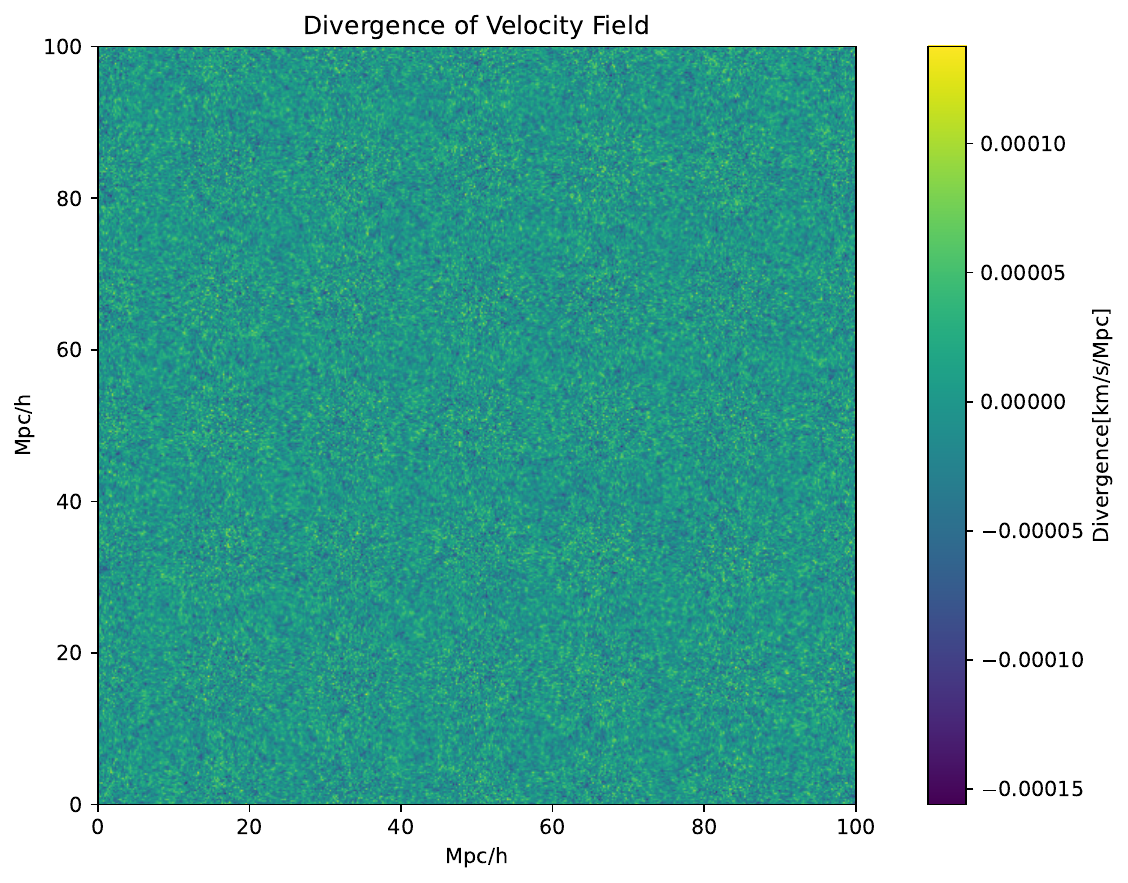}
\includegraphics[width=70mm,height=50mm]{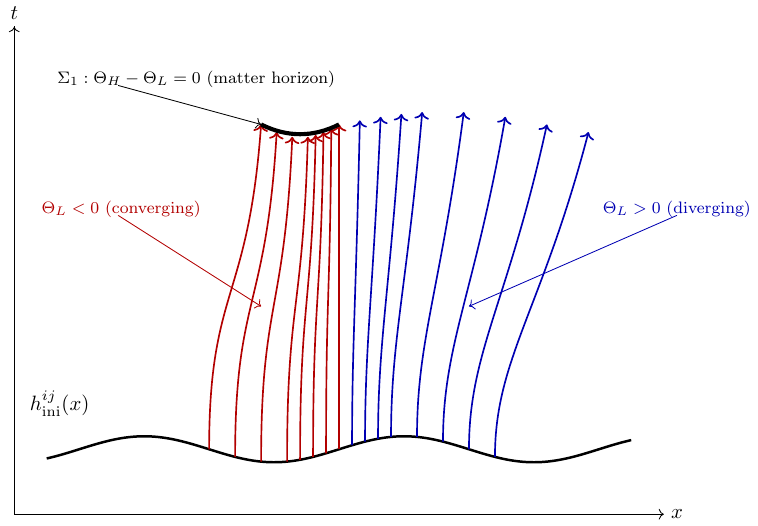}
\caption{
The left panel shows the divergence of the initial relative velocity vector field, it has both negative and positive values. We evaluated the initial data at a redshift of $z =10$.  
Right panel shows two one-parameter families of geodesics emanating from a deformed initial hypersurface $h^{ab}_{\rm{ini}}(x)$. The blue curves denote the geodesics that started out from an initial over-density (initial negative local expansion scalar $\Theta_{L}<0$.)  tend to converge/focus in the future. The red curves denote the geodesics that start out from an under-dense region tend to de-focus $\Theta_{L} >0$.  
For the red curves, there exists a finite time in the future where $\Theta_{H} + \Theta_{L} = 0$. 
} 
\label{fig:initial_diveregcne}
\end{figure}

The initial condition for $\Theta_{H}$ is restricted to positive values, while the initial conditions for $\Theta_{L}$, can be positive or negative. To appreciate this, we estimate $\Theta_{L\rm{ini}}$  following the standard Zeldovich approximation for expressing the initial conditions for the formation of the large-scale structures in the universe~\cite{Zeldovich:1969sb}.  The local expansion is proportial to the divergene of the revlative velocity ${\D}_{i} v^i$,
where $v^i_{\rm{rel}}(t,{\q})  = H(t)f(t)\Psi^i(t,{\q})$ is the relative velocity between the expanding background spacetime and the massive particle.  $\Psi^i$ is the displacement vector field, it is given in terms of the density contrast  $\delta({\k})$: $ \Psi({\k},t_{\rm{ini}}) =  - i {\k} \tilde{\delta}(t_{\rm{ini}},{\k})/{k^2}$. The Fourier transform of the density contrast  $\delta({\k})$ is given by  the matter power spectrum $P(k)$~\cite{Malik:2008im}
$
\tilde{\delta}_{m}({\k}) = \sqrt{P(|{\k}|) } \mu(|{\k}|)\,,
$
where  $P(k)  = \Delta^2 k^{n_s-1} T^2(k)$, where $ \Delta^2$ and $n_s$ are the amplitude and the spectral index of the initial density fluctuation, respectively. These two values are determined by the physics of the early universe.  $T(k)$ is the transfer function, and $\mu(r)$   is white noise, which can be sampled from a Gaussian distribution with zero mean and unit variance. The real space over density is obtained by performing an inverse Fourier transform of $\tilde{\delta}_{m}({\k}) $: $\tilde{\delta}_{m}({\r})  = \mathcal{T}({\r}) \star\mu({\r})$.
The map of the initial local expansion ${\D}_{i} v^i_{\rm{ini}}$  estimated deep into the matter-dominated universe is shown in figure \ref{fig:initial_diveregcne}

To gain more physical intuition into what is going on, we use the  Euler equation to relate the local expansion  $\Theta_{L} = {\D}_{i} v^i$ to the rate of change of the fluctuation in matter density with respect to the coordinate time ${\D}_{i} v^i \sim - \partial \delta_m/\partial t =  -\delta_m'$.
Regions with negative local expansion correspond to regions with a positive rate of density perturbation; they are collapsing regions. Therefore, regions with negative initial local expansion, i.e $\Theta_{L\rm{ini}}$ over time, will grow to more negative values in a finite time $\Theta_{L} \to-\infty $ according to equation \eqref{eq:focusing_theorem2}. 
This needs to be understood in the context of the expanding background spacetime, because the global expansion is positive $\Theta_{H} >0$.
This has consequences for the overall geometry of the manifold since $\Theta = \Theta_{H}  + \Theta_{L}$, hence, there must exist a time and space where a matter horizon forms, i.e $\Theta = 0$ ~\cite{Ellis:2010fr,Umeh:2026ajv}. 
 The matter horizon is a type of dynamical horizon \cite{Ashtekar:2003hk}. They are horizons that emerge due to the local curvature of spacetime. Examples include the apparent horizon for null integral curves. Apparent horizon is well-known in the context of black hole mechanics~\cite{Penrose:1969pc}.   The concept of matter horizon is fairly recent; unlike the apparent horizon,  they are associated with timelike integral curves, hence more relevant in describing the particle nature of matter and structure formation in the universe~\cite{Ellis:2010fr}.  They are dynamical in nature,  that is, they evolve as matter and curvature of spacetime change through merger,  accretion and any other viable physical process~\cite{Ashtekar:2003hk}.

\begin{figure}[h]
\centering
\includegraphics[width=70mm,height=50mm] {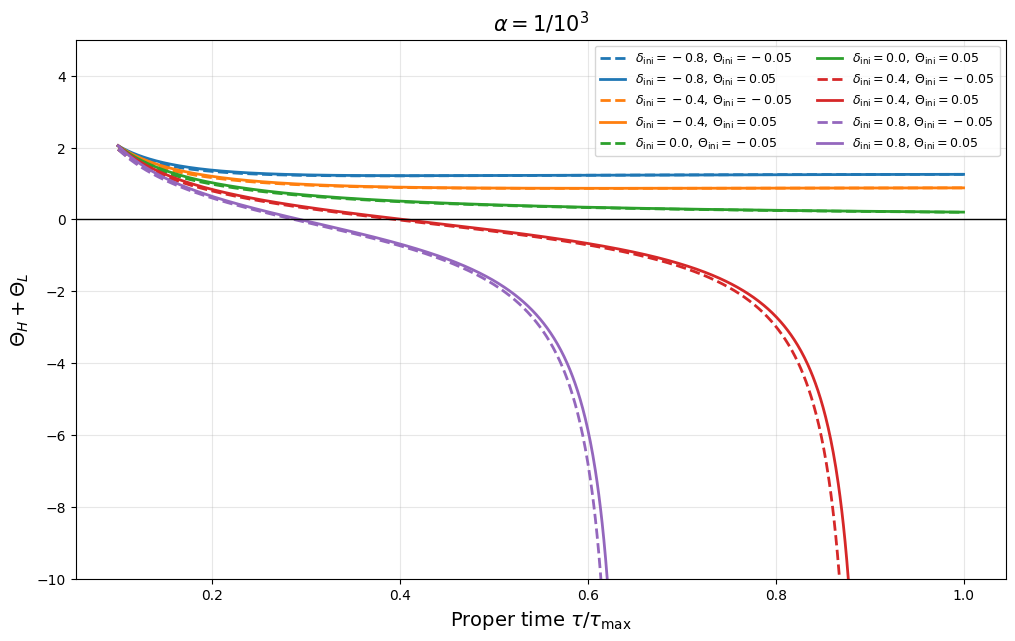}
\includegraphics[width=70mm,height=50mm]{ 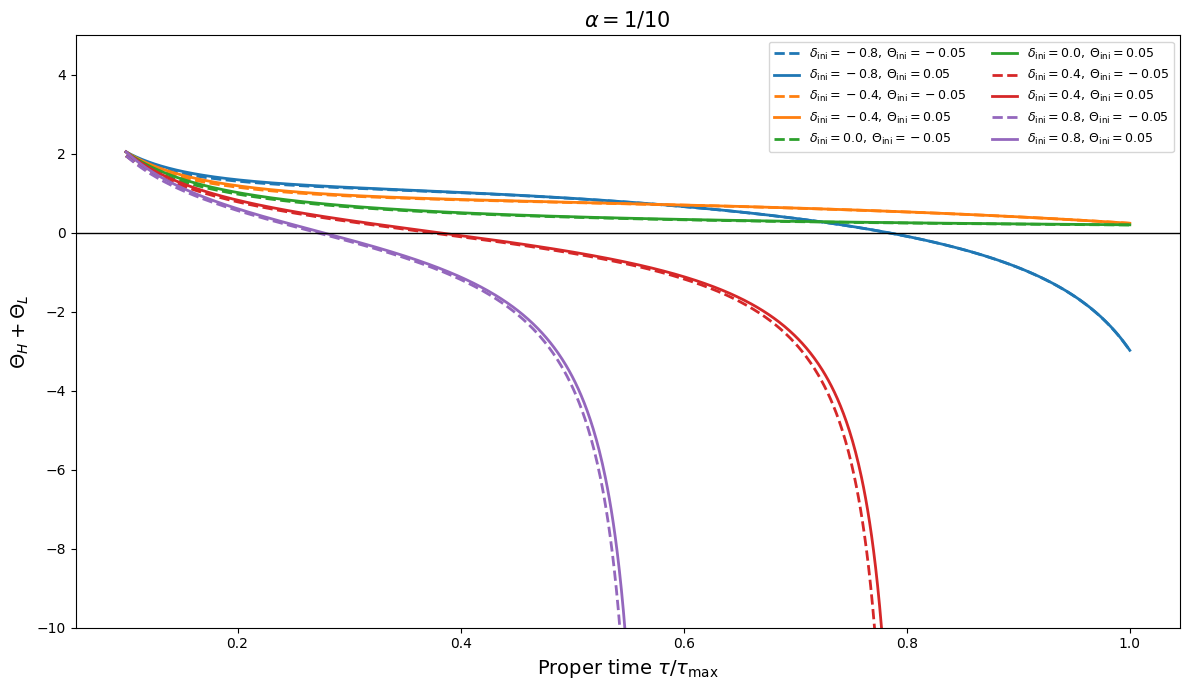}
\includegraphics[width=70mm,height=50mm]{ 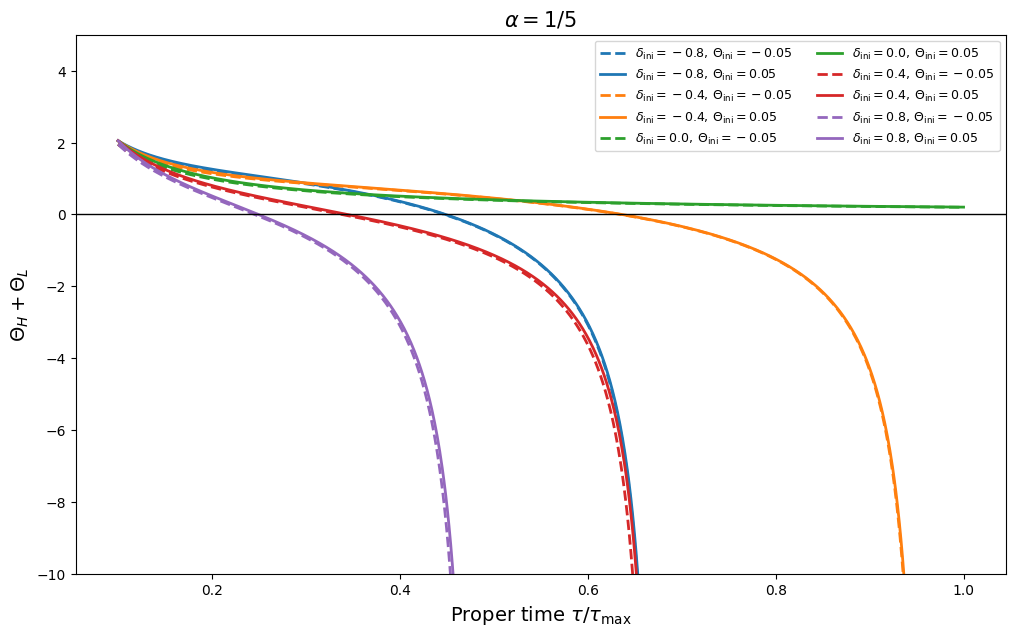}
\includegraphics[width=70mm,height=50mm]{ 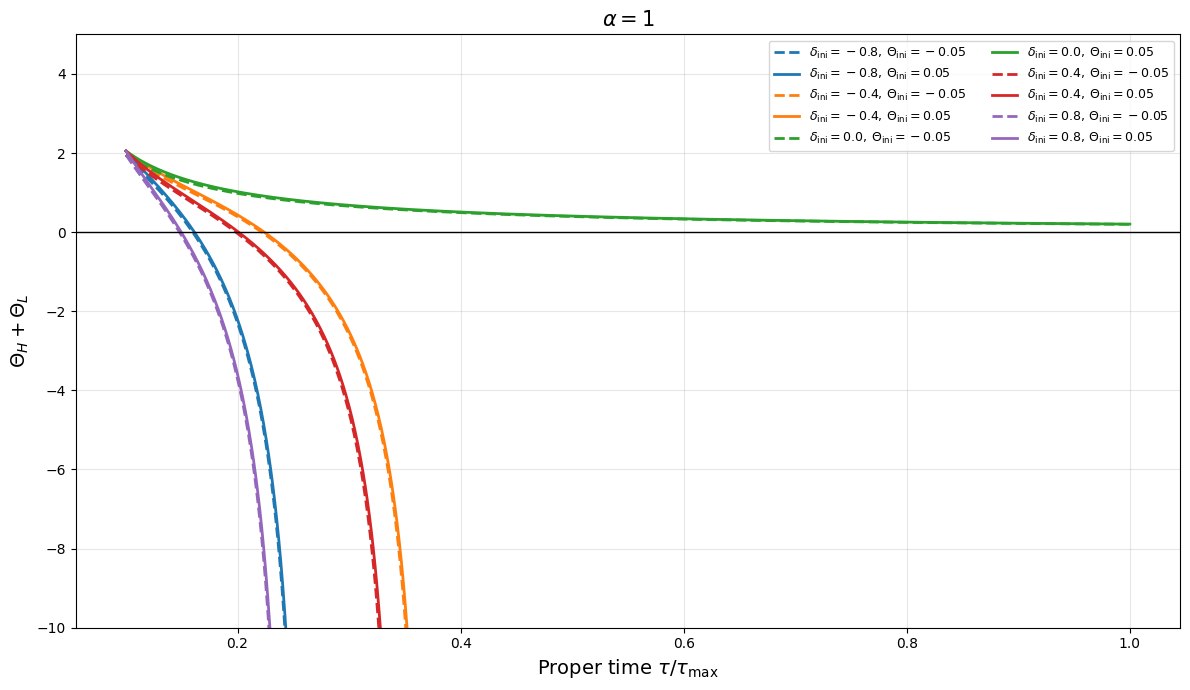}
\caption{The figure shows the plot of the expansion scalar as a function of proper time for various values of the density contrast. The decoupling timescale from the Hubble flow, $\Theta_{H}$, is very sensitive to the initial density contrast. 
} 
\label{fig:matter_horizon}
\end{figure}

We can solve equation \eqref{eq:local-expansion},  in the matter-dominated era within standard cosmology and under some mild assumptions on the nature of the shear scalar contribution. 
In the matter-dominated era, $\Theta_{H} = 3 H = 2/\tau$ and the matter density contrast is given by $\delta_{m} = \delta\rho/\bar{\rho} = \delta_{\rm{ini}}\left( {\tau}/{\tau_{\rm{ini}}}\right)^2$, where $\delta_{\rm{ini}}$ is the initial density contrast. 
If we approximate the shear scalar with ${\sigma}_{ab}{\sigma}^{ab}  =2\alpha \delta_m^2 \Theta^2_{H}/3$ , then equation \eqref{eq:local-expansion} reduces to 
\begin{equation}\label{eq:simplified_ODE}
\frac{ {\rm{D}} {{\Theta_L}} }{{\rm{D}} \tau} 
 =   - \frac{1}{3}{\Theta_L}^2-\frac{2}{3} \Theta_{H}\Theta_{L} - \frac{2}{3}\left(\alpha\delta_{m} \Theta_H\right)^2
-\frac{1}{6}\left({ \Theta^2_{H}} \right)  \delta_{m}  \,,
\end{equation}
where we made use of the Hamilitonian constraint(Friedmann equation): to  relate  the average density to $\Theta_{H}$:  ${ \Theta^2_{H}} =3 \kappa \bar{\rho} $.  $\alpha$ is a parameterisation factor for regulating the amplitude of the shear scale contribution. 
Under this approximation, we can solve equation \eqref{eq:simplified_ODE} numerically for a set of given initial values of $\Theta$ and $\delta_{m}$.  
The plot of $\Theta$ as a function of time is given in figure \ref{fig:matter_horizon} for different values of $\Theta_{\rm{ini}}$ and $\delta_{\rm{ini}}$. 
The timescale of decoupling from the Hubble flow or the timescale for the formation of the matter horizon is mainly determined by the initial matter density contrast and the rate of tidal deformation. 
In the spherically symmetric limit, i.e $\sigma_{ab}\sigma^{ab}\approx 0$($\alpha = 1/1000$), the matter density contrast determines when a local region decouples from the Hubble flow. The regions with large initial density contrast decouple first from the Hubble flow. The under-dense regions remain in the Hubble flow.   These changes in the presense of tidal deformation. The underdense region decouples too from the Hubble flow, but at much later times; however, our modelling of the shear tensor in the underdense region can't be trusted.   The regions with zero density contrast remain in the Hubble flow as expected.


There are many consequences of the matter horizon, i.e $\Theta =0$, in determining the distribution of matter in the universe. One obvious one is that the matter horizon forms first before caustics. This is contrary to the case in Newtonian gravity see equation \eqref{eq:mass_conservation}.  
In \cite{Umeh:2023lbc,Umeh:2026ajv} we showed that the particle trajectory ceases to be geodesic at the matter horizon.  Caustics form when the trajectories that have ceased to be geodesic are extended beyond the matter horizon.  
We reiterate the proof here for completeness. One simple way to see this is to  consider an infinitesimal extension of the geodesic at the matter horizon:
\begin{eqnarray}\label{eq:extension}
\tau  = \tau ' + \Delta\tau \,,
\end{eqnarray}
where  $\tau'  = \tau_{\rm{MH}}$ is the proper time at the matter horizon, i.e $\Theta =0$ hypersurface and $ \Delta\tau$ is a small extension in $\tau$. $\Theta$ is related to the normalised derivative of ${\rm{det}[\bf{\mathcal{J}}]}$ with respect to  the proper time, according to ~\cite{Umeh:2023lbc}
\begin{eqnarray}\label{eq:J_theta}
\frac{1}{{\rm{det}[\bf{\mathcal{J}}]}} \frac{\d {\rm{det}[\bf{\mathcal{J}}]}}{\d\tau} = \Theta\,.
\end{eqnarray}
Putting equation \eqref{eq:J_theta} in  equation  \eqref{eq:expansion} and evalauting at the mattter hgorizon gives
    \begin{eqnarray}\label{eq:translation3}
 \frac{1}{ {\rm{det}[{\mathcal{J}}(\tau)]} } \frac{\d^2 {\rm{det}[ \bf{\mathcal{J}}]}}{\d\tau^2} \bigg|_{\tau =\tau'}&=&   -\left[ {\sigma}_{ab}{\sigma}^{ab} 
  + {R}_{ab} {u}^a {u}^b\right]\,.
 \end{eqnarray}
Now consider a small extension of the Jacobian ${\rm{det}}\left[{\mathcal{J}}(\tau)\right]$, beyond $\tau'$ hypersurface makes caustics formation inevitable  in a finite time~\cite{Umeh:2023lbc}
\begin{eqnarray}\label{eq:Jacobian}
{\rm{det}}\left[{\mathcal{J}}(\tau)\right] \approx{\rm{det}}\left[
 {\mathcal{J}}(\tau')\right] \big[1-   \frac{1}{2}\left[ {\sigma}_{ab}{\sigma}^{ab} 
  + {R}_{ab} {u}^a {u}^b\right](\Delta \tau)^2\big]\,.
\end{eqnarray}
 For $\Delta \tau \neq0$ and imposing the weak energy condition implies that $ R_{ab}  u^{a} u^{b}\ge0$, and noting that the product of the shear tensors $ \sigma_{ab}\sigma^{ab}>0$ is positive definite implies that everything in the square bracket is positive, hence ${\rm{det}}\left[{\mathcal{J}}(\tau)\right]\to 0$. 
 
\subsection{Nature of spacetime geometry around massive particle}\label{subsec:astrophysicalmatter}

The matter horizon defines the limit where the gravitational influence of a gravitationally bound system decouples from an expanding FLRW spacetime.  We made use of the Raychaudhuri equation to estimate the proper time or the characteristic time at which a gravitationally bound system decouples from the Hubble flow.  Within the spherical collapse model, this is sometimes referred to as the turnaround time~\cite{Herrera:2017epn}.  There is also a corresponding notion of turnaround radius, which is defined as the maximum radius reached by a gravitationally collapsing spherical region in an expanding universe~\cite{Faraoni:2021ocg}. It defines a point where the inward pull of gravity is balanced by the outward expansion of the universe. 

One could appreciate this concept beyond the spherical collapse model by considering the spacetime geometry around a gravitationally bound system. The metric tensor of the spacetime geometry is given by Fermi coordinates~\cite{Delva:2011abw}:
\begin{eqnarray}
\d s^{2} = {g}_{ab}^{F}(x_{F}) \d x^a_{F}\d x^b_{F}
= {g}_{00}^{F} (x^{a}_{F})  \d \tau^{2}_{F}+2   {g}^{F}_{0i}(x^{a}_{F}) \d \tau_{F} \d x^{i}_{F}+ {g}_{ij}^{F}(x^{a}_{F})\d x^{i}_{F} \d x^{j}_{F}\,
\end{eqnarray}
where $x^a _{F}$ are the local Fermi coordinates, $ {g}_{ab}^{F}$ is the local Fermi metric tensor, $\tau_{F}$ is the proper time measured by the Fermi observer \cite{Delva:2011abw}. 
The components of the metric tensor are 
\begin{eqnarray}\label{eq:local_metricA}
g_{00}^{F} (x^i_{F}) &=&\left[ -1 - {R}^{F}_{0l0m} \bigg|_{P}x_F^l x_F^m\right]\,,
\\ 
g^{F}_{0i}(x^i_{F})&=&  \left[ -\frac{2}{3} {R}^F_{0lim} \bigg|_{P}x_F^l x_F^m \right]\,,
\\ 
g_{ij}^{F}(x^i_{F})&=& \left[ \delta_{ij} - \frac{1}{3} {R}^F_{iljm} \bigg|_{P} x_F^l x_F^m\right]\,,
\end{eqnarray}
The Riemann tensor is projected onto the hypersurface with the spatial metric $h_{ab}$: 
\begin{figure}
\includegraphics[width=80mm,height=55mm] {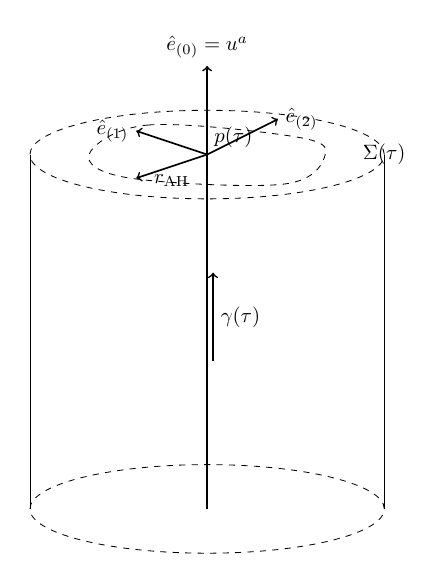}
\caption{We show a typical local geodesic coordinate system. The vertical line at the centre is the central geodesic $\gamma(\tau)$ parametrised by the proper time $\tau$.  At a given hypersurface, $\Sigma$, we have a spatial triad. The time coordinate is synchronised with the observer's proper time along the worldline.  The circle in the middle around the central geodesics denotes the astrophysical matter horizon.   
  } 
\label{fig:FermiCord}
\end{figure}
Using the Gauss decomposition  of the Riemann tensor~\cite{xAct,Gourgoulhon:2007ue} in equation \eqref{eq:local_metricA} and expressing the covariant derivatives of the four-velocity in terms of the irreducible observables gives the metric tensor in terms of the physical observables that an observer can measure
\begin{eqnarray}
 g^{F}_{00}&=&-1-\bigg\{\frac{1}{3} u^b u^dR_{bd} x^c_{F}x^{F}_{c} +E_{lm}  x^{l}_{F} x^{m}_{F}\bigg\}\bigg|_{P}
 \label{eq:g00}
\,,
\\
 g^{F}_{0i} &=&-\frac{2}{3} \bigg\{ \frac{1}{3}\left(x^{l}_{F} x^{F}_{i} {\D}_{l} \Theta - x^l_{F} x^{F}_{l} {\D}_{i}\Theta \right) 
+ x^l_{F} x^m_{F} \left[ {\D}_{m} \sigma_{il} - {\D}_{i} \sigma_{lm}\right] \bigg\} \bigg|_{P}\,,
\label{eq:gio}
\\ \nonumber
 g^{F}_{ij}&=&\delta_{ij} + \bigg\{-\frac{1}{3}R_{iljm}^{(3)}x^l_{F} x^m_{F}  - \frac{1}{9} \bigg[ \sigma_{ij} - \frac{1}{3} h^{F}_{ij} \Theta \bigg] \Theta x^l_{F} x_{l}^{F}+
\frac{1}{27}\Theta^2 x^{F}_{i} x^{F}_{j}
\\ &&
- \frac{1}{3} \bigg[
\sigma_{lm} \sigma_{ij} - \sigma_{il}\sigma_{jm}  \bigg]x^l_{F} x^m_{F} 
+\frac{1}{9}\Theta \bigg[2\sigma_{(il} x^l_{F} x^{F}_{j)} - h^{F}_{ij} \sigma_{lm} x^l_{F} x^m_{F}  
\bigg]\bigg\}\bigg|_{P} \,.
\label{eq:gij}
\end{eqnarray}
where  ${E}_{ab} =  {C}_{acbd} {u}^c{u}^d $ is the electric part of the Weyl tensor.  Our target is to use these equations to show that test particles in the neighbourhood of a gravitationally bound system perceive time differently when compared to a test particle in the Hubble flow. 
To obtain this information, we calculate the time dilation experienced by fiducial observers at various distances from the central geodesic. 
\begin{eqnarray}
\frac{\d \tau_{\rm{obs}}}{\d \tau_{F}}
&\approx&  \sqrt{-g^{F}_{00} } \left(1 - \frac{1}{2}\frac{g^{F}_{0i}}{g^{F}_{00} }\frac{v^i}{c}-\frac{1}{2}\frac{g^{F}_{ij}}{g_{00} } \frac{v^i}{c}  \frac{v^j}{c} +\cdots\right)\approx    \sqrt{-g^{F}_{00} } \left(1 - \mathcal{O}\left(10^{-3}\right)+ \cdots\right)\,,
\\
&\approx& 1+ \frac{1}{6} u^b u^dR_{bd} x^c_{F}x^{F}_{c} +\frac{1}{2}E_{lm}  x^{l}_{F} x^{m}_{F}\,,
\label{eq:time_delay}
\end{eqnarray}
where we have taken a slow-motion limit.  
  The time-time component of the metric tensor receives a contribution from both the  local matter distribution ($u^b u^dR_{bd} x^c_{F}x^{F}_{c})$ and also from the matter elsewhere ($E_{lm}  x^{l}_{F} x^{m}_{F}$). 
 Using the Einstein field equation, we can relate $R_{bd} $ to the matter density  $\rho_F$ and the cosmological constant
 $
  u^b_{F} u^d_{F} R_{bd} \approx\kappa^2\rho_F/{2}  - \Lambda \,. 
$
  The impact of the external matter distributions comes from the electric part of the Weyl tensor:
$ E_{ij}\approx  \left[ \partial_{(i} \partial_{j)} \Phi -\frac{1}{3}\delta_{ij}\partial^2 \Phi \right]
$. We neglect the vector and tensor contributions for simplicity. 
Using the Poisson equation in local coordinates  $\nabla^2 \Phi = \delta \rho_{F}$, equation \eqref{eq:g00} reduces to 
\begin{eqnarray}\label{eq:simpg00}
g^{F}_{00}&=&-\left[1 -{H^2 r_F^2}   \Omega_{\Lambda}\right]- \frac{1}{2}  {H^2 r_F^2}  \Omega_{m}   
-x^{i}_{F} x^{j}_{F}\partial_{(i} \partial_{j)} \Phi \,,
\end{eqnarray}
where $r^2_{F} = x_{i}^{F} x_F^{i} $ is the radial distance from the centricl geodesic,  $\Omega_\Lambda\equiv{\Lambda}/{3{ H}^2} $ and  $\Omega_{ m} = {\kappa^2 \overline{\rho}}/{3{H}^2} $ are matter density parameters.
The last term in equation \eqref{eq:simpg00} can be simplified further by expanding in Fourier space (see Appendix \ref{sec:Fourier})
\begin{eqnarray}
  { x^{i}_{F} x^{j}_{F}}\partial_{(i} \partial_{j)} \Phi  =  - \int \frac{\d^3 k}{(2\pi)^3}\frac{\partial}{\partial k^i} \frac{\partial}{\partial k^j}\left(k_i k_{j}\Phi(k)\right) e^{i{\k}\cdot {\x}} =  \mathcal{F}^{-1} \left[  \mathcal{Z}(k)\Phi(k) \right]\,.
\end{eqnarray}
where $ \mathcal{Z}(k) $ is a scale-dependent weighting function that depends on the quantum mechanics of the early and the physical process of gravitational structure formation
\begin{eqnarray}
 \mathcal{Z}(k) = - \bigg[  \mathcal{Y}  +
 \left(\mathcal{X}  -3\right) \frac{\d \log T(k)}{\d \log k} 
+\frac{1}{ T(k) } \frac{d^2T(k)}{d\log k^2} + \frac{7}{ T(k) } \frac{d T(k)}{d\log k}   \bigg]\,.
\end{eqnarray}
$\mathcal{X} $ and $\mathcal{Y} $ depend on the spectral index and the running of the spectral index of the initial density field. 
To gain more quantitative insights, we consider a scale-invariant initial power spectrum and  a simplified form of the transfer function where all the cosmology dependence drops out~\cite{Umeh:2013UCT}
\begin{equation}\label{eq:Transfer_function_simp}
  T(k) \propto \left\{
  \begin{array}{@{}ll@{}}
    1, \qquad &  k \le k_{\rm{eq}} \\
    \left(\frac{k}{k_{\rm{eq}}}\right)^{-2}, \qquad  &k \ge k_{\rm{eq}}
  \end{array}\right.
  \qquad  \frac{\d \log T(k)}{\d \log k}  \approx -2 \,,
 \qquad
\frac{1}{ T(k) } \frac{d^2T(k)}{d\log k^2}   \approx 6\,,
\end{equation} 
where $k_{\rm{eq}}$ denotes the equality scale. 
In this limit, the log derivatives of the transfer function lead to real numbers and the spectral index is exactly one: $  \mathcal{Y} = \mathcal{X}   = 0 $, hence, $ \mathcal{Z}(k) = 2\,.$

Therefore, every term in $\mathcal{Z}(k) $ is independent of $k$, thus the inverse Fourier transform of $ \mathcal{F}^{-1} \left[  \mathcal{Z}(k)\Phi(k) \right]$  is trivial: $  \mathcal{F}^{-1} \left[  \mathcal{Z}(k)\Phi(k) \right] = \Phi(x^i)$.
Assuming that the universe is populated by point particles, in this limit, the density profile is given by $\delta\rho_F(r) = M \delta^{(3)}(r)$ and the gravitational potential becomes $\Phi = - GM/r $. $M$ is the mass of the particle.
Since the first set of terms in equation \eqref{eq:simpg00} is geometrical (de Sitter term),
The second set is cosmological, and the third term corresponds to the local gravitational potential. 
It is possible to find the radius where the gravitational influence of the local environment equals  the cosmological expansion by setting the cosmological term equal to the gravitational potential
\begin{eqnarray}\label{eq:matter_horizon}
 r_F  \approx \left[\frac{(4 G M)} {\left(\Omega_{m} H^2\right)} \right]^{\frac{1}{3}}\bigg|_{P} \equiv r_{\rm{MH}}\,.
 \end{eqnarray}
 Again, this is obtained by setting the cosmological and the local gravitational potential term to each other and finding the radial distance where the condition holds, that is  $r_{\rm{MH}}$. 
Including the cosmological constant term does not give a real number.   
 In the spherical collapse model in the Newtonian limit to the turnaround radius ~\cite{RoyChoudhury2021,Benisty:2025mfv,Benisty:2025eyc} $
R_{\rm{ta}} = \left[({2 G M})/{\Delta_{\rm{ta}} \Omega_{m}H^2} \right]^{\frac{1}{3}} \,,
$
  indicates that they differ by a factor of  $2/\Delta_{ta}$, where $\Delta_{\rm{ta}}$ is the nonlinear density contrast at turnaround, a number found by solving the spherical collapse dynamics (e.g., $\Delta_{\rm{ta}} \approx 5.55$ in a flat matter-only 
universe)~\cite{Paraskevas:2024ytz}. 
Using the Newtonian gravity  result to estimate the comoving radius where the cosmic expansion velocity $v_{\text{exp}} = H r$ equals the escape velocity $v_{\text{esc}} \approx \sqrt{{2 G M}/{r}}$ from mass $M$ gives  $r_{\rm{esc}} \lesssim \left[ \frac{2 G M}{H^2} \right]^{1/3}$.
 It is well known that the deflection angle in the Newtonian limit differs by a factor of 2  as well from the GR expression \cite{2005physics...8030S}.

The key lesson is that clocks associated with spatially displaced fiducial observers around a massive particle geodesic in an expanding universe tick at different rates depending on the distance from the central geodesics 
\begin{eqnarray}\label{eq:timedilation}
\frac{\d \tau_{\rm{obs}}}{\d \tau_{F}} \approx \sqrt{-g^{F}_{00} }= \eta_{00}
\begin{cases}
    \left(1  +  \frac{1}{2\eta_{00}}\mathcal{F}^{-1} \left[  \mathcal{Z}(k)\Phi(k) \right]\right)  \approx\left[1 - \frac{1}{\eta_{00}}\frac{GM}{r_{F}}\right] \qquad    & \text{for } \quad  r_{F} <r_{\rm{AH}} 
         \\
        1 &   \text{for} \qquad  r_{F}  =r_{\rm{AH}} 
         \\
      \left(1 +\frac{1}{4\eta_{00}} {H^2 r_F^2} \left[ \Omega_{m} \right]\right)  & \text{for } \quad \,\,  r_{\rm{AH}} <r_{F} <r_{\HH}
      \\
\end{cases}
\end{eqnarray}
where $\eta_{00} =\left[1 -{ r_F^2}/\ell_{\rm{dS}}^2\right]$ with $\ell_{\rm{dS}} = 1/H \sqrt{\Omega_{\Lambda}}$ is the de Sitter length.
 In the region $ r_{F} <r_{\rm{AH}} $, the tick rate of the fiducial observer's clock  tick much slower when compared to the Minkowski or de Sitter observer.
 The fiducial observer clock in $r_{F}  =r_{\rm{AH}} $ region ticks at the same rate as in a static patch of the de Sitter spacetime.  While in  $r_{\rm{AH}} <r_{F} <r_{\HH}$, we have an expanding FLRW spacetime.

Finally, in the neighbourhood of the matter horizon, the metric of the spacetime taken the following form 
\begin{eqnarray}\label{eq:maximal_nearby}
 g^{F}_{00}&=&-1
\,,
\qquad
 g^{F}_{0i}
\approx \mathcal{B}_{i}\,,
\qquad
 g^{F}_{ij}
 \approx \delta_{ij} -\frac{1}{3}R_{iljm}^{(3)} x^l_{F} x^m_{F}  - H_{ij}\,.
 \label{eq:bhij}
\end{eqnarray}
where we have defined the vector and tensor contribution
\begin{eqnarray}\label{eq:vectors}
 \mathcal{B}_{i}   &=&- \frac{2}{3}  \bigg[ \frac{1}{3}\left[x^{l}_{F} x^{F}_{i} {\D}_{l} \Theta - x^l_{F} x^{F}_{l} {\D}_{i}\Theta \right]  \bigg]\bigg|_{P} 
\\
H_{ij}&=&  \frac{1}{3}\bigg[
\sigma_{lm} \sigma_{ij} - \sigma_{il}\sigma_{jm}  \bigg]x^l_{F} x^m_{F} \bigg|_{P} 
\end{eqnarray}
The non-vanishing tensor perturbation $H_{ij}$ is given by the shear deformation tensor, which is proportional to the integral of the electric part of the Weyl tensor. The electric part of the  Weyl tensor is related to the gravitational wave~\cite{Goswami:2019fyk}.
Although $\Theta$ vanishes at the boundary, the spatial derivatives of $\Theta$;  $x_{F}^l{\D}_{l}\Theta = x^l{\D}_{l} {\D}_{j} v^j$ do not,  It indicates contributions from the vector perturbations~\cite{Barrow:2006ch}. 
The contribution of the projected Riemann curvature tensor $\mathcal{R}^{(3)}_{iljm} $, that is, the intrinsic curvature of the hypersurface; its components are locally determined by the geometry or metric tensor of the hypersurface or by the topological fields living on the boundary~\cite{Witten:2010cx}.

In addition to estimating the finite extent of a gravitationally bound system using Fermi coordinates, it is possible to arrive at the same decoupling argument using a coordinate-independent approach that mirrors the $1+3$ decompcsition formalism for the timelike four-vector discussed in sub-section \ref{sec:matterhorizon}. This approach is sometimes discussed on the $1+1+2$ covariant decomposition formalism in some limit~\cite{Clarkson:2007yp}.  In this, we define a spacelike 4-vector $r^{a} = \d x^{a}/\d \lambda$, so that the projected covariant derivative  can be decomposed into irreducible units as
\begin{eqnarray}\label{eq:decomposera}
\nabla_{a} r_{{b}} =  r_{a}\tilde{A}_{b} + \gamma_a{}^c\gamma_b{}^d\nabla_c r_d
=r_{a}\tilde{A}_{b} + \tfrac{1}{3}\gamma_{ab}\,\tilde\Theta 
+ \tilde\sigma_{ab} + \tilde\omega_{ab},
\end{eqnarray}
where $\gamma_{ab}$ is the metric on the hyper surface, 
 $\tilde\Theta \equiv \gamma^{ab}\nabla_a r_b$ is the (projected) expansion,
 $\tilde\sigma_{ab} = \tilde\sigma_{\langle ab\rangle} =  \gamma_{\<a}{}^c\gamma_{b\>}{}^d\nabla_c r_d$ is the symmetric tracefree shear,
$\tilde\omega_{ab}=\gamma_a{}^c\gamma_b{}^d\nabla_{[c}r_{d]}$ is the antisymmetric vorticity. where $r^c\nabla_c r_b$ is the acceleration of the congruence $\tilde{A}_b \equiv r^c\nabla_c r_b, $ it is orthogonal to $r^a$:  $\tilde{A}_b r^b=0$.  Using the Ricci identity, the propagation equations for $\tilde\Theta$, $\tilde\sigma_{ab} $ and $\tilde\omega_{ab}$ can be derived, and they have a similar structure as equations \eqref{eq:expansion}, \eqref{eq:shear} and \eqref{eq:vorticity} respectively. 

However, just as in the case for the timelike geodesic, we are interested mainly in the propagation equation   $\tilde{\Theta} $, which we can quantify within standard cosmology using the perturbed FLRW spacetime given in \eqref{eq:metric_tensor}.  We assumed that $r^{a}$ is geodesic , i.e $r^c\nabla_c r_b = 0$. On the FLRW background spacetime, we have $\bar{r}^a = (0, \bar{r}^i)$.  Spacelike geodesics have geometric meaning as the straightest paths connecting two spacelike separated points in curved spacetime; however, they have limited physical meaning in terms of direct observable motion because traversing them would require travelling faster than the speed of light  $(\Delta r>c\Delta t$). 
For concreteness, we consider spacelike geodesics as paths following test particles orbiting a more massive barycenter. 
In this case, they represent the paths along which physical distances between comoving objects (like galaxies in a cluster or stars in a galaxy) are measured.

Imposing the spacelike normalisation  condition ${r}^a {r}_a = 1$, we find  on the background spacetime that $\bar{r}_a \bar{r}^a = a^2 \delta_{ij} \bar{n}^i \bar{n}^j = 1$, where $\bar{n}^i $ is a spacelike unit vector . Therefore, it is convient to  express $\bar{r}^i$  as a function of  $n^i$: $\bar{r}^i = a^{-1} n^i$. At the perturbed order, we have $r^a = \bar{r}^a + \delta r^a = (0 + \delta r^0,\, \bar{r}^i + \delta r^i).$
Using the geodesic equation, we can find $ \delta r^0$ and $\delta r^i$
\begin{eqnarray}
\delta r^0& =&  \frac{1}{a} \int \left( 2 \HH  (\Phi + \Psi) +  \Phi' \right) \d R\,,
\qquad
\delta r^k =   \frac{1}{a} \int \left( 2 \bar{r}^k (\bar{{r}}^{i} \partial_{i} \Phi) -  \partial^k \Phi \right) \d R
\end{eqnarray}
where we set $ \delta r^0_{\text{initial}} = \delta r^k_{\text{initial}} = 0$ and $ \d R =  \d \lambda/a$..   The normalisation condition gives  $ \bar{r}_i \delta r^i = -\Phi$,  Focusing on the leading order contribution to $\tilde{\Theta}  $, that is  $\tilde{\Theta} \approx \bar{\tilde{\Theta}} + \delta \tilde{\Theta}$, where 
$ \bar{\tilde{\Theta}}_{\pm}=\nabla_{a}\bar{r}^{a} = \partial_{0} \bar{r}^{0}+\partial_{i} \bar{r}^{i}+ 3 H \bar{r}^{0}= 3/R$, with define $\bar{r}^{i} =( x//R, y/R, z/R)$, where $R$ is radially constant: $R =ar = a \sqrt{x^{2} + y^{2} +x^{3}}$ on the FLRW background spacetime.
The leading order approximation of $\tilde{\Theta} =  \gamma^{ab}\nabla_a r_b  =\left[ g^{ab}- r^{a}r^{b}\right] \nabla_a r_b$ is given by.
\begin{eqnarray}
\delta \tilde{\Theta} &=& \partial_{0}\left(\delta r^{0}\right)+ 3 \HH \delta r^{0} + \partial_{i}\left(\delta r^{i}\right) + \bar{r}^{i}\partial_{i}\left( \Phi - 3\Psi\right) \,.
\end{eqnarray}
We focus on the leading order contribution, that is  $\delta \tilde{\Theta}  \approx  \partial_{i}\left(\delta r^{i}\right) + \bar{r}^{i}\partial_{i}\left( \Phi - 3\Psi\right) $ and further decompose $\delta r^{i}$ into radial and orthogonal components $\delta r^k = n^{k} \delta r_{||} + \delta r^k_{\bot }$, where 
\begin{eqnarray}
\delta r_{||}  =  \frac{1}{a}   \int \left(  \bar{{r}}^{i} \partial_{i} \Phi  \right) \d R\,, \qquad 
 \delta r^k_{\bot }   =  -  \frac{1}{a}  \int \left(  \partial^k_{\bot} \Phi \right) \d R\,.
\end{eqnarray}
Putting this together in the limit of vanishing anisotropic stress $\Phi = \Psi$, we find 
\begin{eqnarray}\label{eq:thetaFRW}
\tilde{\Theta} &\approx &
\frac{3}{R}- \frac{\d \Phi}{\d r} - \int_{0}^{R}\frac{r}{R} \nabla^2_{\bot }{ \Phi }\d r
\,,
\end{eqnarray}
where the term $-{d\Phi}/{dr}$ is the magnitude of the gravitational acceleration, $a_g(r) = - {d\Phi}/{dr}$ in the point source limit.  Assuming regularity at the centre for simplicity,  ${\d\Phi}/{\d r} $  can be expressed in terms of the enclosed mass, $M(<R)$:
$ {\d\Phi}/{\d r} = {G  M(<R)}/{r^{2}} 
$. 
The gravitational acceleration and the orbital speed $v_c$ are related according to   that is $a_g = v_c^2/r$, we find $v_c(R) = \sqrt{{G M(<R)}/{R}}$.
For an ensemble of particles with velocities $v_{i}$
the velocity dispersion is  $\sigma_{v}^2 = \<\left( v- \<v\>\right)\>^{2}$ , where $\<v\>$ is the  mean bulk velocity (e.g., Hubble flow + coherent  motions).  In general, $\sigma_{v}^2 $  measures the internal random motions created by gravitational collapse.
For an isotropic system,  $v_{c}^2$ is the sum of the variances in 3 directions:
 $v_{c}^2 = \sigma_x^2 + \sigma_y^2 + \sigma_z^2 = 3\sigma_{v}^2$

\begin{figure}
\includegraphics[width=70mm,height=60mm] {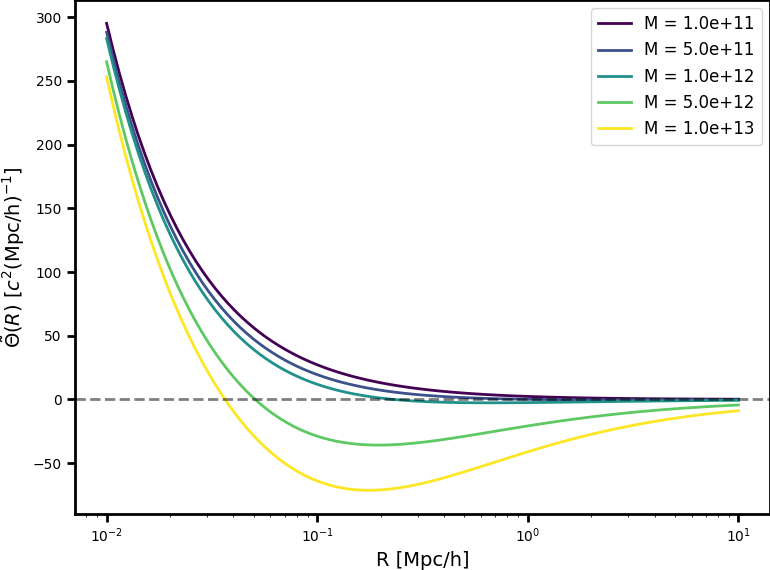}
\includegraphics[width=70mm,height=60mm] {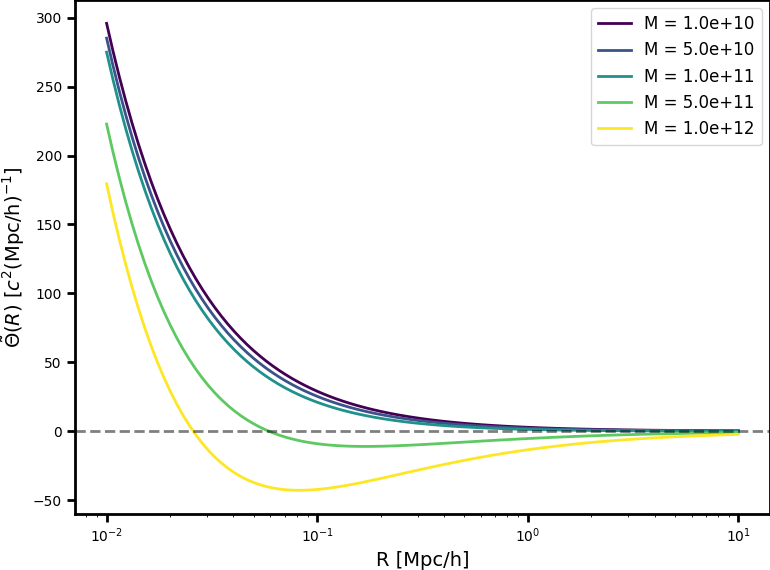}
\caption{\label{fig:local_group}  
We show the convergence of the spacelike geodesics as a function of the physical distance from the centre of a gravitational system. Here $\sigma_{v}^{2 } = 200$[ km/s].  We made use of the NFW halo density profile.  The left panel was estimated at $z =0.5$, while the right panel was estimated at $z=10.5$.}
\end{figure}

The second term in the integrand in  equation \eqref{eq:thetaFRW} can be expressed in terms of the projected mass density field using  a projected Poisson equation, $ \nabla^2_{\bot} \Phi = \Sigma$~\cite{DES:2018kma}, where  $ \Sigma$
 is the projected mass density field. 
 The projected mass density, $\Sigma(R)$, for the NFW profile is given by~\cite{Navarro:1995iw,DES:2018kma} 
$
 \Sigma(R) = 2 \int_{R}^{\infty} \frac{\rho(r) r}{\sqrt{r^2 - R^2}} \d r\,,
$
 where $\rho(r)$ is the 3D NFW density profile: $\rho(r) = {\rho_0}/\left({(r/r_s) (1 + r/r_s)^2}\right)$. 
 Putting all these together, we find that $\tilde{\Theta}$  may be expressed in terms of the projected mass density  according to
\begin{eqnarray}\label{eq:theta_Sigma}
\tilde{\Theta} (R)  \approx  \frac{3}{R}\left[1 +\frac{\sigma_{v}^2}{c^{2}}- \frac{1}{3 c^{2}}\int_{0}^{R} \d  r '{r'} \Sigma( r')\right]\,.
\end{eqnarray}
$\tilde{\Theta} $ vanishes at a finite distance $R_{\star}$ when
$
\frac{1}{c^{2}} \int_{0}^{R_{\star}} \d  r' {r' }\Sigma( r')  = 3\left(1 +\frac{\sigma_{v}^2}{c^{2}}\right)
$.
The  dependence of $\tilde{\Theta} $ on $R$ is shown in figure \ref{fig:local_group}.  The radial  distance where $\tilde{\Theta}  = 0$  is proportional to the astrophysical matter horizon $R_{\star} = r_{\rm{AM}}$. 
A one-parameter family of spacelike geodesics ceases to be geodesic at $\tilde{\Theta} = 0  $. The equivalent discussion in sub-section \ref{sec:matterhorizon} for the timelike geodesics applies in the case of the spacelike geodesics as well. This can easily be seen by using the Ricci identity to derive the corresponding Raychaudhuri equation 
 \begin{eqnarray}\label{eq:Raychaudhurieqn}
 \frac{ {\rm{D}} {\tilde{\Theta}} }{{\rm{D}} \lambda} &=& - \frac{1}{3}{\tilde{\Theta}}^2 - \tilde{\sigma}_{ab}\tilde{\sigma}^{ab} 
 - {R}_{ab} {r}^a {r}^b\,,
 \end{eqnarray}
 The one-parameter family of spacelike geodesics with initial condition at $0$ cannot be extended beyond $R_{{\star}}$.  This can be seen by performing an infinitesimal extension of the spacelike geodesics beyond $R_{{\star}}$: $
 R = R_{{\star}} +\Delta R
$.
 Extending the volume of the hypersurface, $V_{T}(R)$, beyond  $R_{{\star}} $ givesn
 \begin{eqnarray}
 V_{T}(R)
 &=& V_{T}(R_{\star})  \left[ 1 - \frac{1}{2} \left(\tilde{\sigma}_{ab}\tilde{\sigma}^{ab} 
 + {R}_{ab} {r}^a {r}^b\right)  \left(\Delta R\right)^{2}  \right] 
  \end{eqnarray}
 where we made use of 
  \begin{eqnarray}
  \frac{1}{ V_{T}} \frac{d V_{T}}{d\lambda} =  \frac{1}{2} \gamma^{ab} \frac{\d \gamma_{ab}}{\d \lambda} = \tilde{\Theta}
   \end{eqnarray}
   and equation \eqref{eq:Raychaudhurieqn}.  The shear scalar $\tilde{\sigma}_{ab}\tilde{\sigma}^{ab} $ is positive definition and using the Einstein field equation, we find that the contraction of the Ricci tensor with the spacelike 4 vecor is given by $ R_{ab}r^{a}r^{b}  = \kappa\left[  T_{ab}r^{a}r^{b} +  {1 \over 2} T\right]-3\Lambda$.  Assuming the dust energy-momentum tensor or the large N limit of   \eqref{eq:EMT_particles} and vanishing cosmological constant, we find that ${R}_{ab} {r}^a {r}^b >0$. This implies that  $V_{T} \to 0$ at finite $R$ when the spacelike geodesic is extended beyond $R_{\star}$.  Irrespective of the composition of the energy-momentum tensor, provided $\tilde{\sigma}_{ab}\tilde{\sigma}^{ab}  > {R}_{ab} {r}^a {r}^b >0$, caustics is inevitable when the spacelike geodesic is extended beyond the matter horizon.  This shows that  the  local spacetime region is bounded.

\section{Global dynamics from local geodesic domains}\label{sec:FLRW_discrete_symmetry}

In Sections \ref{sec:matterhorizon} and \ref{subsec:astrophysicalmatter}, we demonstrated that the matter horizon decouples local sub-regions from the Hubble flow. In this section, we glue these decoupled regions to the Hubble flow, ensuring a continuous geodesic evolution from early times to the present. Building on \cite{Umeh:2026ajv}, where we showed that a matter horizon implies a multi-sheeted spacetime, we model this transition as follows: A one-parameter family of massive particles originates on an expanding manifold $\mathcal{M}^{+}$ (metric $g_{ab}^{+}$) at $\tau_{\rm{ini}}$. These particles follow geodesics until $\tau_{\star}$, where the expansion scalar $\Theta$ vanishes and the region decouples. Because the description within $\mathcal{M}^{+}$ becomes invalid for the subsequent collapsing phase ($\Theta < 0$), we excise the domain $\Theta \leq 0$ from $\mathcal{M}^{+}$ and replace it with a manifold $\mathcal{M}^{-}$ of opposite orientation. This gluing construction allows the geodesics to extend continuously beyond $\tau_{\star}$.
%
The sub-manifolds of ${\cal M}^{+}$ and ${\cal M}^{-}$ are decomposed as 
\begin{equation}
\partial\mathcal{M}_{+} = \Sigma_{+ \rm{ini}} \cup \Sigma_{+ } \cup B_{+}  \,. \qquad \qquad 
 \partial\mathcal{M}_{-} = \Sigma_{- } \cup \Sigma_{- \rm{final} } \cup B_{-}\,.
\end{equation}
where $\Sigma_{+ \rm{ini}}$ is initial past spacelike boundary of $\mathcal{M}_{+}$, it denotes the starting point for evolution in $\mathcal{M}_{+}$, $\Sigma_{+}$ is the shared spacelike boundary, it is a minmal hypersurface defined by the vanishing of  $\Theta=0$.  
$\Sigma_{- \rm{final}}$ is the final spacelike boundary of $\mathcal{M}_{-}$.
 Furthermore, the timelike hypersurfaces define the spatial extent of the decoupled sub-region: $B_{+}$ is the timelike boundary of the particle sub-region in  $\mathcal{M}_{+}$. It extends from $\Sigma_{+ \rm{ini}}$ to $\Sigma_{+}$.  It is also a maximal hypersurface defined by the vanishing of $\tilde{\Theta} =0$.
Similarly, $B_{-}$ denotes the timelike boundary of $\mathcal{M}_{-}$. This surface extends from $\Sigma_{-}$ to $\Sigma_{- \rm{final}}$. 
See the left panel of figure \ref{fig:spacetime_diagram} for a schematic illustration. 
In general, the physical universe is described by a union of orientation-preserving manifolds and the boundary, $\mathcal{N}$: $\mathcal{M} =  (\mathcal{M}_{+}\setminus \mathcal{D})\cup_{\phi} \mathcal{M}_{-}  $, where $\mathcal{D}$ is the excised region.

  For concretness, the coordinates in the neighbourhood of a point $p_{+} \in M^{+}$ near the boundary are related to the coordinates around a point $p_{-} \in M^{-}$,  via   diffeomorphism maps, $\phi = \phi_{\Sigma_{\pm}} \;\cup\; \phi_{B_{\pm}}:$  that  reverse coordinate orientation
\begin{eqnarray}\label{eq:discrte_map}
x^a_{+}   = {-x}^a_{-} +X^a(x) \,.
\end{eqnarray}
  The orientation reversing map in equation \eqref{eq:discrte_map} includes coordinate time reversal (T) and spatial coordinate reflection (P).
 Most importantly, the equations of general relativity and Newtonian gravity
are invariant under the discrete transformations of coordinates(i.e equation \eqref{eq:discrte_map}).   In Newtonian gravitational theory, time reversal $( t  \to  - t )$ changes the direction of velocities $( v^{i} \to  - v^{i})$ but keeps accelerations unchanged: $a^i \to a^i$ (i.e particles fall at the same rate irrespective of time orientation).  
The spatial reflection $x^i \to  - x^i$ keeps the Poisson equation unchanged; hence, T and P independently leave the equations of motion unchanged in Newtonian gravity. 
A combination of time reversal and parity, form PT, which is a symmetry of GR. When the geometry is coupled to the external matter fields, the PT symmetry might be enhanced to include charge conjugation, leading to  CPT symmetry (PT symmetry plus charge conjugation, (C) ) is a fundamental symmetry of the standard model of particle physics~\cite{Luders:1957bpq,Colladay:1996iz}.

For more details on how the orientation reversing diffeomorphism helps to avoid the gravitational focusing singularity, see \cite{Umeh:2026ajv}.  We provide a brief discussion of this for completeness. 
For any two instants $ t_1 $ and $ t_2 $, there are two possible directions(forward and backwards) of flow of coordinate time. The forward direction is usually chosen in standard cosmology, even when the backwards direction is equally likely.  Here, we consider the two directions of flow:
\begin{eqnarray} \label{eq:time_direction}
   &&\mathcal{M}_{+}\qquad {\rm{ Forward:}} \qquad  t_1 \to t_2 \qquad {\rm{ if}} \qquad \qquad \Delta t  = t_2-t_1>0  \qquad {\rm{for}}\qquad  t :-\infty \to \infty, \\ \nonumber
   && \mathcal{M}_{-}\qquad { \rm{  Backward: }}  \qquad t_1 \to t_2\qquad{ \rm{if}}   \qquad \qquad \Delta t =t_2-t_1<0  \qquad   {\rm{for }}\qquad  t: \infty \to -\infty\,.
\end{eqnarray}
We chose the direction of flow of coordinate time in $\mathcal{M}_{+}$ and $\mathcal{M}_{-}$ to be opposite to each other. 
A similar definition exists for the spatial separation; however, special care is required to handle the topology~\cite{Kumar:2023ctp,Gaztanaga:2025awe}.  
  The most important feature is that applying the orientation-reversing transformation  does not reverse the flow of proper time, 
\begin{eqnarray}\label{eq:propertime}
\d \tau_{\pm} = \pm \sqrt{\d s^2_{\pm}} = \pm \sqrt{-g^{\pm}_{ab} \frac{\d x^a_{\pm} }{\d \lambda_{\pm}}\frac{\d x^b_{\pm} }{\d \lambda_{\pm}}}\,,
\end{eqnarray}
where $\lambda_{\pm}$ is a parameter, $\d s^2_{\pm}$ is an infinitesimally small distance along the worldline, $g^{\pm}_{ab}$ is is the spacetime metric, $x^a_{\pm}(\lambda_{\pm})$  is a parametrisation of the worldline and $\pm$ is a consequnces of the Fundamental theorem of algebra. $+$ indicates forward flow of coordinate time, while $-$ indicates backward flow of coordinate time. 
$\tau_{\pm} $ is the time measured by a clock that moves along with an object and not $t$. The flow of $\tau_{\pm} $ is asymmetrical. 
Similarly, on any spatial slice, one must make a choice of spatial flow direction: $\Delta x^i = x^i_2-x^i_1$, $\Delta x^i>0$ indicating a choice of forward or positive direction along each coordinate axis and $\Delta x^i<0$ indicating the opposite flow direction. The proper length becomes $ \d R_{\pm}  = \pm \sqrt{g^{\pm}_{ij} \d x^i_{\pm} \d x^j_{\pm}}$, the $\pm$ must be chosen so that the proper length remains positive irrespective of the direction of flow.

\subsection{Metric tensor of the sub-region} \label{sub:metric_tensor}

Since the standard model of cosmology already describes the universe so well on large scales, it is consistent to assume that the metric tensor in $\mathcal{M}_{+}$  takes the form of a perturbed FLRW spacetime. Without loss of generality, we consider  Poisson(conformal Newtonian) gauge for $g_{+}$~\cite{Umeh:2010pr}
\begin{eqnarray}\label{eq:metric}
\d  s^2_{+} &=&a^2_{+}\left[-(1 + 2\Phi_{+})\d \eta^2 _{+}+ \omega_{+i} \d\eta \d x^i_{+} + \left((1-2 \Psi_{+})\delta_{+i j} + \chi_{+ij} \right)\d x^{i}_{+}\d x^{j}_{+}\right]\,.
\end{eqnarray}
where $\delta_{+ij}$ is the spatial metric of the flat background spacetime, with covariant derivative $\bar{\nabla}_{+i}$,  $\Phi_{+}$ and $\Psi_{+}$ are scalar potentials, $\omega_{+i}$ and $\chi_{+ij}$ are vector and tensor perturbations.  The conformal time, $\eta_{+}$ is related to the coordinate time, $t_{+}$, according to $ \d t_{+}= a_{+}\d \eta_{+} $, where $a_{+}$ is the scale factor. We neglect both tensor and vector perturbations for simplicity.
The four-velocity is given by $[u^0_{+}, u^i_{+}] = [1 - \Phi_{+} ,{ \partial}^{i}v_{+}]$, where $v_{+}$ is the velocity potential.

The coordinate time is related to the proper time according to $ {\d\tau_{+}}/{\d t_{+}} = \sqrt{-g^{+}_{00}}\approx a_{+}(1+ \Phi_{+}) $, where we considered the slow-motion approximation,   The metric on the spacelike hypersurface of constant proper time 
$\Sigma: \tau_{+}(t_{+},x^i_{+}) =$ const, i.e $h_{ab} =g_{ab} + u_a u_b$  becomes
\begin{eqnarray}\label{eq:spatialmatric}
\d s^2_{+}\big|_{\Sigma_{+}} \approx a^2_{+}(\tau) \left[ \big(1-2 \Psi_{+}\big)\delta_{ij}  \right]{\d }x^{i}_{+}
{\d} x^{j}_{+}\,,
\end{eqnarray}
Therefore, the metric on  hypersurfaces   is conformally Euclidean $h^{+}_{ab} = \Omega_{+}^2 \delta_{ab}^{+}$, where the conformal factor is given $\Omega^2_{+} =a^2_{+}  \big(1-2 \Psi_{+}\big)\ $, 
 The spacetime metric tensor in $\mathcal{M}_{-}$ has the same structure as the metric in equation \eqref{eq:metric} with  $+\to -$, but the values of the components are different after the coordinate re-mapping(equation \eqref{eq:discrte_map}). 
  \begin{eqnarray}\label{eq:metric_inside}
\d  s^2_{-} &=&a^2_{-}\left[-(1 + 2\Phi_{-})\d \eta^2 _{-}+ \omega_{-i} \d\eta \d x^i_{-} + \left((1-2 \Psi_{-})\delta_{-i j} + \chi_{-ij} \right)\d x^{i}_{-}\d x^{j}_{-}\right]\,.
\end{eqnarray}
The essential difference is that the evolution of the scale factor and other perturbed fields is different since the loca curvature is different.
  Therefore, the metric on the hypersurface $\Sigma_{-}$ is given by   $ h^{-}_{ab} = \Omega_{-}^2 \delta_{ab}^{-} $, 
where $\Omega^2_{-} =a^2_{-}  \big(1-2 \Psi_{-}\big)\ $  
Again, the key difference between the metric tensor associated with  $\Sigma_{+}$  and $\Sigma_{-}$  is that the expansion rates differ because the matter content and local curvature $\mathcal{M}_{+}$  and $\mathcal{M}_{-}$ are different.

Since the conformal Euclidean space  has no matter horizon ($\Theta$ vanish exactly on all scales), the metrics are the same $ \delta_{ab}^{-} = \delta_{ab}^{+}$, which then implies that 
\begin{eqnarray}\label{eq:conformalmap_space}
h^{-}_{ab}  =  \frac{ \Omega_{-}^2}{ \Omega_{+}^2}h^{+}_{ab}  =  \Omega^2_{S} h^{+}_{ab} \,,
\end{eqnarray}
where $\Omega_{S} = { \Omega_{-}}/{ \Omega_{+}}$ is the ratio of the effective scale factor in the universe moving backwards in time to the scale factor of the universe moving forward in time.

 On the surface of constant proper radius $\d R /\d r = \sqrt{g_{rr}}$, the metric on the hypersurface is given by 
 $\gamma_{ab} = g_{ab} - r_{a}r_{b}$, where $r^a$ is a spacelike vector.  Note that $\gamma_{ab}$ is Minkowski 
\begin{eqnarray}\label{eq:metrictimehypersurface}
 d s^2\big|_{B_{+}} \approx  a^2_{+} (\tau)\left[ -(1 + 2\Phi_{+} ) \d\eta^2_{+}  + (1 - 2\Psi_{+} ) r^2_{+}  \d\Omega^2_{+}   \right]
\end{eqnarray}
  where  $ \d\Omega^2_{+} = \d\theta^2_{+}  + \sin^2\theta_{+}  \, \d\phi^2_{+}  $ is the standard metric on the unit 2-sphere. in this case, the metric on the hypersurface of constant proper radius is conformal to a perturbed Minkowski spacetime: $\gamma^{+}_{ab} = a_{+}^2 \gamma^{\rm{Min}}_{+ ab}$,  and  $ \gamma^{-}_{ab} = a_{-}^2 \gamma^{\rm{Min}}_{-ab}
$,  The background expansion vanishes in Minkowski spacetime, hence $\gamma^{\rm{Min}}_{+ab}=  \gamma^{\rm{Min}}_{-ab}$ leading to 
\begin{eqnarray}\label{eq:conformalmap_time}
\gamma^{-}_{ab}  =  \frac{ a_{-}^2}{ a_{+}^2}\gamma^{+}_{ab}  =  \Omega^2_{T} \gamma^{+}_{ab} \,,
\end{eqnarray}
where  $\Omega_{T} = { a_{-}}/{ a_{+}} $. Given these metric tensors, the observables such as the expansion scalar, $\Theta $ will have the same structure in both manifolds $ \mathcal{M}_{+} $ and $ \mathcal{M}_{-}$ 
\begin{equation}\label{eq:expanding}
\text{Expansion} \quad \Longrightarrow \quad
\left\{
\begin{array}{ll}
\Theta^{+}  =\frac{3}{a_{+}}\frac{ \d a_{+}}{\d t_{+} }-\frac{ \d \delta_{m+}}{\d t_{+} }  > 0, & \qquad   t_{+} : -\infty \rightarrow \infty \\
\Theta^{-} =\frac{3}{a_{-}}\frac{ \d a_{-}}{\d t _{-}}-\frac{ \d \delta_{m-}}{\d t_{-} } < 0, &  \qquad  t_{-} : \infty \rightarrow -\infty
\end{array}
\right.
\end{equation}
where we made use of the  Euler equation to relate  ${\D}_{i} v^i $ to the rate of change of the density contrast:  $ {\D}_{i} v^i \sim - \partial \delta_m/\partial t =  -\delta_m'$. 
Similarly,  the contraction phase that leads to the singularity has the following form 
\begin{equation}\label{eq:contraction}
\text{Contraction} \quad \Longrightarrow \quad
\left\{
\begin{array}{ll}
\Theta^{+} =\frac{3}{a_{+}}\frac{ \d a_{+}}{\d t_{+} }-\frac{ \d \delta_{m+}}{\d t_{+} } < 0, &  \qquad  t_{+}: -\infty \rightarrow \infty \\
\Theta^{-}= \frac{3}{a_{-}}\frac{ \d a_{-}}{\d t_{-} }-\frac{ \d \delta_{m-}}{\d t_{-} } > 0, & \qquad   t_{-} : \infty \rightarrow -\infty
\end{array}
\right.
\end{equation}
Therefore, we can avoid the gravitational focusing singularity discussed in sub-sections \ref{sec:matterhorizon} and \ref{subsec:astrophysicalmatter} by cutting off $\Theta_{+} <0$ sub-region in $\mathcal{M}_{+}$ and replacing it with a corresponding $\Theta_{-} <0$  in $\mathcal{M}_{-}$. That is the collapsing region in a manifold with forward flowing coordinate time is replaced with a separate expanding universe with the direction of flow of coordinate time reversed.

Finally, we reiterate that the gravitational focusing singularity is avoided by changing the orientation of the coordinate system after the matter horizon.  This is consistent with the Feynman-Stueckelberg interpretation in particle physics,  where the unbounded energy catastrophe associated with a negative energy particle moving forward in time is resolved by introducing antiparticles as particles moving backwards in time~\cite{Stueckelberg:1941rg,Feynman:1949hz}.
Here, we avoid the gravitational focusing singularity by describing the internal dynamics of a gravitationally bound system as a separate universe with coordinate time moving backwards. 

\subsection{Glueing manifolds at the boundary hypersurface}\label{sec:boundary_metrics}

Now that we have described the type of metric tensors on both manifolds, we will describe how they are glued together at the common boundary.  The central requirement for glueing two spacetimes together at a common hypersurface is to impose the continuity of the induced metric tensors.  The continuity condition for the induced metric tensors has so far focused on the Darmois condition \cite{Darmois1927,Israel:1966rt}
\begin{eqnarray}\label{eq:Darmois}
 h_{ab}^{+} \stackrel{\Sigma}{=} h_{ab}^{-} \,, \qquad \qquad 
 \gamma_{ab}^{+} \stackrel{ B}{=} \gamma_{ab}^{-}
\end{eqnarray}
Equation \eqref{eq:Darmois} is not the only condition on the induced metric tensor that ensures diffeomorphisms at the boundary.  We will show in sub-section \ref{sec:timelike}  and \ref{sec:spacelike} that a different condition exists that leads to more symmetric junction conditions. 
The essential difference is that the decomposition of the Riemann tensor in the distributional sense with a step function becomes ill-defined; however, the variational calculus approach remains consistent, as we will show in sub-section \ref{sec:glueingGR}.

\subsubsection{Spacelike hypersurface}\label{sec:spacelike}

The approach we discuss here leads to a boundary with more symmetries when compared to the traditional Israel formulation~\cite{Israel:1966rt}. 
The neccessary condition for glueing both induced metrics at the spacelike hypersurfaces $\Sigma_{-,\tau}$  and   $\Sigma_{+ ,\tau}$ is that there exists a smooth map, $\phi_{\Sigma}$  that maps nearby points $P_{-} \in  \Sigma_{-,\tau}$, to points $P_{+} \in  \Sigma_{+,\tau}$, i.e $\phi_{\Sigma}: \Sigma_{-,\tau} \to \Sigma_{+,\tau} \,, $
Under this transformation (${x}^{a}_{-} \to - x_{+}^{a} + X^{a}_{\Sigma}$), the metric on the spacelike hypersurface  transforms as 
\begin{eqnarray}\label{eq:inducedmetrictransform}
{h}^{+}_{ab}  - {h}^{-}_{ab}  = \left[X^{c} _{\Sigma}\partial_{\bot c} {h}^{+}_{ab} + {h}^{+}_{ab} \partial_{a} X^{c}_{\Sigma} + {h}^{+}_{ac} \partial_{\bot b} X^{c}_{\Sigma} \right]\,,
\end{eqnarray}
The terms in the square brackets are proportional to the Lie derivative of a spacetime metric tensor  
$
\mathcal{L}_{X_{\Sigma}} h^{+}_{ab} = X^c_{\Sigma} \partial_{c} h^{+}_{ab} + h^{+}_{cb} \partial_{a}X^c_{\Sigma} + h^{+}_{ac}\partial_{b} X^c_{\Sigma}
  = \nabla_{a} X_{{\Sigma}b} + \nabla_{b}X_{\Sigma a}\,.
$
Therefore, the Darmois condition in  \eqref{eq:Darmois} is simply a requirement that $X^a$ must satisfy the  Killing equation $\nabla_{a} X_{{\Sigma}b} + \nabla_{b}X_{a\Sigma} = 0$.  

The Darmois condition is not properly suited for our case, or in a situation where the fundamental difference between the two metrics is the scale factor, see equation \eqref{eq:conformalmap_space}. 
Therefore, the suitable continuity condition can be found by substituting equation \eqref{eq:conformalmap_space} in equation \eqref{eq:inducedmetrictransform} to obtain a  conformal Killing equation 
 \begin{eqnarray}\label{eq:CKE} 
 \mathcal{L}_{X_{\Sigma}} {{h}^{+}_{ab}}(x^{c}) 
= 2 \phi(x^{a}){{{h^{+}_{ab}}}}(x^{c}) \,,
\end{eqnarray}  
where $\phi(x^{c}) \equiv( 1-\Omega^2_{S}(x^{c}) )/2$ is a redefined conformal factor.  
Since the metric  $h^{+}_{ab} $   is conformal to the Euclidean metric : 
 ${h}^{+}_{ab} =\Omega_{+}^2 \delta^{+}_{ab}$, we can further express equation \eqref{eq:CKE}  in terms of the Euclidean metric $ \mathcal{L}_{X} {{\delta}^{+}_{ab}}(x^{c}) 
= 2 \phi'(x^{c}){\delta_{+{ab}}} (x^{c}) $, where $  \phi'(x^{c})   \equiv  \left[ -X^{b}_{\Sigma}\nabla_{b}\Omega_{+}/\Omega_{+} +   \phi(x^{c}) \right]$. The $ {X}^{b}_{\Sigma}$  that solves $ \mathcal{L}_{X_{\Sigma}} {{\delta}^{+}_{ab}}(x^{c}) 
= 2 \phi'(x^{c}){\delta_{+{ab}}} (x^{c}) $ is given by \cite{levine1936}
 \begin{equation}\label{eq:symmetryoffLSS}
 {X}^{b}_{\Sigma}= \alpha^{b} + M^{b}{}_{a} x^{a}_{+} + \lambda x^{b}_{+} + 2 \left(x_{{+}a} \beta^{a}\right) x^{b}_{+} - x_{{+}a} x^{a}_{+} \beta^{b} \,,
 \end{equation}
 where $\alpha^{a}$, $\beta^a$ and $\lambda$ are  spatially constant.  $M^{a}{}_{b} $ and $\alpha^{a} $ are associated with rotations and translations respectively.  $\lambda $ is associated with dilatation and $\beta^{b}$ is associated with the special conformal transformation.   This  is $SO(4,1)$  conformal group.
 The effective conformal factor can be expressed in terms of $X^{i}$ as
\begin{eqnarray}\label{eq:scale_factor}
\Omega^2_{S} = 1 - 2 \frac{\mathcal{L}_{X} \Omega_{+}}{\Omega_{+}} - 2 \phi'(x) = 1 - 2 \frac{\mathcal{L}_{X} \Omega_{+}}{\Omega_{+}} - \frac{2}{3} \left(\lambda + 2 x^{i}_{+} \beta_{i}\right)\,,
\end{eqnarray}
where $\phi'(x^a_{+})= \partial^{a} X_{{\Sigma}a} /3=  \left(\lambda + 2 x^{i}_{+} \beta_{i}\right)/3\,.$

\subsubsection{Time-like hypersurface}\label{sec:timelike}

Similarly, for the timelike hypersurface, the induced  metric tensor  transforms as
\begin{eqnarray}\label{eq:metrictransform}
{\gamma}^{-}_{ab}  - {\gamma}^{+}_{ab}  = \left[X^{c}_{B} \partial_{\bot c} {\gamma}^{+}_{ab} + {\gamma}^{+}_{ab} \partial_{a} X^{c}_{B} + {\gamma}^{+}_{ac} \partial_{\bot b} X^{c} _{B}\right]\,,
\end{eqnarray}
Putting equation \eqref{eq:conformalmap_time} in equation \eqref{eq:metrictransform} gives the conformal Killing equation 
\begin{eqnarray}\label{eq:CKE_time}
\mathcal{L}_{X_{{B}}}  \gamma^{+}_{ab} = \nabla_{a} X_{Bb} + \nabla_{b}X_{Ba} = 2 \varphi(x^{c}) \gamma^{+}_{ab} \,,
\end{eqnarray} 
 The $X^{a}_{{B}}$  that solves equation \eqref{eq:CKE_time} is  called the conformal Killing vector of $\gamma^{-}_{ab} $ and $\phi(x^{c}) =( \Omega^2_{T}(x^{c}) - 1)/2$ is a redefined conformal factor.  
 The metric  $\gamma^{-}_{ab} $  given in equation \eqref{eq:metrictimehypersurface} is conformal to a perturbed Minkowski spacetime: 
 ${\gamma}^{+}_{ab} = a^2_{+}(\eta) \gamma^{+\rm{PM}}_{ab}$,
 where $\gamma^{+\rm{PM}}_{ab}$ is a perturbed  Minkowski spacetime. Acting on this with the Lie derivative gives $ \mathcal{L}_{X} {{\gamma}^{+}_{ab}} =a^2(\eta) \mathcal{L}_{X} {{\gamma}^{+\rm{PM}}_{ab}} + 2a(\eta)X^{\mu}\nabla_{\mu}a(\eta){{\gamma^{+PM}_{ab}}}$. Putting this in equation \eqref{eq:CKE}, the conformal Killing equation transforms to another conformal Killing equation with a different conformal factor
 \begin{eqnarray}\label{eq:CKE2nd}
 \mathcal{L}_{X_{B}} {{\gamma}^{+\rm{PM}}_{ab}}
= 2 \phi'(x^{a}){{{\gamma^{+\rm{PM}}_{ab}}}}\,,
\end{eqnarray}  
where $ \phi'(x^{c})   \equiv  \left[ -X^{b}\nabla_{b}a(\eta)/a(\eta) +   \phi(x^{c}) \right]$. 
 We can write $\gamma^{+PM}_{ab} $  as a perturbation of Minkowski spacetime
$
\gamma^{+\rm{PM}}_{ab}(x^{c})  = \eta_{ab} + \delta \gamma^{+\rm{PM}}_{ab}(x^{c}) \,,
$
where $ \eta_{ab}$ is the background Minkowski metric tensor. 
Neglecting  the   quadratic corrections, equation \eqref{eq:CKE2nd}   reduces to
\begin{eqnarray}\label{eq:CKE3}
  \nabla_{a}\tilde{X}_{Bb}+\nabla _{b}\tilde{X}_{B a}=\frac{2}{3}{\eta}_{ab}\nabla _{c}\tilde{X}^{c} + \mathcal{O} \left( \nabla _{c}X^{c}  \delta \gamma^{+PM}_{ab}\right)
\end{eqnarray}
where the conformal factor is given by $\phi'(x^{\mu})  = \nabla _{c}\tilde{X}^{c} /3$. Neglecting the second term on the RHS or taking the leading order approximation, we find that $ {X}^{b}$  that solves equation \eqref{eq:CKE3} is given by
 \begin{equation}
\tilde{X}^{b}_{B}= \tilde{\alpha}^{b} + \tilde{M}^{b}{}_{a} x^{a}_{+} + \tilde{\lambda} x^{b}_{+} + 2 \left(x_{{+}a} \tilde{\beta}^{a}\right) x^{b}_{+} - x_{{+}a} x^{a}_{+} \tilde{\beta}^{b} \,,
 \label{eq:timelike_map}
 \end{equation}
 where $\tilde{\alpha}^{a}$, $\tilde{\beta}$ and $\tilde{\lambda}$ are  spatially constant.  $\tilde{M}^{a}{}_{b} $ and $\tilde{\alpha}^{a} $ are associated with rotations and translations respectively.  $\tilde{\lambda}$ is related to the dilatation.   $\tilde{\beta}^{b}$ is associated with the special conformal transformation. 
 This is $SO(3,2)$ symmetry group, which is isomorphic to the Anti-de Sitter (AdS) group in 3D.   

 \subsubsection{Cosmological zoom-in perturbation theory}\label{sec:metric_from_plus}
 
 The cosmological zoom-in perturbation theory follows the philosophy of the cosmological zoom-in N-body simulation. In this case, the initial data is evolved according to the metric tensor given in equation \eqref{eq:metric}.   Equation \eqref{eq:metric} describes the universe with sufficient accuracy up to the size of the biggest gravitationally bound system in the universe $R=1/k_{R}$.  The sub-region with size $R$ is decoupled from the Hubble flow. To describe the small-scale mode of wave number $ k > k_{R}$ using the cosmological zoom-in perturbation theory, we split the perturbed variables in equation \eqref{eq:metric} into modes with wave number less $k_{R}$ and those with wave number greater than $k_{R}$
\begin{eqnarray}\label{eq:decompotential}
\Phi_{+}  = \Phi_{-} + \Phi_{L}\,, \qquad \qquad   \Psi_{+} =  \Psi_{-} + \Psi_{L} \,,
\end{eqnarray}
where  $\Phi_{L}$ and $\Psi_{L}$ are the long mode components of the gravitational and scalar curvature perturbations, respectively.  The $ \Phi_{-} $ and $\Psi_{-} $ are the corresponding short wavelength mode.
Using equations \eqref{eq:symmetryoffLSS} and \eqref{eq:timelike_map} and focusing on the dilatation and special conformation transformation (we drop the rotation and translation for simplicity), we obtain a coordinate map between the low-res world $\mathcal{M}_{+}$ and the zoom-in world $\mathcal{M}_{-}$
 \begin{eqnarray}
{\eta}_{+} &=& -\eta_{-}+ \epsilon(\eta_{-}) + {x}_{-j}  {\xi}^j (\eta_{-}) \,,\label{eq:etatransform}
\\
x_{+}^{i} &=&-x_{-}^{i} + {\lambda x^{i}_{-} }+ {2 \left( x_{{-}j} \beta^{j}\right) x^{i}_{-} -  x_{{-}j} x^{j}_{-} \beta^{i}}\,,
\label{eq:xtransform}
\end{eqnarray}
where  $\epsilon(\eta) = \lambda \eta$ and   $\xi^i = 2 \beta^i \eta$.  Applying equation \eqref{eq:etatransform} and \eqref{eq:xtransform}  and \eqref{eq:decompotential} to equation \eqref{eq:metric} focusing on scalar perturbation only leads to the 
metric tensor in $\mathcal{M}_{-}$  
\begin{eqnarray}\label{eq;localmetric}
\d s^2\bigg|_{-} &\approx& a_{-}^2(\eta_{-}) \bigg[ - \left[ 1 + 2 \Phi_{-} \right] \d \eta_{-}^2  
+ \left[ 1 - 2\Psi_{-} \right] {\delta_{ij} \d x_{-}^i \d x_{-}^j} \bigg]\,,
\end{eqnarray}
where  $a_{-}(t_{-}) \equiv a_{+}(-t_{-}) $, $\Phi_{-} \equiv \Phi_{-}(-\eta_{+},-x^{i}_{+})$ and $\Psi_{-}\equiv \Psi_{-}(-\eta_{+},-x^{i}_{+})$  are the short wavelength length part of the gravitational and scalar curvature perturbation in $x^{a}_{-}$ coordinate.  And the long wavelength modes $\Phi_{L} $ and  $\Psi_{L} $ are given by
\begin{eqnarray}
\Phi_{L} &\equiv&-\left[  \varepsilon'(\eta) +  x_{i} {\xi^{i}}'
+  \HH \left(\epsilon(\eta) + {x_{j}}  {\xi^{j}} (\eta) \right)\right]\,,
\\
\Psi_{L} &\equiv&-\left[  \lambda	+ 2 \beta_{j} x^{j} 
+  \HH \left(\epsilon(\eta) + {x_{j}}  {\xi^{j}} (\eta) \right)\right]\,.
\end{eqnarray}
The impact of the long-wavelength modes, that is, modes with $k < k_{R}$, impacts only the scale factor of the metric tensor of the decoupled region
\begin{eqnarray}
{ a_{-}} &=&{ a_{+}}\left[1 - 2 \frac{\mathcal{L}_{X} a_{+}}{a_{+}} - \frac{2}{3} \left(\lambda + 2 x^{i}_{+} \beta_{i}\right)\right]^{\frac{1}{2}}\,.
\end{eqnarray}
That is, the modes longer than the size of the decoupled sub-region do not impact the local gravitational dynamics. This is consistent with similar findings from the separate universe approach to large-scale structure modelling~\cite{Dai:2015jaa,Umeh:2019qyd,Umeh:2019jqg}. 
The massive particles or the short wavelength modes within the decoupled region are moving with velocities which are less than the cosmological expansion rate; therefore, they cannot escape their common gravitational attraction under geodesic motion. 
However, the passage of the long wavelength modes will displace the relative position of the decoupled region in the form of a memory effect associated with the page of gravitational waves~\cite{Tolish:2016ggo}.

\subsection{Backreaction contribution to the energy-momentum tensor}\label{sec:glueingGR}

In \cite{Umeh:2026ajv}, we described how to continue the flow of a one-parameter family of timelike geodesics in two separate manifolds $\mathcal{M}_{+}$ and $\mathcal{M}_{-}$ across a spacelike boundary, $\mathcal{N}$.    Geodesic motion by definition, neglects the backreaction of the spacetime on the particle trajectory. 
%
 Here, we describe how to go beyond PPA to capture the impact of backreaction on the particle propagation following the surgery framework described in sub-section \ref{sub:metric_tensor}. 
   \begin{figure}[h]
\centering 
 \includegraphics[width=70mm,height=60mm]{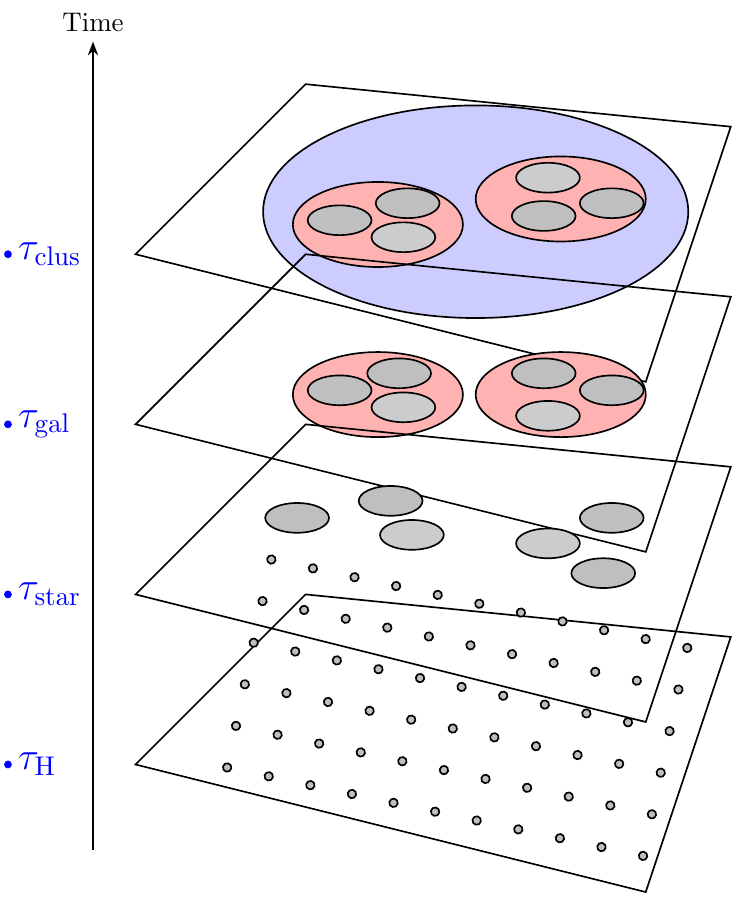}
\includegraphics[width=70mm,height=60mm] {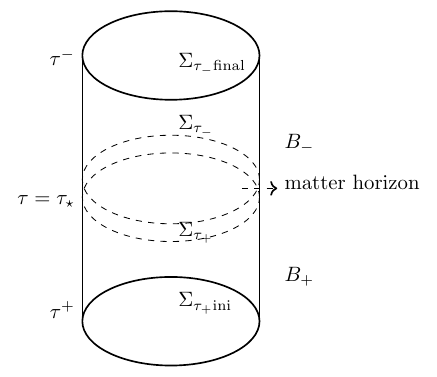}
\label{fig:spacetime_diagram}
\caption{Left panel:  This is a schematic illustration of the timeline of structure formation in the universe, starting from light elements(particles) coming together through gravitational pull to form more massive particles.  The left vertical line has crucial time scales when a collection of these particles decoupled from the Hubble flow:  $\tau_{ \rm{H}}$, $\tau_{ \rm{star}}$,  $\tau_{ \rm{gal}}$ and $\tau_{ \rm{clus}}$. Right panel: The right panel is a zoom-in on one section of the world-tube shown on the left panel in order to illustrate the glueing together of two manifolds along timelike and spacelike hypersurfaces.  
We indicated the initial and final spatial hypersurface surfaces $\Sigma_{\tau_+}^{\text{ini}}$ and $\Sigma_{\tau_-}^{\text{final}}$,  and the common boundary hypersurface at the matter horizon (dashed oval) at $\tau = t_{\star}$. 
The timelike boundaries $B_\pm$ enclose the spatial region.
}
\end{figure}
The starting point is the action of the general gravitational theory describing  the physical universe on ambient spacetime $\mathcal{M} = ( \mathcal{M}_{+}\setminus \mathcal{D}) \cup \mathcal{M}_{-} \cup \mathcal{N} $
\begin{eqnarray}\label{eq:gravity_action}
S_{\rm{Full}}\left[g_{ab}\right] &=& S_g\left[g^{+}_{ab}\right]  + S_g\left[g^{-}_{ab}\right] +   S_{\mathrm{GHY}}\left[h^{{\pm}}_{ab}\right] + S_{\mathrm{GHY}}\left[ \gamma^{{\pm}}_{ab}\right]  
+  S_{\mathrm{Hayward}}\left[N^{{\pm}}_{ab}\right] \,,
\end{eqnarray}
where  $S_g\left[g^{\pm}_{ab}\right] $ is the sum of the Einstein-Hilbert action and the action for the statandard matter fields
$
S_g\left[g^{\pm}_{ab}\right]  = S_{\mathrm{EH}} \left[g_{ab}^{\pm}\right] +S_{\mathrm{M}}\left[g_{ab}^{\pm}\right ]
$, here $ S_{\mathrm{EH}}$ is the Einstein-Hilbert action  and $S_{\mathrm{M}}$ is the action for the standard matter fields.
\begin{eqnarray}
 S_{\mathrm{EH}} \left[g^{\pm}_{ab}\right]   =  \frac{1}{2\kappa}
\int_{\mathcal{M}}\mathcal{R}\left[g^{\pm}_{ab}\right]  \,\sqrt{-g_{\pm}}\d^4 x _{\pm} \,,
\qquad \qquad 
S_{\mathrm{M}}\left[g^{\pm}_{ab}\right]  = 
  \int_{\mathcal{M}}{L}_{\rm{M}}\left[g^{\pm}_{ab}\right]  \,\sqrt{-g_{\pm}}\d^4 x_{\pm} \,,
  \label{eq:matter_action}
  \end{eqnarray}
  where $\mathcal{R}$ is the Ricci scalar and $L_{m}$ is  the Lagrangian for the matter fields.
Furthermore,  $ S_{\mathrm{GHY}}\left[h^{{\pm}}_{ab}\right]$ is the Gibbon-Hawking-York boundary term on the spacelike hypersuface, while $S_{\mathrm{GHY}}\left[ \gamma^{{\pm}}_{ab}\right] $ is a corresponding  Gibbon-Hawking-York boundary term on the timelike hypersuface. Finally, 
$ S_{\mathrm{Hayward}}$ is the Hayward corner term, it depends on the metric on the screen space, $N_{ab}$.

The variation of $S_g\left[g^{+}_{ab}\right]  + S_g\left[g^{-}_{ab}\right] $ with respect to the metric tensors of the ambient manifold $ \mathcal{M} $ foliated as shown in Figure \ref{fig:spacetime_diagram} leads to 
\begin{eqnarray}\nonumber
\delta S_g\left[g_{ab}\right] & =&
\int_{\mathcal{M}^{+}}\frac{1}{2}\left(\frac{1}{\kappa}G^{+}_{ab}+\Lambda g_{ab}-T^{+}_{ab}\right)\delta g^{ab}_{+}\,\sqrt{-g_{+}} d^4x 
 \\ \nonumber& &
+\int_{\mathcal{M}^{-}}\frac{1}{2}\left(\frac{1}{\kappa}G^{-}_{ab}+\Lambda g_{ab}-T^{-}_{ab}\right)\delta g^{ab}_{-}\,\sqrt{-g_{-}} d^4x 
 \\ & &
 +\frac{1}{2\kappa}  \int_\mathcal{M_{+}}  \nabla_{a} \delta V^a_{+} \sqrt{-g_{+}} d^4x_{+}  +\frac{1}{2\kappa}  \int_\mathcal{M_{-}}  \nabla_{a} \delta V^a_{-} \sqrt{-g_{-}} d^4x_{-}\,,
 \label{eq:totalaction_variation}
\end{eqnarray}
where $G^{\pm}_{ab} =  R^{\pm}_{ab} - {1 \over 2} g^{\pm}_{ab} \mathcal{R}^{\pm}$ is the Einstein tensor,  $\delta V^c_{\pm} $ are boundary terms resulting from the variation of the Ricci tensors associated with both manifolds $\mathcal{M}_{\pm}$: $g^{ab}_{\pm} \delta R^{\pm}_{ab} = \nabla_{a} \delta V^a_{\pm}$. It is given by $ \delta V^c_{\pm} =  \left[g^{ab}_{\pm} \delta  \Gamma^c_{\pm ab} - g^{a c}_{\pm} \delta \Gamma^b_{\pm ab}\right]$. Again $T_{ab}^{\pm}$ is the respective energy-momentum tensors for the standard matter(e.g. baryons)
$
T^{{\pm}}_{ab} \equiv -\frac{2}{\sqrt{-g_{\pm}}}
 \frac{\delta S^{\pm}_{\rm{M}}}{\delta g^{\pm}_{ab}}  \,.
$ 
 In  \cite{Umeh:2026ajv}, we showed how to continue the flow of timelike geodesics for point particles across the boundary within the ambient spacetime $\mathcal{M} = ( \mathcal{M}_{+}\setminus \mathcal{D}) \cup \mathcal{M}_{-}  $.
 Essentially,  this involves splitting the proper time integrals in $L_{m}$ into two sectors with the boundary at the matter horizon~\cite{Umeh:2026ajv}. 
  \begin{eqnarray}\label{eq:piecewise_action}
S_{\rm{M}}  &=&\sum^{N}_{\ell} S^{+}_{\ell} +\sum^{N}_{\ell} S^{-}_{\ell}
=-\sum_{\ell}m_{+\ell} \int_{\tau_{\rm{ini}}}^{\tau_{\star}}  \mathrm{d}\tau^{+}_{\ell}   -\sum_{\ell}m_{-\ell} \int_{\tau_{\star}}^{\tau_{\rm{final}}}  \mathrm{d}\tau^{-}_{\ell} 
 \,,
\end{eqnarray}
Note that $m_{-}$ and $m_{+}$ are the masses of the massive particles on different spacetimes. 
The derivation of the energy-momentum tensor follows the same steps, with the only exception being the limits of the proper time integration
 \begin{eqnarray}
 T^{ab}_{+} &=& \sum^N_{\ell}  T^{ab}_{+ \ell}
 = \sum^N_{\ell}  \frac{m_{+\ell}}{\sqrt{-g_{+}}} \int_{\rm{ini}}^{\star} {\d}\tau_{+}\, u^{a}_{+\ell} u^{b}_{+\ell} \delta^4(x_{+} - x_{+\ell}(\tau_{+\ell})) \,.
\\ 
 T^{ab}_{-} &=& \sum^N_{\ell}  T^{ab}_{- \ell}
 = \sum^N_{\ell}  \frac{m_{-\ell}}{\sqrt{-g_{-}}} \int_{\star}^{\rm{final}} {\d}\tau_{-}\, u^{a}_{-\ell} u^{b}_{-\ell} \delta^4(x_{-} - x_{-\ell}(\tau_{-\ell})) \,.
\end{eqnarray}

To simplify the integrals in the third line in equation \eqref{eq:totalaction_variation}, we follow the vertical cylinder in the right panel of figure \ref{fig:spacetime_diagram}. Here we invoke the Stokes theorem to perform the volume integral over $\delta V^a_{\pm}$  as surface integrals over the sub-manifolds~\cite{Booth:1998eh}:
\begin{equation} \label{eq:boundary_term}
  \int_{\mathcal{M}} \nabla_a (\delta V^a_{\pm}) \sqrt{-g_{\pm}}\, d^4x_{\pm}
  = \int_{\Sigma_{\pm}} \delta V^a_{\pm}\, u_{\pm a}\, \sqrt{h_{\pm}}\, d^3x_{\pm}
  + \int_{B_{\pm}} \delta V^a_{\pm}\, r_{\pm a}\, \sqrt{|\gamma_{\pm }|} d^3x_{\pm}\,,
\end{equation}
where  $ \int_{\Sigma_{+}} =  \int_{\Sigma_{+} {\rm{ini}}}- \int_{\Sigma_{+} \star} $ and $ \int_{\Sigma_{-}} =  \int_{\Sigma_{-} {\star}}- \int_{\Sigma_{-} \rm{final}} $, 
 $u^{\pm}_{a} \equiv u^{\pm}_{ \ell a}$  is the normal timelike vector orthogonal to the spacelike hypersurface, we have dropped $\ell$ to reduce cluster, similarly,  $r^{\pm}_{a}  \equiv r^{\pm}_{\ell a}$ is a normal spacelike vector orthogonal to the timelike hypersurface ~\cite{poisson2004,Booth:1998eh,Hawking:1996ww}.   
The normal vector fields $u_a^{\pm}$ and $r^a_{\pm}$ are oriented on their respective side of the hypersurfaces. The directed surface element is related to the normal vectors according to  
$\d\Sigma^{\pm}_b = u^{\pm}_b \,\d\Sigma_{\pm} = u^{\pm}_b \sqrt{h_{\pm}} \d^3 x_{\pm} $  and $\d B_b = r^{\pm}_b \d B = r^{\pm}_b \sqrt{\gamma_{\pm}}\d^3 x_{\pm} $,
where $\d\Sigma$ and and $\d B$ are  the intrinsic volume element on $\Sigma$  and $B$ respectively. 
%
 The induced metrics on the spatial hypersurfaces satisfy the following relation
\begin{equation}\label{eq:Screenspacemtric1}
h^{\pm}_{ab}\equiv g^{\pm}_{ab}+ u^{\pm}_{a}u^{\pm}_{b}, \qquad {h^{\pm}}^a{}_a=3, \qquad  \qquad  h^{\pm}_{ac}{h^{\pm}}^c_b=h^{\pm}_{ab}, \qquad  \qquad h^{\pm}_{ab}{u^{\pm}}^b=0 \,.
\end{equation}
A similar relationship applies to the metric on the timelike hypersurface with the metric $\gamma_{ab}$: $\gamma^{\pm}_{ab} =  g^{\pm}_{ab}- r^{\pm}_{a}r^{\pm}_{b}$ 
\begin{eqnarray}\label{eq:Screenspacemtric2}
\gamma^{\pm}_{ab}\equiv g^{\pm}_{ab}- r^{\pm}_{a}r^{\pm}_{b}, \qquad {\gamma^{\pm}}^a{}_a=3, \qquad  \qquad  \gamma^{\pm}_{ac}{\gamma^{\pm}}^c_b=\gamma^{\pm}_{ab}, \qquad  \qquad \gamma^{\pm}_{ab}{r^{\pm}}^b=0\,.
\end{eqnarray}
Note that   $r^a$ is spacelike $r_a r^a =1$ and $\gamma_{ab} h^{ab} = 2$.
Using the covariant derivatives of $u^{\pm}_{a}$, i.e. $B^{\pm}_{ab} = \nabla_{a}u^{\pm}_{b}$ , we can express  the variation of the  Christoffel symbols in terms of the variations of the covariant derivatives of $u^a$
$
u^{\pm}_{a} \delta \Gamma^{a}_{\pm bc} = \nabla_{b}\delta u_{\pm c} - \delta B^{\pm}_{bc}\,,$
$
u^a_{\pm}\delta \Gamma^{b}_{\pm ba} = - \nabla_{b} \delta u^b_{\pm} +\delta B^{a}_{\pm}{}_{ a}$, then 
projecting $B^{\pm}_{ab}$ into the spacelike hyersurface gives extrinsic curvature tensor of the hypersurface $K_{ab}^{\pm} = h^{c}_{\pm}{}_{a}B^{\pm}_{cb}  =  h^{c}_{\pm}{}_{a}{\nabla}_{c} u^{\pm}_{b}$. Now replacing the contraction of the Christoffel symbols  in  $ \delta V^c_{\pm} =  \left[g^{ab}_{\pm} \delta  \Gamma^c_{\pm ab} - g^{a c}_{\pm} \delta \Gamma^b_{\pm ab}\right]$ leads to 
\begin{eqnarray} \nonumber 
  \frac{1}{2\kappa} \oint _{\Sigma_{\pm}} \sqrt{h^{\pm}}   u_{}{\pm b}\delta V^b_{\pm}\d^3{x}_{\pm}&=&
 \int _{ { {\Sigma}}_{\pm}} \sqrt{h^{\pm}}\bigg\{
  -    \frac{1}{\kappa}\delta u^0_{\pm}  K^{\pm}
+ \bigg[\Pi^{\pm}_{ab}
+L^{\pm}_{(a} u^{\pm}_{b)} \bigg]\delta h^{ab}_{\pm}\bigg\} \d^3{x}_{\pm} 
\\&&
+  \frac{1}{\kappa}  \oint _{\partial { {\Sigma}_{\pm}}}\sqrt{N_{\pm}}  \delta u_{\pm\bot}^a r^{\pm}_{a} \d^2 x_{\pm}
-
 \frac{1}{\kappa} \int _{{ {\Sigma}}_{\pm}}
   \delta\bigg[\sqrt{h^{\pm}} K^{\pm}\bigg] \d^3{x_{\pm}}\,,
\label{eq:boundary_term}
\end{eqnarray}
where $ K^{\pm}$ is the trace of the extrinsic curvature tensor of the spacelike hypersurface.  Note that to obtain the first term in the second line,  we made use of the divergence theorem to relate the divergence of the variation vector field in the spatial hypersurface  to the flux of a vector field through a closed surface
\begin{eqnarray}\label{eq:timelikeboundary}
&&\int _{ {\Sigma}_{\pm}}\sqrt{h^{\pm}}  {\rm{div}} \delta u_{\pm} \d ^3 x_{\pm}=   \oint _{\partial { {\Sigma}_{\pm}}}\sqrt{N_{\pm}}  \delta u_{\pm\bot}^i r^{\pm}_{i} \d^2 x_{\pm}\,,
\end{eqnarray}
where $N_{ab}$ is the metric tensor on the screen space defined with respect to the metric tensor on the hypersurface:  $N^{\pm}_{ab} = h^{\pm}_{ab} - r^{\pm}_{a} r^{\pm}_{b}$ 
on the projected screen space. and $\sqrt{N}$ is the square root of the determinant of the metric tensor. In equation \eqref{eq:boundary_term}, we have defined the following  conjugate momenta on the spacelike hypersurface
\begin{eqnarray}
\Pi^{\pm}_{ab} &=&- \frac{2}{\sqrt{h^{\pm}}}\frac{\delta S^{\pm}_{g}}{ \delta h^{ab}_{{\pm}}} 
 = \frac{1}{\kappa} \bigg[ \left(   \frac{1}{2}h^{cd}_{\pm} B^{\pm}_{cd} +  \frac{1}{2}B^{\pm}_{c}{}^{c} \right)h^{\pm}_{ab}
-B_{ab} 
\bigg]= \frac{1}{\kappa} \bigg[  {K}h^{\pm}_{ab}-K_{ab} \bigg]\,,\label{eq:conjugate_momenta1}
\\
L^{\pm}_{a} &=& -\frac{2}{\sqrt{h^{\pm}}}\frac{\delta S^{\pm}_{g}}{ \delta u^{a}_{\pm}}  = \frac{1}{\kappa}  \bigg[B^{\pm}_{ab}  u^{b}_{\pm} \bigg]=  \frac{1}{\kappa}  \bigg[{K}_{ab}  u^{b}_{\pm} \bigg]\,.
\label{eq:conjugate_momenta3}
\end{eqnarray}
{In general,  $\Pi^{\pm}_{ab}$ is the canonical momentum conjugate to the induced metric $h_{ab}$.  The sign choices define orientation. It is also called the Brown–York quasilocal stress tensor~\cite{York:1972sj,Brown:1992br}. It encodes the surface stress-energy(momentum conjugate to geometry), $L^{\pm}_a$ represents the momentum density of the gravitational field along the direction $u^b$. It encodes the energy flux along the boundary.}

For the time-like hypersurface, we decompose the covariant derivatives of $r^{\pm}_{a}$: $C^{\pm}_{ab} = \nabla_{a}r^{\pm}_{b}$ . Essentailly, we express the variation of the  Christoffel symbols in terms of the variations of the covariant derivatives of $r^a_{\pm}$: $r^{\pm}_{a} \delta \Gamma^{a}_{\pm bc} = \nabla_{b}\delta r_{\pm c} - \delta C^{\pm}_{bc}\,,$
 and $r^a_{\pm}\delta \Gamma^{b}_{\pm ba} = - \nabla_{b} \delta r^b_{\pm} +\delta C^{a}_{\pm}{}_{ a}$.  Then, we 
project $C^{\pm}_{ab}$ into the timelike hyersurface to obtain extrinsic curvature tensor of the hypersurface $\tilde{K}_{ab}^{\pm} = \gamma^{c}_{\pm}{}_{a}C^{\pm}_{cb}  =  \gamma^{c}_{\pm}{}_{a}{\nabla}_{c} r^{\pm}_{b}$. Implementing these in the second term in equation \eqref{eq:boundary_term} gives
\begin{eqnarray}
  \frac{1}{2\kappa} \int _{{B}} \sqrt{\gamma}   r_{b}\delta V^b  \d^3x_{\pm}&=&
\int _{{B}}  \sqrt{\gamma^{\pm}}\bigg\{   \frac{1}{\kappa}
 \delta r^r_{\pm}  \tilde{K}^{\pm}
+ \bigg[\tilde{\Pi}^{\pm}_{ab}
+\tilde{L}^{\pm}_{(a} r^{\pm}_{b)} \bigg]\delta \gamma^{ab}_{\pm}\bigg\} \d^3{x}_{\pm} \,
\\ \nonumber &&+\frac{1}{\kappa}  \oint _{{\partial {}^{(3)}B}_{\pm}}\sqrt{N^{\pm}} \delta r^a u_a \d^2{x}_{\pm}
-
 \frac{1}{\kappa} \int _{{{}^{(3)}B}_{\pm}}
   \delta\bigg[\sqrt{\gamma^{\pm}} \tilde{K}^{\pm}\bigg] \d^3{x_{\pm}}\,,
\end{eqnarray}
where $ \tilde{K}^{\pm}$ is the trace of the extrinsic curvature tensor of the timelike hypersurface.  We define the corresponding  canonical momentum conjugate to the induced metric $\gamma_{ab}$ and also the momentum flux along the timelike boundary 
\begin{eqnarray}
\tilde{\Pi}^{\pm}_{ab} &=&- \frac{2}{\sqrt{\gamma^{\pm}}}\frac{\delta S^{\pm}_{g}}{ \delta \gamma^{ab}_{{\pm}}} 
 = \frac{1}{\kappa} \bigg[ \left(   \frac{1}{2}\gamma^{cd}_{\pm} C^{\pm}_{cd} +  \frac{1}{2}C^{\pm}_{c}{}^{c} \right)\gamma^{\pm}_{ab}
-C_{ab} 
\bigg] = -\frac{1}{\kappa} \bigg[  \tilde{K}\gamma^{\pm}_{ab}
-\tilde{K}_{ab} 
\bigg]  \,,\label{eq:conjugate_momenta4}
\\
\tilde{L}^{\pm}_{a} &=& -\frac{2}{\sqrt{\gamma^{\pm}}}\frac{\delta S^{\pm}_{g}}{ \delta r^{a}_{\pm}}  = \frac{1}{\kappa}  \bigg[C^{\pm}_{ab}  r^{b}_{\pm} \bigg]=  -\frac{1}{\kappa}  \bigg[\tilde{K}_{ab}  r^{b}_{\pm} \bigg]\,,
\label{eq:conjugate_momenta5}
\end{eqnarray}
Furthermore, we made use of the divergence theorem to relate the divergence of the variation vector field in the volume  to the flux of a vector field through a closed surface
\begin{eqnarray}\label{eq:timelikeboundary}
 \int _{ {B}_{\pm}}\sqrt{\gamma^{\pm}}  {\rm{div}} \delta r_{\pm} \d ^3 x_{\pm}=   \oint _{\partial { {B}_{\pm}}}\sqrt{N_{\pm}}  \delta r_{\pm\bot}^i u^{\pm}_{i} \d^2 x_{\pm}\, ,
\end{eqnarray}
where $N^{\pm}_{ab}  =  \gamma^{\pm}_{ab} +u^{\pm}_{a} u^{\pm}_{b}$ on the projected screen space.
Putting all these together, the equation. \eqref{eq:boundary_term} reduces to
\begin{eqnarray}\nonumber
    \frac{1}{2\kappa} \int_{\mathcal{M}} \nabla_a (\delta V^a_{\pm}) \sqrt{-g_{\pm}}\, d^4x_{\pm} &=&
\int _{ { {\Sigma}}_{\pm}} \sqrt{h^{\pm}}\bigg\{
  -   \frac{1}{\kappa} \delta u^0_{\pm}  K^{\pm}
+ \bigg[\Pi^{\pm}_{ab}
+L^{\pm}_{(a} u^{\pm}_{b)} \bigg]\delta h^{ab}_{\pm}\bigg\} \d^3{x}_{\pm} 
\\ \nonumber &&
+\int _{{B}}  \sqrt{\gamma^{\pm}}\bigg\{
  \frac{1}{\kappa}\delta r^r_{\pm}  \tilde{K}^{\pm}
+ \bigg[\tilde{\Pi}^{\pm}_{ab}
+\tilde{L}^{\pm}_{(a} r^{\pm}_{b)} \bigg]\delta \gamma^{ab}_{\pm}\bigg\} \d^3{x}_{\pm}
\\ \nonumber &&
-
 \frac{1}{\kappa} \int _{{ {\Sigma}}_{\pm}}
   \delta\bigg[\sqrt{h^{\pm}} K^{\pm}\bigg] \d^3{x_{\pm}}
-
 \frac{1}{\kappa} \int _{{B}_{\pm}}
   \delta\bigg[\sqrt{\gamma^{\pm}} \tilde{K}^{\pm}\bigg] \d^3{x_{\pm}}
   \\ &&
   +\frac{1}{\kappa}  \oint _{{\partial B}_{\pm}}\sqrt{N^{\pm}} \delta\left( r^a_{\pm} u_{\pm a}\right) \d^2{x}_{\pm} \,.
   \label{eq:decomposed_boundary_term}
\end{eqnarray}
For a  consistent variational principle, the  Gibbon-Hawking-York counter term must be added to the full gravity action(equation \eqref{eq:gravity_action}) \cite{Gibbons:1976ue,Brown:1992br}
\begin{eqnarray}\label{eq:GHY-spacelike}
   S_{\mathrm{GHY}}\left[h^{{\pm}}_{ab}\right] + S_{\mathrm{GHY}}\left[ \gamma^{{\pm}}_{ab}\right]  =   \frac{1}{\kappa} \int _{{ {\Sigma_{\pm}}}}\sqrt{h^{{\pm}}} K^{\pm} d^{3} x_{\pm}+ \frac{1}{\kappa} \int _{{{}^{(3)} {B_{\pm}}}}\sqrt{\gamma^{{\pm}}} \tilde{K}^{\pm}\d^{3} x_{\pm}\,.
\end{eqnarray}
In addition to the Gibbon-Hawking-York boundary term, the  Hayward corner counter term is added to cancel the variation of the inner product of the timelike and spacelike vectors~\cite{Hayward:1993my}
\begin{eqnarray}
  S_{\mathrm{Hayward}}\left[N^{{\pm}}_{ab}\right]  =- \frac{1}{\kappa}  \oint _{{\partial B}_{\pm}}\sqrt{N^{\pm}} \left( r^a_{\pm} u_{\pm a}\right) \d^2{x}_{\pm}\,.
\end{eqnarray}
The Hayward term replaces the variation with respect to angle with variation with respect to the screen space metric. $\delta \sqrt{N^{\pm}} \left( r^a_{\pm} u_{\pm a}\right)  = \delta \sqrt{N^{\pm}} \left( r^a_{\pm} u_{\pm a}\right) +\sqrt{N^{\pm}}  \delta \left( r^a_{\pm} u_{\pm a}\right)$, where $
\delta \sqrt{N^{\pm}}  = \frac{1}{2} \sqrt{N^{\pm}} N^{ab} \delta N_{ab}$.
For orthogonal corner $r^a_{\pm} u_{\pm a} = 0$,  hence, the Hayward term exactly cancels the contribution of the last term in equation \eqref{eq:decomposed_boundary_term}. For non-orthogonal boundaries, see~\cite{Hawking:1996ww,Booth:1998eh}.
 Moreover, the term associated with the trace of the extrinsic curvature tensor vanishes at the matter horizon $ K^{\pm}=\tilde{K}^{\pm}= 0$.
Therefore, putting together, $\delta S_g\left[ g^{\pm}_{ab}\right] = \delta S_{\rm{EH}}[g^{\pm}] +      S_{\mathrm{GHY}}\left[h^{{\pm}}_{ab} + \gamma^{{\pm}}_{ab}\right]  +\delta S_{\rm{M}}[g^{\pm}]  +   S_{\mathrm{Hayward}}\left[N^{{\pm}}_{ab}\right] $,  the non-vanishing terms and performing some algebraic simplification leads to 
\begin{eqnarray}\label{eq:varied_gravity_action}
{\delta S}_{\rm{Full}}\left[ g^{\pm}_{ab}\right]& =&
\int_{\mathcal{M}^{+}}\frac{1}{2}\left(\frac{1}{\kappa}G^{+}_{ab}+\Lambda g_{ab}-T^{+}_{ab}\right)\delta g^{ab}_{+}\,\sqrt{-g_{+}} d^4x 
 \\ \nonumber& &
+\int_{\mathcal{M}^{-}}\frac{1}{2}\left(\frac{1}{\kappa}G^{-}_{ab}+\Lambda g_{ab}-T^{-}_{ab}\right)\delta g^{ab}_{-}\,\sqrt{-g_{-}} d^4x 
\\ \nonumber &&
 +  \int _{ { \Sigma}_{\pm}} \sqrt{h^{\pm}}\bigg\{
Z^{\pm}_{ab} \delta h^{ab}_{\pm}\bigg\} \d^3{x}_{\pm}
+ \int _{{B^{\pm}}}  \sqrt{\gamma^{\pm}}\bigg\{
 \tilde{Z}^{\pm}_{ab} \delta \gamma^{ab}_{\pm}\bigg\} \d^3{x}_{\pm} \,,
\end{eqnarray}
where $ Z^{\pm}_{ab} =  \Pi^{\pm}_{ab}+2L^{\pm}_{(b} u^{\pm}_{a)}$ and $\tilde{ Z}^{\pm}_{ab} = \tilde{ \Pi}^{\pm}_{ab}+2\tilde{L}^{\pm}_{(b} r^{\pm}_{a)}$.
The boundary  terms in the third line of equation \eqref{eq:varied_gravity_action} can be promoted to full spacetime by 
using of equations \eqref{eq:Screenspacemtric1} and  \eqref{eq:Screenspacemtric2} to rewrite the variation of the projected metric tensor in terms of the variation of the spacetime metric tensor
\begin{eqnarray}\label{eq:variationhab}
  \delta   h_{\pm}^{ab} = \delta g_{\pm}^{ab} +\delta u_{\pm}^{a}u_{\pm}^{b} + u_{\pm}^{a}\delta u_{\pm}^{b}\,. \qquad  \delta   \gamma_{\pm}^{ab} = \delta g_{\pm}^{ab} -\delta r_{\pm}^{a}r_{\pm}^{b} - r_{\pm}^{a}\delta r_{\pm}^{b}\,.
\end{eqnarray}
And using the hypersurface metric tensor determinant in terms of the determinant of the full spacetime metric tensor. 
This can be done in general by considering a general decomposition of the metric tensor $\d s^2_{\pm} =
-(N^2_{\pm} + V^{\pm}_i V^i_{\pm} ) \d t^2_{\pm} +  V^{\pm}_i \d t_{\pm} \d x^i_{\pm} +   h^{\pm}_{ij}\d x^i_{\pm} \d x^j_{\pm}$, where $N_{\pm}$ is the lapse function,  $V^a_{\pm}$ is the shift vector, while $h_{ij}$ is the metric on the hypersurface of constant time.  Given the metric decomposition, we find that the proper time is related to the coordinate time according to $ {\d \tau_{\pm}}/{\d t_{\pm}}\ = \sqrt{-g^{\pm}_{00}} = N_{\pm}$ and the determinant of the full spacetime metric tensor is related to the determinant of the metric tensor on the hypersurface according to   $\sqrt{-g_{_{\pm}}}   =  \sqrt{-g^{\pm}_{00}}\sqrt{h^{\pm}} $. 
Using equation \eqref{eq:variationhab}, we can rewrite the terms involving the variation of the spacetime metric  $\delta g_{{\pm}}^{ab}$:  $ Z^{\pm}_{ab} \delta h^{ab}_{\pm} =Z^{\pm}_{ab}\left[\delta g_{\pm}^{ab} +\delta u_{\pm}^{a}u_{\pm}^{b} + u_{\pm}^{a}\delta u_{\pm}^{b}\right]$  in the form
\begin{eqnarray}
\int _{{ {\Sigma}_{\pm}}}\sqrt{h^{\pm}} 
Z^{\pm}_{ab} \delta g_{{\pm}}^{ab} d^3 x_{\pm} 
   &=&  \int \d\tau_{\pm}  \int _{ {{\Sigma}_{\pm}}}  \delta\left( \tau_{\pm}(y) - \tau_{\star}\right)
      Z^{\pm}_{ab} \delta g_{{\pm}}^{ab} \sqrt{h^{\pm}}d^3 y_{\pm} 
      \\
      &=&  \int_{\mathcal{M}^{{\pm}}}\,\delta\left( \tau_{\pm} - \tau_{\star}\right)
      Z^{\pm}_{ab} \delta g_{{\pm}}^{ab}      \sqrt{-g_{_{\pm}}} d^4x_{\pm}\,,
\end{eqnarray}
where $\tau(t)$  is the porper time parameter whose level-sets $\tau =$ const are the matter horizon.
Working in the local coordinates $x^a=(\tau, y^i )$ adapted to the foliation makes it easier to visualise this. 
In the second equality, we made use of the delta distribution to introduce time integration. This allows us to express the integral over the hypersurface as an integral over the spacetime volume element
 $ \int_{\mathcal{M}^{{\pm}}} \sqrt{-g_{_{\pm}}}  d^4x_{\pm} =  \int \d \tau_{\pm} \oint _{\Sigma_{+}}  \sqrt{-g^{\pm}_{00}}\sqrt{h^{\pm}} d^3 x_{\pm}  $. 
In general,  the first term on the third line in equation \eqref{eq:varied_gravity_action}, can be written as
   \begin{eqnarray}
\int _{ {\Sigma^{\pm}}}Z^{\pm}_{ab} \delta h^{ab}_{\pm}\sqrt{h^{\pm}} \d^3 x_{\pm}&=&
   \int_{\mathcal{M}^{\pm}}\,\delta\left( \tau_{\pm}(x^{\pm}) - \tau_{\star}\right)
     Z^{\pm}_{ab}\delta g_{\pm}^{ab}   \sqrt{-g_{\pm}} d^4x_{\pm}  
     +    \int _{ {\Sigma^{\pm}}}\sqrt{h^{\pm}}  \d^3 x_{\pm}  L^{\pm}_{b}\delta u_{\pm}^{b}  \,.
     ~~~~~
\end{eqnarray}
 Similarly, for the radial foliation, the metric tensor decomposition is of the form $ds^2 = \gamma_{ij}\,dx^i dx^j 
+ 2\,\gamma_{ij}\,\beta^j\,dx^i dr 
+ \left( \alpha^2 + \gamma_{ij}\,\beta^i \beta^j \right) dr^2$, where $\alpha$ is the radial Lapse function, $\beta_{i}$ is the corresponding shift vector and $\gamma_{ij}$ is the metric tensor on the hypersurface of constant radius. 
Therefore, we can elevate the timelike hypersurface volume element to full spacetime  volume element $\delta g_{ab}$
\begin{eqnarray}
\int _{{B_{\pm}}} \sqrt{\gamma^{\pm}} 
\tilde{Z}^{\pm}_{ab} \delta g_{{\pm}}^{ab} d^3 x_{\pm} 
   &=&  \int \d r_{\pm}  \int_{{B}}   \sqrt{g^{\pm}_{rr}}  \sqrt{\gamma^{\pm}}  \delta\left( R(r_{\pm}) - R_{\star}\right)
      \tilde{Z}^{\pm}_{ab} \delta g_{{\pm}}^{ab} d^3 x_{\pm} 
      \\
      &=&   \int_{\mathcal{M}^{{\pm}}}\,\delta\left( R_{\pm}(r_{\pm}) - R_{\star}\right)
      \tilde{Z}^{\pm}_{ab} \delta g_{{\pm}}^{ab}      \sqrt{-g_{_{\pm}}} d^4x_{\pm}
\end{eqnarray}
where ${\d R_{\pm}}/{\d r_{\pm}} = \sqrt{g^{\pm}_{rr}}  =  \alpha$ and  $\sqrt{-g_{_{\pm}}}   =  \sqrt{g^{\pm}_{rr}}\sqrt{-\gamma^{\pm}}  =  \alpha \sqrt{-\gamma^{\pm}}$.  Using equation \eqref{eq:variationhab}, we can express the second term in the third line of equation \eqref{eq:varied_gravity_action} as 
\begin{eqnarray}
\int _{{B^{\pm}}} \sqrt{\gamma^{\pm}} 
\tilde{Z}^{\pm}_{ab} \delta \gamma_{{\pm}}^{ab} d^3 x_{\pm}  
 &=&   \int_{\mathcal{M}^{{\pm}}}\,\delta\left( R_{\pm}(r_{\pm}) - R_{\star}\right)
      \tilde{Z}^{\pm}_{ab} \delta g_{{\pm}}^{ab}      \sqrt{-g_{_{\pm}}} d^4x_{\pm}
     +    \int _{ {B^{\pm}}}\sqrt{\gamma^{\pm}}  \tilde{L}^{\pm}_{b}\delta r_{\pm}^{b}   \d^3 x_{\pm}\,.
 ~~~
\end{eqnarray}

We can now glue both manifolds along the oriented spatial and the time-like hypersurface.  Using orientation-reversing diffeomorphism given in equation \eqref{eq:discrte_map}, both the timelike and spacelike four vector fields maps according to $u^a_{+} =-u^{a}_{-}$ and $r^a_{+} = -r_{-}^a$ since the proper time and proper length strickly positive $\d \tau_{+}/\d\tau_{-} >0$ and $\d R_{+}/\d R_{-} >0$ .
The variations of the orthogonal vectors become $\delta u^a_{+} = - \delta u^a_{-}\,$
 and  $\delta r^a_{+} = - \delta r^a_{-}$.  
Putting all these together, the variation of the total gravitational action is given by
\begin{eqnarray}\nonumber 
 \frac{\delta S_{\rm{Full}}\left[ g_{ab}\right] }{\delta g_{ab}}&=&
  \int_{\mathcal{M}^{+}}\frac{1}{2}\left(\frac{1}{\kappa}G^{+}_{ab}+ \Lambda g^{+}_{ab}-\tau^{+}_{ab}\right)\delta g^{ab}_{+}\,\sqrt{-g_{+}} d^4x_{+} 
   \\ \nonumber &&
  +  \int_{\mathcal{M}^{-}}\frac{1}{2}\left(\frac{1}{\kappa}G^{-}_{ab}+ \Lambda g^{-}_{ab}-\tau^{-}_{ab}\right)\delta g^{ab}_{-}\,\sqrt{-g_{-}} d^4x_{-}
 \\ &&
      +   \int _{\Sigma_{+}}  
      \delta u_{+}^{i} \left(\mathcal{L}^{+}_{i}  - \mathcal{L}^{-}_{i}\right) \d^3 x_{+}
          +   \int _{ {B}_{+}} \delta r_{+}^{i} \left(\tilde{\mathcal{L}}^{+}_{i}  -\tilde{ \mathcal{L}}^{-}_{i}\right)
            \d^3 x_{+} \,,
                   \label{eq:varied_action}
\end{eqnarray}
where we have introduced  tensor densities: $\mathcal{L}^{\pm}_{b} = L^{\pm}_{b}\sqrt{h_{\pm}}$, $\tilde{\mathcal{L}}^{\pm}_{b} = \tilde{L}^{\pm}_{b}\sqrt{-\gamma_{\pm}}$.
By imposing proper variation and piece-wise continuity at the boundary, the critical point of the total action vanishes
${\delta S_g\left[ g_{ab}\right] }/{\delta g_{ab}}   = 0$ leading to the following equations of motion
\begin{eqnarray}\label{eq:EoM}
G_{+}^{ab}+\Lambda g_{+}^{ab} = {\kappa} \tau_{+}^{ab} \,,\qquad G_{-}^{ab} +\Lambda g_{-}^{ab}= {\kappa} \tau_{-}^{ab} \,,
\end{eqnarray}
where $\tau_{ab}$ is the effective energy-momentum tensor
\begin{eqnarray}\label{eq:effective_emt}
 \tau^{\pm}_{ab} = \sum_{\ell=1}^{N}\left[T^{\pm}_{\ell ab} + \delta\left( \tau(x_{\pm}) - \tau_{\ell \star}\right)Z^{\pm}_{\ell ab} + \delta\left( R(x_{\pm}) - R_{\ell \star}\right)
      \tilde{Z}^{\pm}_{\ell ab}  \right]. 
 \end{eqnarray}
The boundary energy flux must satisfy
\begin{eqnarray}\label{eq:boundary_conditions}
\sum_{{\ell=1}}^{N}\left[\mathcal{L}_{\ell a}^{+}  -\mathcal{L}_{\ell a}^{-}\right] = 0, \qquad \qquad 
\sum_{\ell =1}^{N}\left[\tilde{\mathcal{L}}_{\ell a}^{+}  -\tilde{\mathcal{L}}_{\ell a}^{-} \right] = 0\,.
\end{eqnarray}

\section{Geometric backreaction  as energy momentum tensor}\label{sec:backreaction}

The backreaction contribution to the effective energy-momentum tensor can be interpreted as an effective fluid with non-trivial contributions.  To appreciate this, we  perform an irreducible covariant decomposition of ${Z}^{ab}_{\ell\pm} $ with respect to the four velocity $u^{a}_{\ell}$
\begin{eqnarray}\label{eq:Sab}
{Z}^{ab}_{\ell \pm} = \rho_{{\ell\pm}} {u}^a_{{\ell\pm}} {u}^b_{{\ell\pm}} + P_{{\ell\pm}}{h}^{ab}_{{\ell \pm}}
  +2 q^{(a}_{{\ell \pm}} {u}^{b)}_{{\ell \pm}} + \pi^{\<ab\>}_{{\ell \pm}} \,,
\end{eqnarray}
where $ \rho_{{\pm}} $, $P_{{\pm}}$, $q_{{\pm} a}$ and $\pi^{\<ab\>}_{{\pm}}$ are the spacelike boundary energy density, boundary pressure,  boundary energy flux vector and boundary aisotropic stress tensor respectively:
\begin{eqnarray}\label{eq:shell_crossing_EMT}
{\rho}_{{\pm}}  ={Z}_{ \pm ab}{u}^a_{\ell\pm} {u}^b_{\ell\pm}\,,
 \qquad P_{\pm}  = \frac{1}{3}h_{\pm ab} {Z}^{ab}_{\pm}\,, \qquad 
q_{{\pm} \<a\>} = -{Z}^{\pm}_{\<a\>b} {u}^b_{\ell\pm}\,, \qquad  \pi^{\pm}_{\<ab\>} = {Z}^{\pm}_{\<ab\>} \,.
\end{eqnarray}
Using the covariant decomposition of $\nabla_{a} u^{{\pm}}_{\ell b}$ is given in equation \eqref{eq:decomposeCDU}  in the expression for the conjugate momenta given in equations  \eqref{eq:conjugate_momenta1} and \eqref{eq:conjugate_momenta3},  ${Z}_{ \pm ab} $ becomes
\begin{eqnarray}
{Z}_{ \pm ab} =  \Pi^{{\pm}}_{ab}+2L^{{\pm}}_{(b} u^{{\pm}}_{\ell a)} =\frac{1}{\kappa} \bigg[  {K}_{\pm}h^{\pm}_{ab}-K^{\pm}_{ab} \bigg] +\frac{2}{\kappa}L^{\pm}_{(b} u^{{\pm}}_{\ell a)}
=\frac{1}{\kappa} \bigg[  \frac{2}{3}\Theta_{{\pm}} h^{{\pm}}_{ab} - \sigma^{{\pm}}_{ab} + 2A^{{\pm}}_{(b} u^{{\pm}}_{\ell a)}  \bigg]\,.
\end{eqnarray}
In terms of these observables, the components of the effective energy-momentum tensor become
\begin{eqnarray}\label{eq:pressure_def}
{\rho}_{\pm}  &=& 0\,, ~~~~~~~~~~~~
P^{\pm}  =  \frac{2} {3\kappa}{ \Theta}^{\pm}  \,, 
~~~~~~~~
q^{\pm }_{\<a\>}  =\frac{1}{\kappa}  A^{{\pm}}_{a} = 0 \,,
~~~~~~~~
 \pi^{\pm}_{\<ab\>} =-\frac{1} {\kappa}{ \sigma}^{{\pm}}_{ab}  \,.
\end{eqnarray}
where ${\rho}_{\pm} $ vanishes because ${Z}^{ab}_{\pm}$ is spatial hypersurface  projected tensor, $q^{\pm }_{\<a\>} $ vanishes because $u^a$ is  geodesic. $P^{\pm}  $ vanishes because $Z_{ab}$ is evaluated at the boundary, $\Theta_{{\pm}}=0$. 
Only the anisotropic stress tensor is non-vanishing at the boundary.

Similarly, we project the components of $\tilde{Z}^{ab}_{\pm}$ to the spatial hypersurface and  perform an irreducible covariant decomposition  with respect to the four velocity of an observer at rest in the gravitational field
\begin{eqnarray}\label{eq:Sab}
\tilde{Z}^{ab}_{\pm} = \tilde{\rho}_{{\pm}} {u}^a_{{\pm}} {u}^b_{{\pm}} + \tilde{P}_{{\pm}}{h}^{ab}_{{\pm}}
  +2 \tilde{q}^{(a}_{{\pm}} {u}^{b)}_{{\pm}} + \tilde{\pi}^{\<ab\>}_{{\pm}} \,,
\end{eqnarray}
where $\tilde{ \rho}_{{\pm}} $, $\tilde{P}_{{\pm}}$, $\tilde{q}_{{\pm} a}$ and $\tilde{\pi}^{\<ab\>}_{{\pm}}$ are the timelike boundary energy density, boundary pressure,  boundary energy flux vector and boundary aisotropic stress tensor respectively:
\begin{eqnarray}\label{eq:shell_crossing_EMT}
\tilde{\rho}_{{\pm}}  =\tilde{Z}_{ \pm ab}{u}^a_{\pm} {u}^b_{\pm}\,,
 \quad \tilde{P}_{\pm}  = \frac{1}{3}h_{\pm ab} \tilde{Z}^{ab}_{\pm}\,, \quad 
\tilde{q}_{{\pm} \<a\>} = -\tilde{Z}^{\pm}_{\<c\>b} h_{\pm a}{}^{c} {u}^b_{\pm}\,, \quad \tilde{ \pi}^{\pm}_{\<ab\>} = h_{\pm a}{}^{c}h_{\pm b}{}^{d}  \tilde{Z}^{\pm}_{\<cd\>} \,.
\end{eqnarray}
Using the definition of the conjuagte momenta given in equations \eqref{eq:conjugate_momenta4} and \eqref{eq:conjugate_momenta5}, we can express $\tilde{Z}^{\pm}_{ab} $ in terms of observerables associated with the irreduciable decomposion of $ \nabla_{a} r_{\pm b}$.
Note that $\gamma_{\pm ab}$ is the projected metric tensor on the timelike hypersurface orthogonal to $r^a_{\pm}$. $\gamma_{\pm ab}$  is  related  to $h^{\pm}_{ab}$ according  
\begin{eqnarray}
h^{\pm}_{ab}=   \gamma^{\pm}_{ab}+r^{\pm}_{a}r^{\pm}_{b} +u^{\pm}_{a}u^{\pm}_{b}   \qquad  \qquad 
 \gamma^{\pm}_{ab} = h^{\pm}_{ab}-r^{\pm}_{a}r^{\pm}_{b} -u^{\pm}_{a}u^{\pm}_{b} \,.
\end{eqnarray}
And the two orthogonal vectors on the two hypersurfaces(timelike and spacelike) are related according to $r^{\pm}_{a} u^a_{\pm} = 0$, $u^{\pm}_a u^a_{\pm} = -1$ and $r^{\pm}_a r^a_{\pm} = 1$.
%
Using these, it is straight-forward to express $ \tilde{Z}^{\pm}_{ab}$ in terms of the geometric quantities
\begin{eqnarray}
 \tilde{Z}^{\pm}_{ab} =  - \frac{1}{\kappa}  \bigg[  \tilde{K}_{\pm}\gamma^{\pm}_{ab}
-\tilde{K}^{\pm}_{ab} \bigg]   -  \frac{2}{\kappa}\tilde{L}^{\pm}_{(b} r^{\pm}_{a)}=  -\frac{1}{\kappa}\bigg[ \bigg[ \frac{2}{3}\tilde{\Theta}_{\pm}\gamma^{\pm}_{ab}  - \tilde{\sigma}^{\pm}_{\<ab\>} \bigg]+2\tilde{A}^{\pm}_{(b} r^{\pm}_{a)}]  \bigg]
\end{eqnarray}
where $\gamma^{\pm}_{ab} u^a_{\pm} = u^{\pm}_b$. It becomes immediately clear that the components of the energy-momentum tensor are given by
\begin{eqnarray}
\tilde{\rho}_{{\pm}}  &=&\tilde{Z}_{ \pm ab}{u}^a_{\pm} {u}^b_{\pm} 
= \frac{1}{\kappa} \bigg[ {u}^a_{\pm} {u}^b_{\pm} \tilde{\sigma}_{\<ab\>} \bigg]\,,
\label{eq:geometric_matter}
\\
\tilde{P}_{\pm} & =&\frac{1}{3}h_{\pm ab} \tilde{Z}^{ab}_{\pm}=\frac{1}{3}\frac{1}{\kappa}\bigg[ u^{\pm}_{a}u^{\pm}_{b}  \tilde{\sigma}^{\<ab\>}\bigg]\,,
\label{eq:geometric_pressurer}
\\
\tilde{q}_{{\pm} \<a\>} &=& -\tilde{Z}^{\pm}_{\<c\>b} h_{\pm a}{}^{c} {u}^b_{\pm}
=\frac{1}{\kappa} \bigg[ - \tilde{\sigma}_{\<cb\>}h^{\pm c}{}_{a} u^{b} \bigg] \,,
\\
 \tilde{ \pi}^{\pm}_{\<ab\>} &=& h_{\pm a}{}^{c}h_{\pm b}{}^{d}  \tilde{Z}^{\pm}_{\<cd\>}
 = h_{\pm a}{}^{c}h_{\pm b}{}^{d}   \frac{1}{\kappa}\bigg[   \tilde{\sigma}_{\<cd\>} \bigg] \,.
\end{eqnarray}
      Note that $\tilde{\rho}_{{\pm}} $, $\tilde{P}_{\pm} $, $\tilde{q}_{{\pm} \<a\>} $ and $ \tilde{ \pi}^{\pm}_{\<ab\>} $ capture the impact of the backreaction of the spacetime on particle propagation.   
      Most importantly, the induced surface pressure is related to the induced surface energy density according to 
$
\tilde{P}_{\pm}= \tilde{\rho}_{\pm} /3\,.
$


 Finally, the full effective energy-momentum tensor for the ensemble of massive particles(baryons) plus the geometric backreaction contribution  is given by
 \begin{eqnarray}\label{eq:total_emt}
 \tau^{ab}_{{\pm}}= \sum_{\ell =1}^{N} \tau_{\pm\ell}^{ab} \approx \sum_{\ell =1}^{N} \bigg[
 \rho^{\pm}_{\ell T}  {u}^a_{\pm \ell} {u}^b_{\pm \ell} + \hat{P}_{\ell \pm} {h}^{ab}_{\pm \ell}   +2 \hat{q}^{(a}_{\ell \pm} {u}^{b)}_{\pm \ell} + \hat{\pi}^{\<ab\>}_{\ell \pm} \bigg]\,,
\end{eqnarray}
where $\rho_{T}$ is the effective matter density field. It is a sum of the matter density for an ensemble of point particles  and the backreaction contribution due to the local  environment:  
\begin{eqnarray}
\rho_{ T \ell} =  \rho_{\ell m \pm} +\hat{\rho}_{\ell \pm}\,,
\end{eqnarray}
where $ \hat{\rho}_{\ell \pm} =  \delta\left( R_{\pm}(x_{\pm}) - R_{\ell \star}\right)\tilde{\rho}_{\ell {\pm}} $ and the baryonic matter density 
\begin{eqnarray}\label{eq:ordinarymatter}
 \rho_{\ell m +}  &=& 
  \frac{ m^{+}_{\ell}}{\sqrt{h^{+}}}\delta^{3}\left(x^i_{+} - \gamma^i_{+ \ell}(t_{+})\right) ~~ {\rm{for}} ~~ 0 \leq \tau(t) \leq \tau_{\star\ell}\,,
\\ 
 \rho_{\ell m -}  &=&   
 \frac{ m^{-}_{\ell}}{\sqrt{h^{-}}}\delta^{3}\left(x^i_{-} - \gamma^i_{- \ell}(t_{-})\right) ~~ {\rm{for}} ~~ \tau_{\star\ell}\leq \tau(t) \leq \tau_{\rm{final}\ell} \,.~~~~~~~
\end{eqnarray}
Note that the density is supported only on spacetime points that lie on the particle worldline.
We performed the delta function integration using the shifting property of the Dirac delta function, 
with the delta function decompostion  $\delta^{(4)}\left( x^i_{\pm}-x_{\ell}(\tau_{\pm\ell})\right) = \delta\left(t_{\pm}- t_{\pm \ell}(\tau_{\pm})\right)\delta^{3}\left(x^i_{\pm} - \gamma^i_{\pm \ell}(t_{\pm}) \right)$, $\sqrt{-g_{\pm }}u^{0 }_{\pm}  = \sqrt{h_{\pm}}$, where  $u^{0 }_{\pm}= {\d t_{\pm}}/{\d \tau_{\pm}}$ for $ t_{\pm}(\tau_{\rm{ini}}) \le t_{\pm}  \le t_{\pm \ell}(\tau_{\star})$ .
The induced pressure/surface tensor  is given by
$
\hat{P}_{\ell \pm}  =  \delta\left( R_{\pm}(r_{\pm}) - R_{\star \ell}\right) \tilde{P}_{{\pm} \ell}  \,.
$
 The  effective energy flux is given by:  
$\hat{q}^{a}_{\pm}  =  \delta\left( R_{\pm}(r_{\pm}) - R_{\star \ell}\right)    \tilde{q}^{a}_{{\pm} \ell} \,. $
The effective  anisotropic stress tensor  is given by
\begin{eqnarray}
  \hat{\pi}^{\<ab\>}_{{\pm}} &=& \left[ \delta\left( \tau^{+}(x^{\pm}) - \tau^{\pm}_{\star}\right)-\delta\left( \tau^{+}(x^{\pm}) - \tau^{\pm}_{\rm{ini}}\right)\right] \pi^{\<ab\>}_{{\pm}\ell}  +  \delta\left( R_{\pm}(r_{\pm}) - R_{\star \ell}\right)   \tilde{\pi}^{\<ab\>}_{{\pm}\ell}\,.
  \\
  &\approx& \delta\left( R_{\pm}(r_{\pm}) - R_{\star \ell}\right)   \tilde{\pi}^{\<ab\>}_{{\pm}\ell}\,.
\end{eqnarray}
In general, the contribution of $\left[ \delta\left( \tau^{+}(x^{\pm}) - \tau^{\pm}_{\star}\right)-\delta\left( \tau^{+}(x^{\pm}) - \tau^{\pm}_{\rm{ini}}\right)\right] \pi^{\<ab\>}_{{\pm}\ell}$ is non-zero, but we have made a simplification approximation to reduce clutter. 
The shear tensor $\tilde{\sigma}_{\pm ab}$, which drives the secondary contribution to the energy-momentum tensor, is a consequence of spatial variations in the local environment.

\subsection{ From microscopic particle picture to macroscople fluids}\label{sec:effectivefluid}

Equation \eqref{eq:total_emt} gives the total microscopic contributions to the energy-momentum tensor. The contribution is labelled by particle position, $\gamma^{i}$ and the matter horizon associated with the particle $R_{\star}$. We are interested in the total stress–energy tensor a coarse-grained observer would infer if they only see averages over finite resolution.
To obtain this, we replace the exact positions of particles with a continuous density distribution. That is, we take all the individual energy-momentum contributions from every point particle and the associated backreaction contribution and sum them up while respecting the curvature of the local spacetime:
\begin{eqnarray}\nonumber
 \tau^{ab}_{\pm\rm{fluid}}
&\equiv& \frac{1}{\Delta V_{\pm}}  \frac{1}{\Delta R_{\pm}}  \sum _{i=1}^{n} \sum _{j=1}^{m}  \tau^{ab}_{\pm}(\gamma^{\pm}_{i}, R_{\star j}) \Delta \gamma^{i}_{\pm} \Delta R_{\star}
\\ &&
 \xrightarrow{\Delta \gamma_{\pm}\rightarrow 0,\Delta R_{\star}\rightarrow 0} 
  \frac{1}{ V_{\pm}}   \frac{1}{ R_{\pm}} \int_{\Sigma_{\pm}}\int_{R_{\pm}}   \tau^{ab}_{\pm} \sqrt{h_{\pm}} \, \sqrt{h^{\pm}_{RR}}  d^3\gamma_{\pm} \d r_{\star}\,,
  \label{eq:fluidEqn}
 \end{eqnarray}
 where $\gamma^{i}$ is the coordinate position of a microscopic particle, $R_{\star}$ is the location of the matter horizon, which could a proxy for the particle's physical size. The sum is over the microscopic stress–energy normalised by a small spatial volume it occupies, and another sum over a small range of $R_{\star}$ normalised by the length element. 
 In the limit $\Delta \gamma^{i}_{\pm} / \Delta V_{\pm} \ll1 $ and $\Delta R_{\star}/{\Delta R_{\pm}}  \ll 1 $, we replace the sum with integrals, where  $R_{\pm}  = \int   \sqrt{h^{\pm}_{RR}} \d r_{\star}$ and $V_{\pm} = \int \d^{3 } \gamma_{\pm}\sqrt{h_{\pm}}$.
 To perform the integrals in equation \eqref{eq:fluidEqn},  we decompose the four velocities of individual microsopic particles $ u^a_{\ell} $ into average and random parts:
  $u^a_{\ell} = u^a_{\rm{bulk}} + w^a_{\ell}$, where  $u^a_{\rm{bulk}} $ is the bulk 4-velocity and  $w^a_{\ell}$ is a random or thermal velocity. Just to reduce clutter,  we use $ u^{a}$ in place of $u^a_{\rm{bulk}} $ in the rest of the discussion. 
The projected metric tensor is similarly split as follows: ${h}^{\pm ab \ell} = {h}^{\pm  ab}+\left[2u^{\pm (a}{w}^{\pm b)}
+{w}^{\pm a}_{\ell} {w}^{\pm b}_{\ell}\right] $, then putting equation \eqref{eq:total_emt}  in equation \eqref{eq:fluidEqn} and performing all the algebraic simplification is straightforward, but the physics involved deserves some explanation.   The first term in equation \eqref{eq:total_emt}  describes the contribution to the total energy-density from N-number of microscopic particles: $ \sum_{\ell =1}^{N} 
 \rho^{\pm}_{\ell T}  {u}^a_{\pm \ell} {u}^b_{\pm \ell} $, where 
$\rho_{ T \ell} =  \rho_{\ell m \pm } +\hat{\rho}_{\ell \pm}$. Although both $\rho_{\ell m \pm } $ and $\hat{\rho}_{\ell \pm}$ contribute to $\rho_{ T \ell} $, they have  different origins.  For $\rho_{\ell m \pm }$ we have 
\begin{eqnarray}\label{eq:quadvelocity}
     \sum_{\ell =1}^{N}  m^{\pm}_{\ell} {u}^a_{\pm \ell} {u}^b_{\pm \ell}     &=&    \sum_{\ell =1}^{N}  m^{\pm}_{\ell}
     u^a u^b +       \sum_{\ell =1}^{N} m^{\pm}_{\ell}u^a w^b_{\ell} +      \sum_{\ell =1}^{N} m^{\pm}_{\ell}w^a_{\ell} u^b +      \sum_{\ell =1}^{N} m^{\pm}_{\ell}w^a_{\ell} w^b_{\ell}  \,.
 \end{eqnarray}
  We group the terms based on their corresponding physical interpretation: The first term denotes the baryonic mass-energy density ($\rho_{m}$). It is given by the sum of the effective masses per unit volume.
  $\rho_{\pm m} \equiv \sum_{\ell \in  V} m_{\ell}/{ V_{\pm}}  = M^{\pm}/V_{\pm}$, where $ M^{\pm} =\sum_{\ell \in  V} m^{\pm}_{\ell} $ . The energy density associated with the bulk coherent motion of the fluid is given by $\rho_{\pm m} u^a_{\pm} u^b_{\pm}$.
The weighted average of the velocity  fluctuations vanishes: $\sum_{\ell =1}^{N}  m_{\ell} w^a_{\ell}  = 0$ by the definition of a mean flow, 
 The fourth terms of the RHS of equation \eqref{eq:quadvelocity} lead to an anisotropic stress-tensor: 
     $   \mathcal{P}^{ab}_{\pm}  =   \frac{1}{ V} \sum_{\ell =1}^{N} m^{\pm}_{\ell}w^a_{\ell} w^b_{\ell} $. This can be decomposed into isotropic and anisotropic parts: $
\mathcal{P}^{ab}_{\pm} = \mathcal{P}_{\pm} h^{ab}_{\pm} + \pi^{ab}_{\pm},$
where $\mathcal{P}_{\pm} = \frac{1}{3} h_{ab} \mathcal{P}^{ab}_{\pm} = \frac{1}{3} \frac{1}{V} \sum_{\ell \in  V} m_{\ell} w^a_{\ell} w^b_{\ell} $  is the isotropic pressure,  and $
 \mathcal{P}^{\langle ab \rangle}_{\pm} = \big( h^{(a}_{\;c} h^{b)}_{\;d} - \frac{1}{3} h^{ab} h_{cd} \big) \mathcal{P}^{cd}_{\pm} $ is the traceless symmetric projection of the anisotropic stress.
 Therefore, the fluid limit of the baryonic matter density  is given by
 \begin{eqnarray}
  \sum_{\ell =1}^{N} 
 \rho^{\pm}_{\ell m}  {u}^a_{\pm \ell} {u}^b_{\pm \ell} \bigg|_{\rm{fluid}} =    u^a_{\pm} u^b_{\pm} \rho^{\pm}_{m \pm}  + \mathcal{P}_{\pm}  h^{ab}_{\pm} + \mathcal{P}^{\langle ab \rangle}_{\pm}  \,.
 \end{eqnarray}
 Similarly, the contribution from the backreaction-induced energy density is given by
  \begin{eqnarray}\label{eq:backreactiondensity}
  \sum_{\ell =1}^{N} 
\hat{ \rho}^{\pm}_{\ell }  {u}^a_{\pm \ell} {u}^b_{\pm \ell} \bigg|_{\rm{fluid}} =
     u^a_{\pm} u^b_{\pm} \hat{\rho}_{\pm}
      + \hat{\mathcal{P}}_{\pm}  h^{ab}_{\pm} + \hat{\mathcal{P}}^{\langle ab \rangle}_{\pm} \,,
 \end{eqnarray}
 where  $\hat{\rho}_{\pm} = \sum_{\ell = 1}^{N}   \tilde{\rho}_{\pm \ell} /R_{\pm}$,  $ \hat{ \mathcal{P}}_{\pm}  $ is the corrresponding isotropic pressure due to random velocity components,   $  \hat{ \mathcal{P}}_{\pm}  =   \frac{1}{ 3R} \sum_{\ell =1}^{N}  \tilde{\rho}_{\pm \ell} w^a_{\ell} w_{a\ell} $  and $ \hat{ \mathcal{P}}^{\langle ab \rangle }_{\pm} $ is the associated anisotropic stress tensor
       $  \hat{ \mathcal{P}}^{ab}_{\pm}  =   \frac{1}{ V} \sum_{\ell =1}^{N}  \tilde{\rho}_{\pm \ell} w^{\<a}_{\ell} w^{b\>}_{\ell} /R$
   and $\sum_{\ell =1}^{N}  \tilde{\rho}_{\pm \ell} v^a_{\ell} /R = 0$ by definition. 

For the  backreaction induced pressure $\hat{P}_{\ell \pm}   =  \delta\left( R_{\pm}(r_{\pm}) - R_{\star \ell}\right) \tilde{P}_{{\pm} \ell}  $, we make use of ${h}^{\pm ab \ell} = {h}^{\pm  ab}+\left[2u^{\pm (a}{w}^{\pm b)}
+{w}^{\pm a}_{\ell} {w}^{\pm b}_{\ell}\right] $ to split the contribution into coherent and random parts
\begin{eqnarray}
    \sum_{\ell =1}^{N} \hat{P}_{\ell \pm} {h}^{ab}_{\pm \ell}  
 \bigg|_{\rm{fluid}}   &=&\hat{P}_{ \pm} {h}^{ab}_{\pm}   + \mathcal{S}^{ab}_{\pm}  = \hat{P}_{ \pm} {h}^{ab}_{\pm} +  \mathcal{S}_{\pm} h^{ab}+  \mathcal{S}^{\langle ab \rangle}_{\pm}
 \end{eqnarray}
 where $\hat{P}_{\ell \pm}  =    \sum_{\ell =1}^{N} \tilde{P}_{{\pm} \ell} /R$ and  $\mathcal{S}^{ab}_{\pm} =     \sum_{\ell =1}^{N} \tilde{P}_{{\pm} \ell}  {v}^{\pm a}_{\ell} {v}^{\pm b}_{\ell}/R  = \mathcal{S}_{\pm} h^{ab}+  \mathcal{S}^{\langle ab \rangle}_{\pm} $. The isotropic pressure is defined as 
 $\mathcal{S} _{\pm} = \frac{1}{3} h_{ab}\mathcal{S}^{ab}_{\pm}   = \frac{1}{3R} \sum_{\ell =1}^{N} \tilde{P}_{\pm \ell}  {v}^{\pm a}_{\ell} {v}^{\pm }_{a \ell}$ and the anisotrpic stress $ \mathcal{S}^{\langle ab \rangle}_{\pm}=
 \mathcal{P}^{\langle ab \rangle}_{\pm} = \big( h^{(a}_{\;c} h^{b)}_{\;d} - \frac{1}{3} h^{ab} h_{cd} \big)\mathcal{S}^{cd}_{\pm} $.
The heat flux  $\hat{q}^{(a}_{\ell \pm} {u}^{b)}_{\pm \ell} $ is given by $    \sum_{\ell =1}^{N} \hat{q}^{(a}_{\ell \pm} {u}^{b)}_{\pm \ell}  \big|_{\rm{fluid}}   = \hat{q}^{(a}_{\pm} {u}^{b)}_{\pm } $, 
   where $\hat{q}^{(a}_{\pm}  =  \sum_{\ell =1}^{N}  \tilde{q}^{a}_{\ell \pm}  /R$ and  anisotropic stress tensor is 
   $  \sum_{\ell =1}^{N} \hat{\pi}^{\langle ab\rangle}_{\ell \pm} \big|_{\rm{fluid}}    = \hat{\pi}^{\langle ab\rangle}_{\pm}$
    where $ \hat{\pi}^{\langle ab\rangle}_{\pm} =    \sum_{\ell =1}^{N}  \tilde{\pi}^{\langle ab\rangle}_{\ell \pm} /R$.

Implementing all these in equation \eqref{eq:total_emt}  immediately leads to an  energy-momentum tensor of the form
 \begin{eqnarray}\label{eq:total_emt_dust}
 \tau^{ab}_{\rm{fluid}} =
 \rho^{\pm}_{T}  {u}^a_{\pm} {u}^b_{\pm} + {P}_{ T\pm} {h}^{ab}_{\pm}   +2 {q}^{(a}_{T \pm} {u}^{b)}_{\pm} + {\pi}^{\<ab\>}_{T\pm} \,,
\end{eqnarray}
where $ {P}_{ T\pm}  =  \hat{P}_{ \pm}  +\mathcal{P}_{\pm}  + \hat{\mathcal{P}}_{\pm}+\mathcal{S}_{\pm} $, $ {q}^{(a}_{T \pm}  =  \hat{q}^{(a}_{ \pm} $ and ${\pi}^{\<ab\>}_{T\pm}  = \hat{\pi}^{\<ab\>}_{\pm} + \mathcal{P}^{\langle ab \rangle}_{\pm}  
+ \hat{\mathcal{P}}^{\langle ab \rangle}_{\pm}  +\mathcal{S}^{\langle ab \rangle}_{\pm} $. 
 These are well-defined observables without any singularities in the large N limit.  
 Furthermore, viscosity is a macroscopic manifestation of the exchange of momentum by microscopic particles through collisions.
The viscosity has two distinct physical effects: Bulk Viscosity ($\zeta$), which describes the resistance to uniform expansion or collapse and shear viscosity ($\eta$), which describes the resistance to the sliding motion of fluid layers (shape deformation).
For the bulk viscosity, we parameterise the induced pressure as $\hat{P}_{\pm} =\zeta
\sum_{{\ell = 1}}^{N}  \frac{1}{3}\frac{1}{\kappa}\big[ u^{\pm}_{ \ell a}u^{\pm}_{\ell b}  \tilde{\sigma}^{\<ab\>}_{\ell}\big]$
which then implies that the equation of state for the induced  isotropic  pressure  and the induced density is given by
$\hat{P}_{ \pm}  = \frac{1}{3} \zeta \hat{\rho}_{ \pm} $. For the shear viscosity, we have 
$ \hat{\pi}^{\<ab\>}_{\pm}  = \eta \sum_{{\ell = 1}}^{N}  \tilde{\sigma}^{\<ab\>}$.

Finally, the total pressure contribution can be simplified further by defining an observable-weighted average of the squared random velocity as
\begin{eqnarray}
\langle w^2 \rangle_{\rm{m}} = \frac{\sum m_{\pm \ell} | w_\ell|^2}{\sum m_{\pm\ell}} = \sigma_{m \rm{tot}}^2\,,
\quad 
\langle {w}^2 \rangle_{\hat{\rho}_{\pm}} = \frac{\sum \tilde{\rho}_{\pm \ell}  | v_\ell|^2}{\sum \tilde{\rho}_{\pm \ell} } = \sigma_{\hat{\rho} \rm{tot}}^2
\quad 
\langle {w}^2 \rangle_{\hat{P}} = \frac{\sum \tilde{P}_{\pm \ell}  | w_\ell|^2}{\sum \tilde{P}_{\pm \ell} } = \sigma_{\hat{P} \rm{tot}}^2.~~~~
\end{eqnarray}
Therefore, the total pressure is given by
\begin{eqnarray}\label{eq:pressureterm}
 {P}_{ T\pm}  
= \frac{1}{3} \zeta \hat{\rho}_{ \pm}  + \rho_{ m \pm } \sigma_{1D}^2 + \hat{\rho} \sigma_{\hat{\rho} 1 D}^2 + 
  \frac{1}{3} \zeta \sigma_{\hat{P} 1 D}^2\hat{\rho}_{ \pm}  
    \,,
\end{eqnarray}
where we have made a common approximation in cosmology $\sigma_{Xtot}^2 = 3\sigma_{X1D}^2$, since most cosmological surveys detect sources along a line of sight.

\subsection{Darmois-Israel  boundary stress tensor is a limit}

The additional contribution to the bulk energy-momentum tensor shown in equation \eqref{eq:effective_emt}, that is  $ \delta\left( \tau(x_{\pm}) - \tau_{\star}\right)Z^{\pm}_{ab} + \delta\left( R(x_{\pm}) - r_{\star}\right)
      \tilde{Z}^{\pm}_{ab} $ is a consequence of the backreaction of the spacetime due to the fact that any massive particle has a finite size. 
This contribution is a re-expression of the boundary stress-energy tensor that arises in Israel's formulation of Junction conditions in GR, i.e see equation \eqref{eq:Darmois}~\cite{Israel:1966rt}. The key difference, which we have already discussed in detail in sub-section \ref{sec:boundary_metrics}  is that the Darmois continuity condition on the boundary metric tensor does not consistently capture the type of boundaries that arise in the case of the matter horizon in an expanding universe.  In an expanding universe, the local environment responsible for star formation decouples from the Hubble flow because it cannot catch up with the expansion rate of the Hubble volume. This is discussed in detail in \cite{Ellis:2010fr}.

To show that our approach is consistent,  we show below how to recover the thin shell stress energy tensor associated with the Darmois continuity condition. The straight-forward way to achieve this is by adding  a boundary term to the bulk matter action  $S_{M}\left[g_{ab}\right]  +    S_{\mathrm{BM}}\left[h_{ab}\right]+  S_{\mathrm{BM}}\left[\gamma_{ab}\right]$ (equation \eqref{eq:matter_action}) 
\begin{eqnarray}\label{eq:boundary_action}
   S_{\mathrm{BM}}\left[h_{ab}\right] = \int _{\Sigma}  
   \mathcal{L}_{\rm{BT}} \left[h_{ab} \right]  \d^3 x_{\pm} 
   \qquad \qquad    \tilde{S}_{\mathrm{BM}}\left[\gamma_{ab}\right] = \int _{B}  
  \tilde{ \mathcal{L}}_{\rm{BT}} \left[\gamma_{ab} \right] \d^3 x_{\pm}\,,
\end{eqnarray}
where   $\mathcal{L}_{\rm{BT}}  =L_{\rm{BT}} \sqrt{h}$ and $\tilde{\mathcal{L}}_{\rm{BT}}  =L_{\rm{BT}} \sqrt{\gamma}$ is the Lagrangian density that depends on the boundary metric $h_{ab}$ and $\gamma_{ab}$ respectively.  
And then imposig Darmois condition for the boundary metric tensor  \eqref{eq:Darmois} , i.e $h^{-}_{ab} = h^{+}_{ab} = h^{ab}$ and $\gamma^{-}_{ab} = \gamma^{+}_{ab} = \gamma^{ab}$
and varying equation \eqref{eq:boundary_action} gives 
\begin{eqnarray}
 \delta    S_{\mathrm{BM}}\left[h_{ab}\right] 
  = -\frac{1}{2}\int_{\Sigma}S_{ab} \delta h^{ab}_{\pm}  \d \Sigma \,.
  \qquad
    \delta   \tilde{ S}_{\mathrm{BM}}\left[h_{ab}\right]  = -\frac{1}{2}\int_{ B}\tilde{S}_{ab} \delta \gamma^{ab}_{\pm}  \d B\,,
\end{eqnarray}
where $ S_{ab} =-2 {\delta \mathcal{L}_{\rm{BT}} }/({\sqrt{h}\delta h^{ab}_{\pm}} )$ and $ \tilde{S}_{ab} =-2 \delta \tilde{\mathcal{L}}_{\rm{BT}} /({\sqrt{\gamma}\delta \gamma^{ab}_{\pm}} )$ are boundary stress-energy tensors. 
Putting this back into the varied total gravity action (equation \eqref{eq:varied_gravity_action})  gives
\begin{eqnarray}\nonumber 
 \frac{\delta S_{\rm{Full}}\left[ g_{ab}\right] }{\delta g_{ab}}&=&
  \int_{\mathcal{M}^{+}}\frac{1}{2}\left(\frac{1}{\kappa}G^{+}_{ab}+ \Lambda g^{+}_{ab}-T^{+}_{ab}\right)\delta g^{ab}_{+}\,\sqrt{-g_{+}} d^4x_{+} 
   \\ \nonumber &&
  +  \int_{\mathcal{M}^{-}}\frac{1}{2}\left(\frac{1}{\kappa}G^{-}_{ab}+ \Lambda g^{-}_{ab}-T^{-}_{ab}\right)\delta g^{ab}_{-}\,\sqrt{-g_{-}} d^4x_{-}
 \\ &&
      +   \int _{\Sigma_{+}}  
      \delta h_{+}^{ab} \left[ \left(Z^{+}_{ab}  - Z^{-}_{ab}\right)  -S_{ab}  \right]\d \Sigma_{+}
          +   \int _{B_{+} }\delta \gamma_{+}^{ab}\left[ \left(\tilde{Z}^{+}_{ab}  -\tilde{Z}^{-}_{ab}\right)- \tilde{S}_{ab} \right]
            \d B_{}{+} \,.
                   \label{eq:varied_action_BT}
\end{eqnarray}
Imposing proper variation of the total action and piece-wise continutity at the boundary lead to the Einstein field equation $G^{\pm}_{ab}+ \Lambda g^{\pm}_{ab} = \kappa T^{\pm}_{ab}$  and the expression for the boundary stress-energy tensor 
\begin{eqnarray}\label{eq:Israel_limit}
S^{ab}=\frac{1}{\kappa}
\left[  Z^{+}_{ab}-Z^{-}_{ab}\right] \,,
\qquad \tilde{S}^{ab}=\frac{1}{\kappa}
\left[  \tilde{Z}^{+}_{ab}-\tilde{Z}^{-}_{ab}\right] \,.
\end{eqnarray}
This is the well-known second Israel junction condition for the boundary stress tensor \cite{Israel:1966rt},  This is usually written in this form $\kappa S_{ab} =  [K_{ab}] - [K]h_{ab} 
$, 
where $[K_{ab}] = K_{ab}^{+} - K_{ab}^{-}$ is the jump in extrinsic curvature,
and $ [K] = K_{+} - K_{-}$ is the jump in trace of $[K_{ab}] $.

\subsection{Energy-momentum conservation equation}

The invariance of the  total action $ S_{\rm{EH}}[g^{+}_{ab}] +S_{\rm{M}}[g^{+}] + S_{\mathrm{GHY}}\left[h^{+}_{ab}\right]  + S_{\mathrm{GHY}}\left[\gamma^{+}_{ab}\right]  +  S_{\mathrm{Hayward}}\left[N^{{+}}_{ab}\right]+ 
S_{\rm{EH}}[g^{-}_{ab}] + S_{\rm{M}}[g^{-}]  +  S_{\mathrm{GHY}}\left[h^{-}_{ab}\right]  + S_{\mathrm{GHY}}\left[\gamma^{-}_{ab}\right]  
+   S_{\mathrm{Hayward}}\left[N^{{-}}_{ab}\right]$ under diffeomorphism transformation $x^a_{\pm} \to x^a_{\pm}(x)$  leads to a covariant conservation of the energy-momentum tensor.  To appreciate this, we consider infinitesimal diffeomorphisms
$
x^a_{\pm} \to {x'_{\pm}}^a(x) =   x^a_{\pm} + \xi^a_{\pm}(x)\,,
$
where $\xi^a_{\pm}(x)$ is an infinitesimal generator.  
The metric tensors on both manifolds transform under the infinitesimal coordinate transformation as $\delta g^{ab}_{\pm} = - 2 \nabla^{(a} \xi^{b)}_{\pm}\,,$  and a four-vector transforms as $\delta u^a_{\pm} = \xi^{b}_{\pm} \partial_{b} u^{a}_{\pm} - u^{b}_{\pm} \partial_{b} \xi^a_{\pm} = \mathcal{L}_{\xi} u^a_{\pm} = \xi^{b}_{\pm} \nabla_{b} u^{a}_{\pm} - u^{b}_{\pm} \nabla_{b} \xi^a_{\pm} $ and  $\delta r^a_{\pm}  =   \mathcal{L}_{\xi} r^a_{\pm} =\xi^{b}_{\pm} \partial_{b} r^{a}_{\pm} - r^{b}_{\pm} \partial_{b} \xi^a_{\pm} = \xi^{b}_{\pm} \nabla_{b} r^{a}_{\pm} - r^{b}_{\pm} \nabla_{b} \xi^a_{\pm} $.
The components of the infinitesimal variation of the vector can be written as 
\begin{eqnarray}
\delta u^0_{\pm} =  \mathcal{L}_{\xi_{\pm}} u^0_{\pm}, \qquad\qquad \delta u^i_{\pm}  =  \mathcal{L}_{\xi_{\pm}} u^i_{\pm},  \qquad\qquad \delta u^{A}_{\pm}  =  \mathcal{L}_{\xi_{\pm}} u^{A}_{\pm}.
\end{eqnarray}
Similarly for $\delta r^a$.
The projected terms are defined as 
 \begin{eqnarray}
  \mathcal{L}_{\xi_{\pm}}  u^i= h^i{}_{a}  \mathcal{L}_{\xi_{\pm}} u^a\qquad \qquad   \mathcal{L}_{\xi_{\pm}}  u^A= N^A{}_{a}  \mathcal{L}_{\xi_{\pm}} u^a \qquad {\rm{and}}\qquad 
   \mathcal{L}_{\xi_{\pm}} u^0_{\pm}, = u_a  \mathcal{L}_{\xi_{\pm}} u^a_{\pm},
 \end{eqnarray}
 We reqiure that $\xi^a$ preserves foliations, that is  $ \mathcal{L}_{\xi_{\pm}} u^a =0$ and $ \mathcal{L}_{\xi_{\pm}} r^a =0$ vanishes. That is, we require that the  $u^a$ and $r^a $  are "Lie-dragged" along the flow lines without changing its geometric properties. 
 Using integration by parts, it is possible to perform the 4D integration of equation \eqref{eq:varied_action}.
\begin{eqnarray}
 \frac{\delta S_{\rm{Full}}\left[ g_{ab}\right] }{\delta g_{ab}}  &=& \int_{\mathcal{M} _{+}}\xi^{b}_{+} \nabla^{a} \tau^{+}_{ab} \d^4 x_{+}  
 +  \int_{\mathcal{M} _{-}}\xi^{b}_{-} \nabla^{a}  \tau^{-}_{ab} \d^4 x_{-}  +
 \int_{\Sigma_{+}}u^a_{+}\bigg[  \xi^b_{+}   \mathcal{T}^{+}_{ab}  -
  \xi^b_{-}   \mathcal{T}^{-}_{ab} \bigg]\d^3 x_{+}
  \\ \nonumber &&
  + \int_{{}^{(3)}B_{+}}r^a_{+}\bigg[  \xi^b_{+}  \tilde{ \mathcal{T}}^{+}_{ab}  -
  \xi^b_{-}   \tilde{\mathcal{T}}^{-}_{ab} \bigg]\d^3 x_{+}\,,
\end{eqnarray}
where we introduced appropriate tensor densities $\mathcal{T}^{\pm}_{ab} = \left[G^{\pm}_{ab}+\Lambda g^{\pm}_{ab} -\kappa\tau^{\pm}_{ab}\right] \sqrt{h^{\pm}}$ and  $\tilde{\mathcal{T}}^{\pm}_{ab} =\left[\tilde{G}^{\pm}_{ab}+\Lambda g^{-}_{-ab} -\kappa \tilde{\tau}^{-}_{ab} \right]\sqrt{-\gamma^{-}}$ . 
We imposed the Bianchi identities: $ \nabla^a G^{\pm}_{ab} = 0 $ and the metricity condition: $ \nabla_{a} g_{bc} = 0$. These are geometric properties of the Lorentzian(pseudo-Riemannian) spacetime. 
The $\xi_{\pm}^{a}$  is  a geometric object defined independently of the volume form, hence  $ \xi^i_{+}   =  \xi^i_{-}$.
\begin{eqnarray}
 \frac{\delta S_{\rm{Full}}\left[ g_{ab}\right] }{\delta g_{ab}}&=& \int_{\mathcal{M} _{+}}\xi^{b}_{+} \nabla^{a} \tau^{+}_{ab} \sqrt{-g_{+}}\d^4 x_{+}  
 +  \int_{\mathcal{M} _{-}}\xi^{b}_{-} \nabla^{a}  \tau^{-}_{ab} \sqrt{-g_{-}}\d^4 x_{-}  +
  \int_{\Sigma_{+}}\xi^b_{+}  u^a_{+}
 \bigg[ \mathcal{T}^{+}_{ab}-  \mathcal{T}^{-}_{ab}\bigg]\d^3 x_{+}
 \\ \nonumber &&
 +  \int_{{}^{(3)}B_{+}}\xi^b_{+} r^a_{+}
 \bigg[ \tilde{ \mathcal{T}}^{+}_{ab}- \tilde{\mathcal{T}}^{-}_{ab}\bigg]\d^3 x_{+}\,.
\end{eqnarray}
 Imposing piece-wise continuity, we find that the diffeomorphism invariance leads to the  conservation of the energy-momentum tensor on-shell
\begin{eqnarray}\label{eq:conservation_equation}
\nabla^{a}  \tau^{+}_{ab}  = 0\,, \qquad \qquad  \nabla^{a}  \tau^{-}_{ab} = 0\,, 
\qquad \qquad \big[ \mathcal{T}^{+}_{ab} -  \mathcal{T}^{-}_{ab}\big]\bigg|_{\Sigma}  = 0  =\big[\tilde{ \mathcal{T}}^{+}_{ab} -\tilde{\mathcal{T}}^{-}_{ab}\big]\bigg|_{B}\,.
\end{eqnarray}
   The temporal and spatial components of the effective energy-momentum tensor (equation \eqref{eq:total_emt_dust}) in the fluid limit are
\begin{eqnarray}
 \dot{{\rho}}_{T\pm}  +\left({P}_{T\pm} +{\rho}_{T\pm} \right){\Theta}_{\pm} &=&-\left[  2{A}^a_{\pm} {q}_{T\pm a} + {\D}_{a} {q}^a_{T\pm}+ {\Pi}^{ab}_{T\pm} {\sigma}_{T\pm ab}  
 \right]\,,
 \label{eq:conservation_eqn1}
 \\
  \dot{{q}}^a_{T\pm}+ \frac{4}{3} {q}^a_{T\pm} {\Theta}   &=&- \bigg[
    {\D}^a {P}_{T\pm} +  {\D}_{c} {\pi}^{ca}_{T\pm} 
+ {A}^a_{\pm}\left( {P}_{T\pm}  +  \rho_{T\pm}\right)
+{ q}^b_{T\pm}\left({\sigma}^{a}{}_{\pm b} -{\omega}^{a}{}_{\pm b} \right)
   \bigg]\,.
    \label{eq:conservation_eqn2}
\end{eqnarray}
where $ {\pi}^{\<ab\>}_{T{\pm}} u_{\pm a} = 0= {q}^{\<a\>}_{T{\pm}}u_{\pm a}  $ since  both ${\pi}^{\<ab\>}_{T{\pm}} $  and ${q}^{\<a\>}_{T{\pm}} $ are full projected. 

\subsection{Matter frame and CMB frame transformation}

The conservation equations given in equations \eqref{eq:conservation_eqn1} and \eqref{eq:conservation_eqn2} are expressed with respect to the matter four velocity or in the frame where the observer is at rest in the matter field. 
In the case of the universe, the observer is not in the Hubble flow,  therefore,  we suppose that the matter 4-velocity in the Hubble flow is related to the four-velocity of the observer, $\bar{u}^a_{\pm}$, according to ~\cite{Umeh:2010pr},
\begin{eqnarray}\label{eq:relativevelocity}
{u}^a_{\pm}=\gamma\left(\bar{u}^a_{\pm}+v^a_{\pm}\right) \approx \bar{u}^a_{\pm}+v^a_{\pm}\,,\quad v^a_{\pm}{u}^{\pm}_a=0  \quad {\rm{and}}\quad \gamma_{\pm}={{(1-v^2_{\pm})}}^{-\frac{1}{2}} \,,
\end{eqnarray}
where $v^a_{\pm}$ is the relative velocity between the matter and observer frames and  $\gamma_{\pm}$ is the Lorentz boost factor.   The projection tensor to matter hypersurface is given by $g^{\pm}_{ab} = h^{\pm}_{ab}- u^{+}_{a}u^{+}_{b} = \bar{h}^{+}_{ ab}- \bar{u}^{+}_{ a}\bar{u}^{+}_{ b} $ and at the leading order in $v^a_{+}$ is given by
$
{h}^{\pm}_{ab} \approx \bar{h}^{\pm}_{ ab}+\left[2u^{\pm}_{(a}{v}^{\pm}_{b)}
+{v}^{\pm}_{a} {v}^{\pm}_{b}\right]  + \mathcal{O} \left(\epsilon v^2_{\pm}\right) \,.
$
  The decomposition of the  full spacetime covariant derivative of $v^a_{+}$ with respect to $u^a_{+}$ is given by 
\begin{eqnarray}
\nabla_{a}v_{\pm b}&=&-\dot u_{ \pm c}  v^c_{\pm}\, u_{\pm a} u_{\pm b} - u_{\pm a}\dot v_{\pm \langle b\rangle}
+\left( \frac{1}{3}\Theta_{+} v_{\pm a}+\sigma_{\pm ac}v^c_{\pm }\right)u_{\pm b}
\\ \nonumber &&
+\frac{1}{3}\left({\D}_{c} v^{c}_{\pm}\right) \bar{h}^{\pm}_{ab}+{\D}_{\<a} v_{\pm b\>}+{\D}_{[a}v_{\pm b]}\,.
\end{eqnarray}
where ${\D}_{\<a} v_{\pm b\>} = {\D}_{(a} v_{\pm b)}  - {\D}_{c} v^{c}_{\pm}h^{\pm}_{ab}/3$ is the  symmetric trace-freee part of ${\D}_{a} v_{\pm b}$,, ${\D}_{[a}v_{\pm b]} = ({\D}_{a}v_{\pm b]} - {\D}_{a}v_{\pm b]})/2$ is the anti-symmetric part  and  ${\D}_{c} v^{c}_{\pm}$ is the trace. 
 At the leading order in $v^a_{\pm}$ these observable quantities in both frames are related according to~\cite{Maartens:1998xg}
\begin{eqnarray}
{\Theta}_{\pm} &\approx& \bar{\Theta}_{\pm} + {\D}_a v^a_{\pm}\,,
~~~~~~~~~~~~~
{\sigma}_{\pm ab} \approx \bar{ \sigma}_{\pm ab} + {\D}_{\<a} v_{\pm b\>}\,,
\label{eq:sigma_transformation}
\\
{\omega}_{\pm ab} &\approx&\bar{ \omega}_{\pm ab} + {\D}_{[a} v_{\pm b]}\,,
~~~~~~~~~~
{A}^a_{\pm } \approx \bar{A}^a_{\pm }+ \frac{1}{3}\Theta v^a_{\pm } + \dot{v}^a_{\pm }\,,
\label{eq:A_transformation}
\end{eqnarray}

The leading order transformation between the components of the  energy-momentum tensor between these frames is given by
\begin{eqnarray}\label{eq:energyflux}
{q}^a_{T\pm} &\approx& \bar{q}^a_{T\pm}
\qquad
{\rho}_{T\pm} \approx \bar{\rho}_{T\pm} \,,
\qquad
{P}_{T\pm} \approx \bar{P}_{T\pm}\,, 
\qquad {\pi}^{ab}_{T\pm} \approx \bar{\pi}^{ab}_{T\pm} \,.
\end{eqnarray}
Implementing the transformation given in equations \eqref{eq:sigma_transformation} and \eqref{eq:A_transformation} leads to the matter and momentum conservation equation at the leading order in relative velocity
\begin{eqnarray}  \label{eq:conservation_eqn3}
 \dot{\rho}_{T\pm}  +\left({\rho}_{T\pm}+{P}_{T\pm}\right) \left( \bar{\Theta}_{\pm} 
 + {\D}_a v^a_{\pm}\right)  &=& -   {\Pi}^{ab}_{T\pm }   {\D}_{\<a} v_{\pm b\>} - {\D}_{a} {q}^a_{T\pm}
 \\ \nonumber &&
  - 2 {q}_{T \pm a}  \left[  \bar{A}^a_{\pm}+ \frac{1}{3} \left( \bar{\Theta}_{\pm} 
 + {\D}_a v^a_{\pm}\right)v^a_{\pm} + \dot{v}^a_{\pm}\right]\,,
\\ \nonumber 
\dot{v}_{\pm a} + \frac{1}{3}  \left( \bar{\Theta}_{\pm} 
 + {\D}_{b} v^{b}_{\pm}\right) v_{\pm a} + \bar{A}_{\pm a}   + \frac{{\D}_a  {P}_{T\pm}}{(\rho_{T \pm } +  {P}_{T\pm})} &=&-
 \frac{1  }{(\rho_{T \pm} +  {P}_{T\pm})} \bigg[ h^{\pm}_{ab}{\D}_{c}  {\Pi}^{cb}_{T\pm} + \frac{4}{3} \bar{\Theta}_{\pm}  {q}_{T\pm a} 
  \\ &&
 + \dot{ {q}}_{T\pm a}
 +
 \left({\D}_{(a} v_{\pm b)} - {\D}_{[a} v_{\pm b]}\right)  {q}^b_{T\pm}
 \bigg]\,,
  \label{eq:conservation_eqn4}
\end{eqnarray}
where we have assumed that in the frame of observer $\bar{\omega}_{\pm ab} =0$ and $\bar{\sigma}_{\pm ab} = 0$. 
In the absence of backreaction, equations \eqref{eq:conservation_eqn3} and \eqref{eq:conservation_eqn4} reduce to the traditional continuity and Euler equations for the standard matter fields
\begin{eqnarray}
 \dot{\rho}_{m\pm}  +{\rho}_{m\pm} \left( \bar{\Theta}_{\pm} 
 + {\D}_a v^a_{\pm}\right)  &=& 0 
   \label{eq:dust_energy_conservation_eqn}
\,,
\\
\dot{v}^a_{\pm} + \frac{1}{3}  \left( \bar{\Theta}_{\pm} 
 + {\D}_{b} v^{b}_{\pm}\right) v^a_{\pm} + {\D}^a \Phi_{\pm}   &=&0
   \label{eq:dust_momentum_conservation_eqn}
\end{eqnarray}
where $\bar{A}^a_{\pm} = {\D}^a \Phi_{\pm}$.

\subsection{ Gravitational stability in hydrostatic equilibrium}\label{sec:dark-matter}

In this sub-section, we study the stability of the sub-region in hydrostatic equilibrium. 
 For simplicity, we consider the perfect fluid limit of  equations \eqref{eq:conservation_eqn3} and \eqref{eq:conservation_eqn4}, that is, the limit where the terms on the right-hand  vanish (the spacelike hypersurface projected shear tensor and energy flux vanish)
\begin{eqnarray}
 \dot{\rho}_{T\pm}  +\left({\rho}_{T\pm}+{P}_{T\pm}\right) \left( \bar{\Theta}_{\pm} 
 + {\D}_a v^a_{\pm}\right)   = 0\,,
   \label{eq:fluid_limit_energy_conservation_eqn}
 \\
 \dot{v}_a + \frac{1}{3}  \left( \bar{\Theta}_{\pm} 
 + {\D}_{b} v^{b}_{\pm}\right) v_{\pm a} + {\D}_{a} \Phi_{\pm}  + \frac{{\D}_a {P}_{T\pm}}{(\rho_{T\pm} + {P}_{T\pm})} =0\,.
    \label{eq:fluid_limit_momentum_conservation_eqn}
\end{eqnarray}
Even in the limit of vanishing energy flux and anisotropic stress tensor, the equations \eqref{eq:fluid_limit_energy_conservation_eqn} and \eqref{eq:fluid_limit_momentum_conservation_eqn}  still differ from the standard continuity and Euler equations (equations \eqref{eq:dust_energy_conservation_eqn} and \eqref{eq:dust_momentum_conservation_eqn}) in two ways. 
\begin{itemize}
\item The first difference is the composition of the energy density.  There is a backreaction contribution to the total energy density:  ${\rho}_{T\pm} ={ \rho}_{m\pm} +    \hat{\rho}_{\pm} $, where ${ \rho}_{m\pm}$ is the contribution to the energy density from the standard matter(baryonic matter) and $ \hat{\rho}_{\pm}  $ is the backreaction contribution due to the finite size of the gravitationally bound systerm. 
\item  The second difference is the pressure contribution. Pressure vanishes in the standard dust limit; however, backreaction induces a pressure contribution which survives even in the limit of vanishing random velocity where it behaves as radiation:  $ \hat{P}_{\pm} =  \zeta \hat{\rho}_{\pm} /3$. Therefore, the inward gravitational pull in the decoupled region is supported by the induced pressure preventing continued collapse to a singularity. 

\end{itemize}
We would like to solve equation \eqref{eq:fluid_limit_momentum_conservation_eqn} in the hydrostatic equilibrium limit. In the limit, at every point within the gravitationally bound system, the inward pull of gravity is perfectly balanced by the outward pressure gradient force.  The balance leads to a stable, non-collapsing, and non-expanding gravitationally bound system over long timescales. In spherical symmetric limit, equation \eqref{eq:fluid_limit_momentum_conservation_eqn} reduces to 
\begin{equation}\label{eq:hydrostaticlimit}
{\D}_{a} \Phi_{\pm} + \frac{{\D}_a {P}_{T\pm}}{(\rho_{\pm T} +{ {P}_{T\pm}})} =0 \Rightarrow 
\frac{d\Phi_{\pm}}{dr} = -\frac{\frac{d}{dr} \left( \sigma_{1D}^2 \rho_{m\pm} + \sigma_{T\hat{\rho}1D}^2 \hat{\rho}_{\pm} \right)}{\rho_{m\pm} + \hat{\rho}_{\pm} + \sigma_{1D}^2 \rho_{m\pm} + \sigma_{T\hat{\rho}1D}^2 \hat{\rho}_{\pm}} \,,
\end{equation}
From equation \eqref{eq:pressureterm}, we can write the total pressure  in a more compact form 
$
 {P}_{ T\pm}  
   =   \sigma_{1D}^2 \rho_{ m \pm }  +  \sigma_{T\hat{\rho} 1 D}^2  \hat{\rho} 
$, where  $\sigma_{T\hat{\rho} 1 D}^2  =  \frac{1}{3} \zeta +\sigma_{\hat{\rho} 1 D}^2 +  \frac{1}{3} \zeta \sigma_{\hat{P} 1 D}^2$. Note also that the  total density is given by $\rho_{T\pm} = \rho_{m\pm} + \hat{\rho}_{\pm}$.
 The general solution to equation \eqref{eq:hydrostaticlimit} in the  isothermal limit ($ \sigma_{1D}^2 =  \langle w^2 \rangle \equiv {3} k_B T/M$),  velocity dispersions $\sigma_{1D}^2$ and $\sigma_{T\hat{\rho}1D}^2$ are constant throughout the volume is given by
  \begin{eqnarray}\label{eq:integralsolution}
 \Phi_{\pm}(r_{\pm}) = \Phi_{\pm}(r_{\text{ref}}) - \int_{r_{\text{ref}}}^{r_{\pm}} \frac{\sigma_{1D}^2 \frac{d\rho_{m\pm}}{d r'} + \sigma_{T\hat{\rho}1D}^2 \frac{d\hat{\rho}_{\pm}}{d r'}}{\rho_{m\pm}(r')(1 + \sigma_{1D}^2) + \hat{\rho}_{\pm}(r')(1 + \sigma_{T\hat{\rho}1D}^2)} d r' \,,
   \end{eqnarray}
Assuming that the baryonic matter density $\rho_m$ traces the backreaction density $\hat{\rho}$ such that $\rho_{m \pm}(r) = X_{\rho_{\pm}} \hat{\rho}_{\pm}(r)$ for some constant $X_{\rho_{\pm}} $  then  the integral in equation \eqref{eq:integralsolution} can be performed analytically
  \begin{eqnarray}
 \Phi_{\pm}(r) -\Phi_{ini}
 = -{\Upsilon_{\pm}} \ln \left[ \frac{\hat{\rho}_{\pm}(r)}{\hat{\rho}_{\pm}(r_{ini})} \right] 
 \qquad {\rm{with}} \qquad 
 {\Upsilon}_{\pm} 
 = \frac{\left( \sigma_{1D}^2 X_{\rho_{\pm}} \  + \sigma_{T\hat{\rho}1D}^2  \right) }{X_{\rho_{\pm}} (1 + \sigma_{1D}^2) + (1 + \sigma_{T\hat{\rho}1D}^2)} \,.
 \label{eq:hydrophi}
    \end{eqnarray}
 Here, we have fixed the constant of integration $C$ by introducing the boundary gravitational potential $\Phi(r_{ini}) = \Phi_{ini}$.
To obtain the density profile $ \hat{\rho}$, we  have to solve the Poisson equation 
 $\nabla^2 \Phi = 4 \pi G \left( \rho_{T} + \frac{3P_{T}}{c^2} \right) = 4 \pi G \left( \rho_{T} + \frac{3\sigma_{v}^2  }{c^2}\rho_{T} \right)  =4 \pi G \rho_{T} \left(1 +\frac{3\sigma_{v}^2  }{c^2}\right) $.
Without loss of generality, we focus on the non-relativistic limit:
$\nabla^2 \Phi_{\pm}  =  4\pi G\left(\rho^{\pm}_{m} + \hat{\rho}\right) =4\pi G\hat{\rho}_{\pm}  \left(X_{\rho_{\pm}}+ 1 \right)$ .
In the spherically symmetric limit, it is given by
\begin{eqnarray}\label{eq:poisson_eqn}
\frac{\d^2  \Phi^{\pm}  }{\d r^2_{\pm}} + \frac{2}{r_{\pm}} \frac{d \Phi^{\pm} }{dr_{\pm}}= 4\pi G\hat{\rho}_{\pm}  \left(X_{\rho_{\pm}}+ 1 \right) \,.
\label{eq:poisson}
\end{eqnarray}
Given equation \eqref{eq:hydrophi},  the density profile, $\hat{\rho}_{\pm}$  that solves equation \eqref{eq:poisson} is given by
\begin{eqnarray}\label{eq:energy_density2}
\hat{\rho}_{\pm}(r_{\pm}) = \frac{\Upsilon_{\pm}}{2\pi G (X_{\rho_{\pm}} + 1) r_{\pm}^2}  
= \frac{\sigma^{2}_{\pm v}}{ 2\pi G } \frac{1}{r_{\pm}^{2} }
\qquad {\rm{with}} \qquad 
 \sigma^{2}_{\pm v}  =\frac{\Upsilon_{\pm}}{(X_{\rho_{\pm}} + 1) } \,.
\end{eqnarray}
This is a density profile for a Singular Isothermal Sphere (SIS) in galactic gynamics  ~\cite{Tran:2024hry,Robertson:2020pxj} $  \hat{\rho}_{\pm}(r_{\pm}) = {\sigma^{2}_{v}}/{2\pi G} {r^2_{\pm} } $,
where $\sigma_{v} $ is one-dimensional velocity dispersion, comparing with equation \eqref{eq:energy_density2}, it is determined by the coefficient of bulk viscosity.  The density profile has local power-law slope $ ({d \ln \hat{\rho}})/({d \ln r_{\pm}} )= -2 $. 
The first integral of equation \eqref{eq:poisson_eqn} gives the enclosed mass due to the density profile (equation \eqref{eq:energy_density2})
\begin{eqnarray}\label{eq:accn}
\frac{\d\Phi_{\pm}}{\d r_{\pm}} &=&\frac{r_{\pm\rm{ini}}^2}{r^2_{\pm}}\frac{\d\Phi_{\pm}}{\d r_{\pm}} \bigg|_{r_{\pm}=r_{\pm\rm{ini}}}  +\frac{G M_{\pm}(r_{\pm \rm{ini}}< r_{\pm})}{r^2_{\pm}}  \,,
\end{eqnarray}
where $M_{\pm}(r_{\pm}) $ is the total mass
\begin{eqnarray}\label{eq:massterm}
M_{ \pm }(r_{\pm})  = 4\pi  \left(X_{\rho_{\pm}}+ 1 \right)\int_{r_{\pm\rm{ini}}}^{r_{\pm}} \hat{\rho}_{\pm}(r'_{\pm}) r'^2_{\pm} \d r'_{\pm}   =\frac{2 {\Upsilon_{\pm}}}{G} (r_{\pm} - r_{\pm\rm{ini}}) \,.
\end{eqnarray}
If the rotation velocity is calculated from $v_{\phi} = \sqrt{ r_{\pm}\frac{\d\Phi}{\d r_{\pm}} }$ with $\frac{\d\Phi_{-}}{\d r_{-}} \big|_{r_{-}=r_{-\rm{ini}}}  = 0$, it immediately leads to a flat rotation curve on the ootuskirts since the total mass is proportional to $r_{\pm} $: $M_{ \pm }(r_{\pm}) \propto (r_{\pm} - r_{\pm\rm{ini}})$. However, this is mathematically inconsistent since equation \eqref{eq:energy_density2}  was derived in the limit of vanishing velocity.  A consistent expression for the rotation curve is derived in sub-section \ref{sec:subsectionreotationcurve}.

\subsection{The bactreaction density from averaging over microscopic sources}

Our plan is to evalaute $\tilde{\rho}_{{\pm} \ell}  = \big[ {u}^a_{\pm \ell } {u}^b_{\pm \ell } \tilde{\sigma}_{\<ab\>  \ell} \big]/{\kappa} $ within standard cosmology so that we can estimate how the backreaction terms impacts the rotation curves from first principle.  
We make use of the Ricci identity $r^c_{\pm}\nabla_c\nabla_m r_{\pm n} - \nabla_m(r^c_{\pm}\nabla_c r_{\pm n}) 
= R_{\pm cmn}{}^{e}\, r^c_{\pm} r_{\pm e},$  to derive the propagation equation for  $ \tilde{\sigma}_{\pm \<ab\>}$  
\begin{eqnarray}\nonumber
r^c_{\pm}\nabla_c \tilde\sigma_{\pm \langle ab\rangle}
&=& -\tfrac{2}{3}\,\tilde{\Theta}_{\pm}\,\tilde\sigma_{\pm ab}
- \tilde{\sigma}_{\pm ac}\tilde\sigma^c{}_{\pm b}
- \tilde{\omega}_{\pm ac}\tilde{\omega}^c{}_{\pm b}
+ \mathcal{D}_{\langle a}  \tilde{A}_{\pm b\rangle}
+ \tilde{A}_{\pm \langle a}  \tilde{A}_{\pm b\rangle} 
\\&&
- \gamma_{\pm a}{}^m\gamma_{\pm b}{}^n\big( C_{\pm m c n d}\, r^c_{\pm} r^d_{\pm} \big)
+ \tfrac{1}{2}\,\gamma_{\pm a}{}^m\gamma_{\pm b}{}^n
\Big[ R_{\pm mn} - \tfrac{1}{3}\gamma_{\pm mn}R_{\pm pq}\gamma^{pq}_{\pm}\Big] .
\label{eq:shear_timehyper}
\end{eqnarray}
where
 $\tilde{\mathcal{D}}_{\pm a} \equiv \gamma_{\pm a}{}^b\nabla_b$ is the projected derivative on the orthogonal time hypersurface.
$C_{\pm abcd}$ is the Weyl tensor; $C_{\pm m c n d}\, r^c_{\pm} r^d_{\pm}$ encodes the tidal (free gravitational) field. The first term on the RHS
 $\tfrac{2}{3}\tilde\Theta\,\tilde\sigma_{\pm ab}$ denotes damping/amplification from expansion.
 $\tilde\sigma_{\pm ab}\tilde\sigma^{ab}_{\pm}$ and $\tilde\omega_{\pm ab}\tilde\omega^{ab}_{\pm}$ denote nonlinear self-interaction due to the shear tensor and vorticity. Furthermore, 
 $C_{\pm acbd} r^c_{\pm} r^d_{\pm}$ denotes the tidal field due to the free gravitational field (Weyl curvature) and 
 $\tilde{\mathcal{D}}_{\langle a} \tilde{A}_{\pm b\rangle}$ denotes the shearing effect sourced by acceleration gradients of the congruence.
 The last term on the RHS of equation \eqref{eq:shear_timehyper} captures the geometric symmetric trace-free Ricci contribution.

 To solve for $\tilde{\sigma}_{\pm ab}$, we assume that  $r^a_{\pm}$ is geodesic, therefore $\tilde{A}_{\pm a}=0$, also for consistency, we assume that $r^a_{\pm}$ is hypersurface orthogonal and by Frobenius' theorem, the vorticity must vanish  $\tilde\omega_{\pm ab}=0$.    Imposing all these in equation \eqref{eq:shear_timehyper} in the linear regime (that is, neglecting the quadratic  shear tensor terms), leads to
\begin{eqnarray}\nonumber
r^c_{\pm}\nabla_c \tilde\sigma_{\pm ab} 
&+&\tfrac{2}{3}\tilde\Theta\,\tilde\sigma_{\pm ab}
+ \gamma_{\pm a}{}^m\gamma_{\pm b}{}^n\big( C_{\pm m c n d}\, r^c_{\pm} r^d_{\pm} \big)
\\  &&
- \tfrac{1}{2}\gamma_{\pm a}{}^m\gamma_{\pm b}{}^n
\Big( R_{\pm mn} - \tfrac{1}{3}\gamma_{\pm mn}R_{\pm pq}\gamma^{pq}_{\pm}\Big) \approx  0\,.
\end{eqnarray}
Again $ \bar{\tilde{\Theta}}_{\pm} = \partial_{i} r^i_{\pm} = 3/(ar) = 3/r_{\pm}$. On an exact FLRW spacetime, the Weyl tensor vanishes  $C_{\pm abcd}=0$ and there is no anisotropic stress, so the shear tensor also vanishes. 
At leading order in cosmological perturbation theory, the Weyl term and anisotropic stress act as a source for the shear tensor
\begin{eqnarray}
\tilde\sigma_{ \pm ab}& =& \frac{\tilde{\sigma}_{\pm{\rm{ini}}ab}}{r^2_{\pm}} - \frac{1}{r^2_{\pm}} \int_{0}^{r} {r'_{\pm}}^2 \d r_{\pm}\left[ \mathcal{E}_{\pm ab} - \frac{1}{2} \mathcal{R}_{\pm ab}\right]\,,
\end{eqnarray}
where $\tilde{\sigma}_{0ab}$ is the shear tensor at the initial position, $\mathcal{E}_{ab}=  \gamma_{\pm a}{}^m\gamma_{\pm b}{}^n\big( C_{\pm m c n d}\, r^c_{\pm} r^d_{\pm} \big)$
 and $\mathcal{R}_{ab}=\gamma_{\pm a}{}^m\gamma_{\pm b}{}^n
\Big( R_{\pm mn} - \tfrac{1}{3}\gamma_{\pm mn}R_{\pm pq}\gamma^{pq}_{\pm}\Big)$.  At leading order in cosmological perturbation theory, we find
\begin{eqnarray}
\mathcal{E}_{\pm mn} 
&\approx&\frac{1}{2} \left[ h_{\pm mn} - \frac{2}{3} \gamma_{\pm mn}\right]{\D}^2\left( \Phi_{\pm} + \Psi_{\pm}\right)
+
\gamma^{c}_{\pm}{}_{(m} {\D}_{c}{\D}_{n)} \left( \Phi_{\pm} + \Psi_{\pm}\right)
\\ \nonumber &&
+
\left[\frac{1}{2}g_{\pm mn}\gamma^{ab}_{\pm}- \gamma^{ab}_{\pm} h_{\pm mn}\right]{\D}_{a} {\D}_{b}\left( \Phi_{\pm} + \Psi_{\pm}\right)
-{\D}_{(m}{\D}_{n)}\left( \Phi_{\pm} + \Psi_{\pm}\right) \,.
\\
\mathcal{R}_{\pm ab} 
  &\approx&{\D}_{a}{\D}_{b} \Phi_{\pm} - {\D}_{a}{\D}_{b}\Phi_{\pm}
+\frac{2}{3} \gamma^{cd}_{\pm} \gamma_{\pm ab} \left({\D}_{a} {\D}_{b} \Phi_{\pm} - {\D}_{a}{\D}_{b}\Phi_{\pm}\right)
+g_{ab} \gamma^{ab}_{\pm} \left( {\D}_{a}{\D}_{b} \Phi_{\pm} -{\D}_{a}{\D} \Psi_{\pm}\right)
\\ \nonumber &&
+ \gamma^{c}{}_{(b} \left( {\D}_{a)}{\D}_{c} \Psi_{\pm} - {\D}_{a} {\D}_{c} \Phi_{\pm}\right)
+ \left[-g_{\pm ab} + \frac{4}{3} g_{\pm ab} + h_{\pm ab}\right]{\D}^2 \Phi_{\pm}
+ \left(h_{\pm ab} + \frac{2}{3} \gamma_{\pm ab}\right){\D}_{c}{\D}^c \Psi_{\pm} \,.
\end{eqnarray}
 Our interest is to calculate 
 $\tilde{\rho}_{{\pm}}  =  \big[ {u}^a_{\pm} {u}^b_{\pm} \tilde{\sigma}_{\<ab\>} \big]/{\kappa}$, therefore,   contracting $ \tilde{\sigma}_{\pm\<ab\>}$ with $u^a_{\pm}$  gives 
\begin{eqnarray}
 {u}^a_{\pm} {u}^b_{\pm} \tilde\sigma_{\pm ab}&\approx&\frac{ \left[{u}^a_{\pm} {u}^b_{\pm}\tilde{\sigma}_{ab}\right]_{\rm{ini}}}{r^2_{\pm}}   + \frac{1}{r^2_{\pm}} \int_{r_{\pm\rm{ini}}}^{r} {r'_{\pm}}^2 \gamma^{ab}_{\pm} \left[{ \D}_{a}{\D}_{b} \Phi_{\pm} \right]\d r_{\pm}\,,
\end{eqnarray}
where  we have set $\Psi_{\pm}= \Psi_{\pm}$ for simplicity. We can  decompose the double spatial derivatives of $\Phi$ further   into irreducible units  using 
$ {\D}_{i} {\D}_j = n_{\pm i} n_{\pm j} \nabla_{\p}^2 + 2 n_{\pm (i} \nabla_{\bot j} \nabla_{\p} + \frac{1}{ r_{\pm}} N_{\pm ij}  \nabla_{\p}  + \nabla_{\bot i}\nabla_{\bot j}$. where $n^i_{\pm}$  is a spatial 3 vector and $\nabla_{\p} = {\d } /{\d} r_{\pm}$ denotes the directional  derivative along $n^i_{\pm}$ and $\nabla_{\bot}$ denotes the angular derivative.,   Note that  $ \gamma^{\pm}_{ab}  = N^{\pm}_{ab} - u^{\pm}_{a} u^{\pm}_{b}$ and $u^a_{{\pm}} n_{\pm a} =0$. The non-vanishing components contributing to the total energy density induced by the backreaction effect become
\begin{eqnarray}
\tilde{\rho}_{\pm} &=&\frac{1}{\kappa}
\bigg[
 \frac{2}{r^2_{\pm}} \int_{r_{\pm\rm{ini}}}^{r} {r'_{\pm}}^2\nabla_{\bot}^2 \Phi_{\pm}  \d r_{\pm}  +\frac{2}{r^2_{\pm}}\left[ \left[ r_{\pm}\Phi_{\pm}(r_{\pm}) \right]^{r}_{r_{\pm\rm{in}}} - \int_{r_{\pm\rm{ini}}}^r \Phi(r'_{\pm}) \d r'_{\pm}\right]\bigg]\,,
\label{eq:backreation_energy_density}
\end{eqnarray}
where we set $\left[\tilde{\sigma}_{ab}\right]_{\rm{ini}} = 0 $ to zero for simplicity. 
The surface energy density given in \eqref{eq:backreation_energy_density}  consists of local and radially integrated terms. This structure implies that the impact of $\tilde{\rho}$ on the total energy density is sensitive to both local fluctuations and global radial constraints. Further detials on this later. 

Finally, from equation \eqref{eq:backreactiondensity}, the backreaction energy density is given by
\begin{equation}
\hat{\rho}_{\pm} =\frac{1}{R_{\pm}} \sum_{\ell = 1}^{N}   \tilde{\rho}_{\pm \ell} 
\xrightarrow{N \to \infty} \; \frac{\langle \tilde{\rho}_{\pm} \rangle}{R_{\pm}}.
\end{equation}
where we have invoked the Law of Large Numbers  (LLN) by treating  $\tilde{\rho}_{\pm \ell}$ as independent and identically distributed (i.i.d.) random variables and that the sample average converges to the finite mean:$
 \sum_{\ell=1}^{N} \bar{\rho}_{\pm \ell} \xrightarrow{N \to \infty} \langle \tilde{\rho} \rangle
$. 
Applying this to equation \eqref{eq:backreation_energy_density}, we find that the first term vanishes on average, leading to 
\begin{eqnarray}\label{eq:energy_densitymicro}
\langle \tilde{\rho}_{\pm} (r_{\pm})\rangle&\approx &\frac{1}{\kappa}
\frac{2}{r^2_{\pm}}\bigg[
 \left[ r_{\pm} \Phi_{\pm}(r_{\pm}) \right]^{r}_{r_{\rm{in}}} - \int_{r_{\pm\rm{ini}}}^r \Phi_{\pm}(r_{\pm}) \d r'_{\pm}\bigg]  
 = \frac{1}{\kappa}
\frac{2}{r^2_{\pm}}Q_{\pm} (r_{\pm}) \,,
\end{eqnarray}
where the bulk density is given by   
\begin{eqnarray}\label{eq:backreactiondensityprofile}
\hat{\rho}_{\pm}  =\frac{1}{\kappa}
\frac{2}{r_{\star}r^2_{\pm}}Q_{\pm} (r_{\pm})\,,
\qquad 
Q_{\pm} (r_{\pm}) = 
 \left[ r_{\pm} \Phi_{\pm}(r_{\pm}) \right]^{r}_{r_{\rm{in}}} - \int_{r_{\pm\rm{ini}}}^{r_{\p,}} \Phi_{\pm}(r_{\pm}) \d r'_{\pm}\,.
\end{eqnarray}
On the background spacetime, $h_{RR} =a$, thus $ R  = a\int \d r_{\star} = r_{\star} $.

\subsection{ Impact of backreaction on rotation curves: microscopic picture}\label{sec:subsectionreotationcurve}

\begin{figure}[h]
\includegraphics[width=70mm,height=50mm] {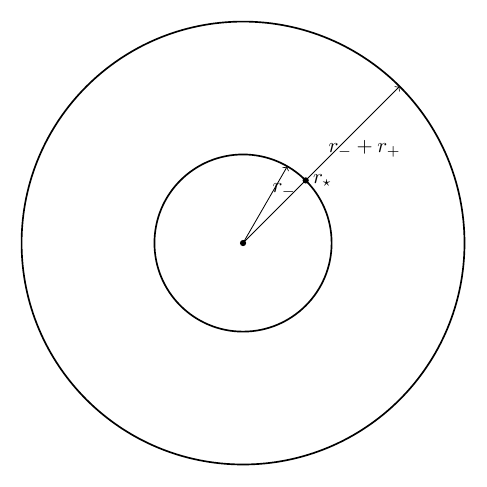} 
\caption{A simplified geometric set-up of a gravitationally bound system in an expanding spacetime. $r_{\star}$ denotes the location of the matter horizon. 
 }
\label{fig:enclosed_mass}
\end{figure}

The galaxy rotation curve in the steady state approximation
from equation \eqref{eq:fluid_limit_momentum_conservation_eqn} 
\begin{eqnarray}\label{eq:quais-statics-approximation}
 v^b_{\pm}{\D}_b v_{\pm} a  +{\D}_a\Phi_{\pm}  + \frac{{\D}_a {P}_{T\pm}}{(\rho_{\pm T} + \frac{{P}_{T\pm})}{c^{2}}} =0\,,
\end{eqnarray}
where we defined the steady state limit as the limit where $ {\partial v^a_{\pm}}/{\partial t_{\pm}}   =  \bar{\Theta}_{\pm} = 0= {\D}_{b} v^{b}_{\pm} $.
Assuming that the motion is mostly in $\phi$-direction: $v^a_{\pm}= (0,0, v^{\phi}_{\pm})$. In the  isothermal limit,  equation \eqref{eq:quais-statics-approximation} reduces to 
\begin{eqnarray}\label{eq:rotationcurve}
 \frac{{v^{\pm}}_{\phi}^2(r_{\pm})}{r_{\pm}} = \frac{\d \Phi_{\pm}}{\d r_{\pm}} + \frac{1}{(\rho_{\pm T} + {P}_{T}/c^{2})}\frac{\d {P}_{T\pm} }{\d r_{\pm}}  = \frac{\d \Phi_{\pm}}{\d r_{\pm}} +
 \frac{1}{r_{\pm}} \left[ \mathcal{Z}_{\pm}
\frac{\d \ln \hat{\rho}_{\pm}}{\d \ln r_{\pm}}  +  \mathcal{Y}_{\pm}
\frac{\d \ln {\rho}_{m\pm}}{\d \ln r_{\pm}} 
\right]
 \,,
\end{eqnarray}
where we have introduced a function of dispersion velocity and  bias parameters
\begin{eqnarray}
 \mathcal{Z}_{\pm}= \frac{ \sigma_{T\hat{\rho}_{\pm}1D}^2  }{X_{\rho_{\pm}} (1 + \sigma_{1D\pm}^2/c^{2}) + (1 + \sigma_{T\hat{\rho}_{\pm}1D}^2/c^{2})} \,,
 \quad
  \mathcal{Y}_{\pm}= \frac{\sigma_{1D\pm}^2  }{(1 + \sigma_{1D\pm}^2/c^{2}) + \left(1 + \sigma_{T\hat{\rho}_{\pm}1D}^2/c^{2}\right)/X_{\rho_{\pm}} } \,.
 \end{eqnarray}
There are two major contributions to the rotation velocity: the gradient of the gravitational potential, ${\d \Phi_{\pm}}/{\d r_{\pm} }$ and the gradient of the induced pressure.
We consider a simple case of a concentric sphere (see the geometry illustrated in Figure \ref{fig:enclosed_mass}.  

In this setup, the rotation velocity of a  gravitationally bound system is a piecewise function of the gravitational acceleration and the gradient of the induced pressure
  \begin{equation}\label{eq:rotationvelo}
v_{\phi}(r_{\pm}) 
=
\begin{cases}
\displaystyle
v^{-}_{\phi}(r_{-}) =\sqrt{\,r_{-}\,\frac{\d\Phi_{-}}{dr_{-}}  +\mathcal{Z}_{-}
\frac{\d \ln \hat{\rho}_{-}}{\d \ln r_{-}}  +  \mathcal{Y}_{-}
\frac{\d \ln {\rho}_{m -}}{\d \ln r_{-}} 
},
& r_{-} < r_\star, \qquad 
\\[10pt]
\displaystyle
 v^{+}_{\phi}(r_{+})=\sqrt{\,r_{+}\,\frac{\d\Phi_{+}}{dr_{+}}  +\mathcal{Z}_{+}
\frac{\d \ln \hat{\rho}_{+}}{\d \ln r_{+}}  +  \mathcal{Y}_{+}
\frac{\d \ln {\rho}_{m+}}{\d \ln r_{+}} 
 }\,
& r_{+} \ge r_\star  \qquad  
\end{cases}
\end{equation}
The Poisson equation, $\nabla^2 \Phi \approx  4\pi G(\rho_{m} + \hat{\rho})$ is linear and the linearity allows to separate the potential $\Phi_{\pm}$ as the sum of two independent potentials:$\Phi_{\pm}(r) = \Phi_{m}^{\pm}(r) + \hat{\Phi}_{\pm}(r)$, where $\Phi_{m}^{-}(r_{\pm}) $ is the gravitational potential due to baryonic matter and $\hat{\Phi}_{-}(r_{\pm})$ is the gravitational potential induced by backreaction due to the presence of the matter horizon. 
This implies that we can solve two separate Poisson equations:$\nabla^2 \Phi_{m\pm}^{\pm} = 4\pi G \rho_{m \pm}^{\pm}$ and $\nabla^2 \hat{\Phi}_{\pm} = 4\pi G \hat{\rho}_{\pm}$ for $\Phi_{m\pm}$ and $\hat{\Phi}_{\pm}$. 
The corresponding radial gradient of gravitational potential splits as well:
 $\frac{\d \Phi_{\pm} } { \d r_{\pm}} = \frac{d\Phi_{m}^{\pm}  }{dr_{\pm}}+ \frac{d\hat{\Phi}_{\pm} }{dr_{\pm}}$.
 The first integral of the baryonic Poisson equation $\nabla^2 \Phi_{m\pm} = \frac{1}{r^2_{\pm}} \frac{d}{dr_{\pm}} \left( r^2_{\pm} \frac{d\Phi_{m\pm}}{dr_{\pm}} \right)= 4\pi G  \rho_{\pm m}$ gives.
\begin{eqnarray}\label{eq:accn}
\frac{\d\Phi_{m-}}{\d r_{-}} &=&\frac{r_{\rm{ini}}^2}{r^2_{-}}\frac{\d\Phi_{m-}}{\d r_{-}} \bigg|_{r_{-}=r_{\rm{ini}}}  +\frac{G M_{bk -}(r_{-}< r_{\star})}{r^2_{-}} \,,
\\
\frac{\d\Phi_{m+}}{\d r_{+}} &=&\frac{r_{\star}^2}{r^2_{+}}\frac{\d\Phi_{m+}}{\d r_{+}} \bigg|_{r_{+}=r_{\star}}  +\frac{G M_{bk +}( r_{+}>r_{\star})}{r^2_{+}} \,.
\end{eqnarray}
We impose the following standard  initial condition for spherical mass distribution for the inner region and the outer region initial condition following from the continutity condition for the flux
\begin{eqnarray}\label{eq:standardinitalcondition}
\frac{\d\Phi_{m-}}{\d r_{-}} \bigg|_{r_{-}=r_{\rm{ini}} }  =0 \,,
\qquad
\frac{\d\Phi_{m+}}{\d r_{+}} \bigg|_{r_{+}=r_{\star}  } =\frac{\d\Phi_{m-}}{\d r_{-}} \bigg|_{r_{-}=r_{\star} } \,. 
\end{eqnarray}
The second equation describes the matching of the field gradients at the boundary $r_{\star}$ 
This condition ensures the continuity of the gravitational acceleration across the interface at $r_{\star}$.
$M_{\rm{bk}\pm} $ is interpreted as the corresponding mass contained within a region  with radius $r_{\pm}$:
\begin{eqnarray}
M_{bk\pm}(r_{\pm},r_{\rm{ini}} ,a_{-}) &=&M_{\pm} a_{\pm} \left[ \frac{2r_{\rm{ini}} + a_{\pm})}{(r_{\rm{ini}} + a_{\pm}))^2} - \frac{2r_{\pm} + a_{\pm})}{(r_{\pm} + a_{\pm}))^2} \right]\,.
\end{eqnarray}
where we have adopted the Hernquist density profile  for simplicity \cite{Hernquist:1990ApJ} 
 \begin{eqnarray}\label{eq:rhomatter}
 \rho_{ m \pm }(r_{\pm}, a_{\pm})  =  \frac{M_{\pm}}{2\pi} \frac{a_{\pm}}{r_{\pm}(r_{\pm} + a_{\pm})^3}
 \end{eqnarray}
 $"a_{\pm}"$  is a free scale parameter and $M_{\pm}$ is the total mass.  
For the backreaction density field,  we start with the corresponding Poisson equation $\nabla^2 \hat{\Phi}_{\pm} = 4\pi G \hat{\rho}_{\pm}$  and substitute for the bulk density $\hat{\rho}$ using equation \eqref{eq:energy_densitymicro} leading to a second-order Integro-differential equation 
 \begin{eqnarray}\label{eq:intrgro-diff}
 \frac{\d^2  \hat{\Phi}_{\pm}  }{\d r^2_{\pm}} + \frac{2}{r_{\pm}} \frac{d \hat{\Phi}_{\pm}  }{dr_{\pm}}
 =  
 \frac{1}{r^2_{\pm} r_{\star}}\bigg[
 \left[ r_{\pm} {\Phi}_{\pm}  (r_{\pm})-r_{\rm{in}}  {\Phi}_{\pm}  (r_{\rm{in}} )  \right] - \int_{r_{\pm\rm{ini}}}^r {\Phi}_{\pm}  (r'_{\pm}) \d r'_{\pm}\bigg] \,,
 \end{eqnarray}
 where we have substituted for $\hat{\rho}$ using equation \eqref{eq:energy_densitymicro}.   Using equation \eqref{eq:rhomatter} in the Poisson equation gives $\Phi^{\pm}_{m}(r_{\pm})=-\frac{GM_{\pm}}{r_{\pm}+a_{\pm}}.$ 
 Equation \eqref{eq:intrgro-diff}  can be reduced to an ordinary differential equation by a change of variables $g_{\pm}  (r):=r^{2}\hat{\Phi}'_{\pm}  (r)$.  
\begin{eqnarray}
 \frac{\d^2  g_{\pm}   }{\d r^2_{\pm}} -\frac{1}{r_{\pm}\,r_{\star}}\,g_{\pm}  (r_{\pm})= \frac{r_{\pm}}{r_{\star}}\Phi_{m\pm}'(r_{\pm})=\frac{GM_{\pm} r_{\pm}}{r_{\star}(r_{\pm}+a)^2}.
\end{eqnarray}
This is a modified Bessel equation, which becomes obvious after the introduction of dimensionless variables
\begin{eqnarray}
y=2\sqrt{\frac{r}{r_{\star}}}\qquad\Bigl(r=\frac{r_{\star}}{4}y^{2},\;
\frac{d}{dr}=\frac{2}{r_{\star}y}\frac{d}{dy}\Bigr).
\end{eqnarray}
Setting $g(r)=yh(y)$ leads to modified Bessel equation of order 1.
\begin{eqnarray}
y^{2}h''(y)+y h'(y)-(1+y^{2})h(y)=0,
\end{eqnarray}
The homogeneous equation  has a unique  solution
$
h(y)=A I_{1}(y)+B K_{1}(y),
$
where $I_{1},K_{1}$ are the modified Bessel functions and $A,B$ are arbitrary constants,
The  particular solution is determined by the method of Variation of Parameters in the presence of the source term  
\begin{eqnarray}\label{eq:solng}
g_{\pm}(r_{\pm}) = A_{\pm}g_{1\pm}(r\pm)  + B_{\pm} g_{2\pm }(r_{\pm})  +g_{\pm p}(r_{\pm}) \,, 
\end{eqnarray}
where the two linearly independent solutions are
\begin{eqnarray}
g_{1\pm}(r_{\pm}) = 2\sqrt{\frac{r_{\pm}}{r_{\star}}} I_1\left( 2\sqrt{\frac{r_{\pm}}{r_{\star}}} \right), 
\quad g_{2\pm}(r) =2\sqrt{\frac{r_{\pm}}{r_{\star}}}K_1\left( 2\sqrt{\frac{r_{\pm}}{r_{\star}}} \right)
\end{eqnarray}
We made use of the Wronskian  $W = g_{1\pm} g_{2\pm}' - g_{2\pm} g_{1\pm}'$ which is determined with the help of the standard Bessel identity $I_1(x)K_0(x) + I_0(x)K_1(x) = 1/x$: $W(r) = -{2}/{r_{\star}}$ to derive the particular solution
\begin{eqnarray}
g_{p\pm }(r_{\pm}) = \frac{GM_{\pm}}{2}\Bigl[ g_{1\pm}(r_{\pm})\int_{r_{\rm{ini}} }^{r_{\pm}} \frac{g_{2\pm}(r'_{\pm})\,r'}{(r'_{\pm}+a_{\pm})^{2}}\,dr'_{\pm} - g_{2\pm}(r_{\pm})\int_{r_{\rm{ini}} }^{r_{\pm}}  \frac{g_{1\pm}(r'_{\pm})\,r'_{\pm}}{(r'_{\pm}+a_{\pm})^{2}}\,dr'_{\pm} \Bigr],
\end{eqnarray}

Using equation \eqref{eq:solng}, we immediately obtain the gradient of the gravitational potential: $\hat{\Phi}'_{\pm}  (r_{\pm}) = {g_{\pm} (r_{\pm})}/{r^{2}_{\pm}}$.
\begin{eqnarray}\label{eq:backreactionAccen}
      \frac{d\hat{\Phi}_{\pm}}{dr_{\pm}} &=&   \frac{1}{{r^{2}_{\pm}}}
\Bigl[A_{\pm}g_{1\pm}(r_{\pm}) +B_{\pm}\,g_{2\pm}(r_{\pm}) 
       + g_{p\pm}(r_{\pm}) \Bigr]\,.
     \end{eqnarray}
The gravitational potential is then obtained by performing another integration $\Phi(r)=\Phi_{0}+\int\frac{g(r)}{r^{2}}\,dr$:
\begin{eqnarray}\label{eq:gensolution}
\hat{\Phi}_{\pm}(r_{\pm}) &=& \hat{\Phi}_{\pm 0}+ 
A_{\pm} \mathcal{E}_{1\pm}(r_{\pm}) +B_{\pm}\,\mathcal{E}_{2\pm}(r_{\pm}) +\mathcal{E}_{p\pm}(r_{\pm})  \,,
\end{eqnarray}
where the integrals can be performed in terms of the generalised Hypergeometric functions or the Meijer G-function
\begin{eqnarray}
 \mathcal{E}_{ 1\pm}(r_{\pm})  &=&  \int_{r_{\rm{ini}} }^{r_{\pm}}  \frac{1}{{r^{'2}_{\pm}}}
 g_{1\pm}(r'_{\pm}) 
 dr'_{\pm}  = \frac{4}{\sqrt{r_{\star}}} \left[ \frac{\sqrt{r_{\pm}}}{r_{\star}} {}_1F_2 \left( \frac{1}{2}; \frac{3}{2}, 2; \frac{r_{\pm}}{r_{\star}} \right) - \frac{\sqrt{r_{\rm{ini}}}}{r_{\star}} {}_1F_2 \left( \frac{1}{2}; \frac{3}{2}, 2; \frac{r_{\rm{ini}}}{r_{\star}} \right) \right]\,,
 \\
  \mathcal{E}_{ 2\pm}(r_{\pm}) &=& \int_{r_{\rm{ini}} }^{r_{\pm}}  \frac{1}{{r^{'2}_{\pm}}}
 g_{2\pm}(r'_{\pm}) 
 dr'_{\pm} = \frac{4}{r_{\star}} \left[ \frac{1}{x} G_{1,3}^{2,1} \left( \frac{x^2}{4} \bigg| \begin{matrix} 1 \\ \frac{1}{2}, -\frac{1}{2}, -1 \end{matrix} \right) - \frac{1}{x_{\rm{ini}}} G_{1,3}^{2,1} \left( \frac{x^2_{\rm{ini}}}{4} \bigg| \begin{matrix} 1 \\ \frac{1}{2}, -\frac{1}{2}, -1 \end{matrix} \right) \right]  \,,
 \ \\
  \mathcal{E}_{ 3\pm} (r_{\pm}) &=&\int_{r_{\rm{ini}} }^{r_{\pm}}  \frac{1}{{r^{'2}_{\pm}}} g_{p\pm}(r'_{\pm})  dr'_{\pm} 
  ~~~
   \end{eqnarray}
Now we need to determine the initial value of the integration $\hat{\Phi}_{0\pm}$  and the two arbitrary constants $A_{\pm}$ and $B_{\pm}$.  In general, the value of $\hat{\Phi}_{0\pm}$ is not independent of $A_{\pm}$ and $B_{\pm}$. This condition is usually obtained by demanding that the general solution (equation \eqref{eq:gensolution}) satisfies the original equation (i.e equation \eqref{eq:intrgro-diff}) say at $r = r_{\rm{ini}}$, this may allow to express  $\hat{\Phi}_{0\pm}= \hat{\Phi}_{0\pm}(A_{\pm},B_{\pm})$. However, in our case(equation \eqref{eq:intrgro-diff}), the source term vanishes exactly at $r = r_{\rm{ini}}$, therefore, the value of $\hat{\Phi}_{0\pm}$ does not influence the dynamics, hence, we determine it independently by imposing regularity condition at the origin.  Note that this is consistent with the equivalence principle. 

Since equation \eqref{eq:gensolution} is a solution to the Poisson equation, which is an elliptic equation, the values of the constant $A_{\pm}$ and $B_{\pm}$. must be chosen such that $\hat{\Phi}$ is regular within the entire domain.  That is, the solution at any point is influenced by every other point in the domain at a given time. We do not go into detail on how spacelike-separated events can influence each other; however, it is important to note that all the particles within the domain have a common origin and that they can also influence each other is they form an EPR pair~\cite{Einstein:PhysRev.48.73,Maldacena:2013xja,Gaztanaga:2025awe}.

Our target really is to ensure that these solutions satisfy a set of physically plausible boundary conditions at specific edges.  For concreteness, we fix the initial conditions for the interior region by recalling that when we fixed the initial condition for the gravitational field associated with the baryonic density field, we imposed that the gravitational acceleration be regular at the origin(see equation \eqref{eq:standardinitalcondition}). Extending this requirement to the gravitational field associated with the backreaction density field, we immediately must impose that the constant $B_{-}$ must vanish: $B_{-} = 0$ because the modified Bessel function of the second kind blows up at the origin
$K_1(y) = \frac{1}{y} + \frac{y}{2} \left[ \ln\left(\frac{y}{2}\right) + \gamma - \frac{1}{2} \right] + \mathcal{O}(y^3 \ln y)$, where:$\gamma \approx 0.5772$ is the Euler-Mascheroni constant.  $I_{1}$ is regular $I_1(y) \approx \frac{y}{2}$  in the limit $y\to 0$.
Therefore, imposing the corresponding regularity condition at the origin
   \begin{eqnarray}
\hat{\Phi}_{-}(r_{\rm{ini}})  =  0 
\qquad \qquad
\frac{d\hat{\Phi}_{-}}{dr_{-}} \bigg|_{r =r_{\rm{ini}} } =0
\end{eqnarray}
in equations \eqref{eq:backreactionAccen} and \eqref{eq:gensolution}, we find that 
$\hat{\Phi}_{- 0} = 0$ and 
   \begin{eqnarray}
A_{-} = \frac{- g_{p-}(r_{\text{ini}})}{2\sqrt{r_{\text{ini}}/r_{\star}} \, I_{1}\left(2\sqrt{r_{\text{ini}}/r_{\star}}\right)}
\,.
\end{eqnarray}
For the exterior region, there are two possible cases.
From the left panel of  Figure \ref{fig:spacetime_diagram}, there are two proper time hypersurfaces of interest  $\tau_{{\rm{gal}}}$ and$\tau_{{\rm{clusl}}}$.  The difference between the galaxy rotation curves at $\tau_{{\rm{gal}}}$ and $\tau_{{\rm{clusl}}}$ is determined by the boundary condition at the edge of $\mathcal{M}_{+}$. 
\begin{itemize}
\item $\tau_{{\rm{gal}}}$: Galaxy formation time scale, when a galaxy decouples from the Hubble flow, forming a gravitationally bound system. 
For the purposes of calculating the galaxy rotation curve at $\tau_{{\rm{gal}}}$-hypersurface, the boundary of the $\Sigma_{+}$ is at infinity. 
$K_{1}$ vanishes at large $y$: $K_1(y\to\infty) = 0$, that is $K_1(y) \sim \sqrt{\frac{\pi}{2y}} e^{-y}$, while $I_{1}$ grows rapidly  $I_1(y) \sim \frac{e^y}{\sqrt{2\pi y}}$ as $y\to \infty$.  Therefore, to prevent divergence at infinity, we must set $A_{+}= 0$.  
Hence, imposing  the  continuity of the potential and its gradient at: $r = r_{\star}$, 
   \begin{eqnarray}\label{eq:continuitycondition}
\hat{\Phi}_-(r_{\star}) = \hat{\Phi}_+(r_{\star}) \,, 
\qquad \qquad 
  \frac{d{\hat{\Phi}}_{-}}{dr_{-}} \bigg|_{r_{\star}} = \frac{d\hat{\Phi}_+}{dr_{+}} \bigg|_{r_{\star}} 
\end{eqnarray}
leads to $\hat{\Phi}_{+ 0}  = \hat{\Phi}_-(r_{\star}) $ and 
   \begin{eqnarray}
   B_{+} &=& \frac{r_{\star}^2 }{2 K_1(2)}  \frac{d{\hat{\Phi}}_{-}}{dr_{-}} \bigg|_{r_{\star}}  -\frac{ g_{p-}(r_{\star})}{2 K_1(2)}\,,
   \end{eqnarray}

\item $\tau_{{\rm{clusl}}}$: Galaxy cluster formation time scale, when a galaxy cluster decouples from the Hubble flow, forming a gravitationally bound system. 
However, at $\tau_{{\rm{clusl}}}$-hypersurface, the galaxy is now living inside another gravitationally bound system with a compact finite boundary. Given that $K_1(y) \sim \sqrt{\frac{\pi}{2y}} e^{-y}$, decay facts, the general solution is dominated by the contribution from $I_{1}$, therefore, we can set $B_{+}$ to zero and impose the continuity condition(equation \eqref{eq:continuitycondition}) to fix $\hat{\Phi}_{+ 0}  = \hat{\Phi}_-(r_{\star}) $ and $A_{+}$
   \begin{eqnarray}
   A_{+} &=& \frac{r_{\star}^2 }{2 I_1(2)}  \frac{d{\hat{\Phi}}_{-}}{dr_{-}} \bigg|_{r_{\star}}  -\frac{ g_{p-}(r_{\star})}{2 I_1(2)}\,,
   \end{eqnarray}
\end{itemize}

 \begin{figure}[h]
\centering 
\includegraphics[width=75mm,height=60mm] {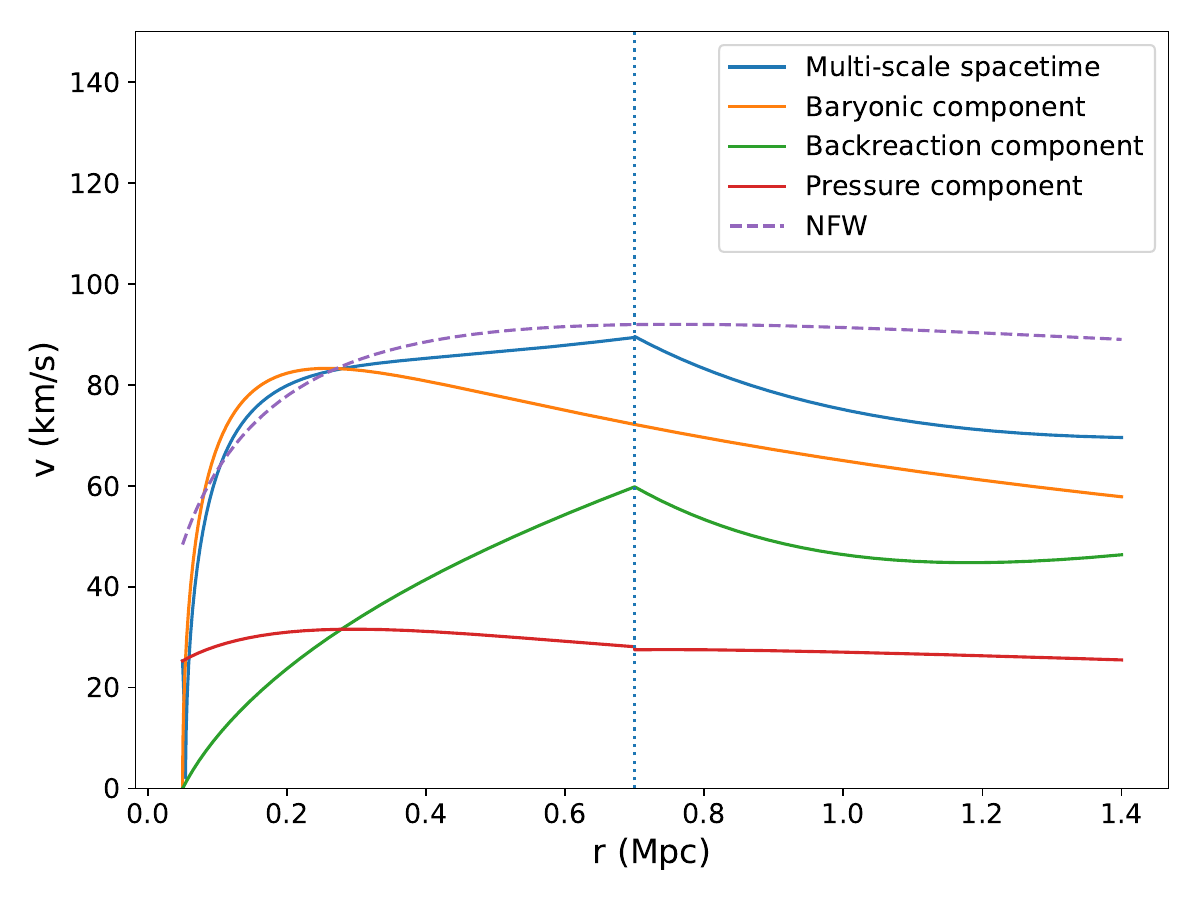}
\includegraphics[width=75mm,height=60mm] {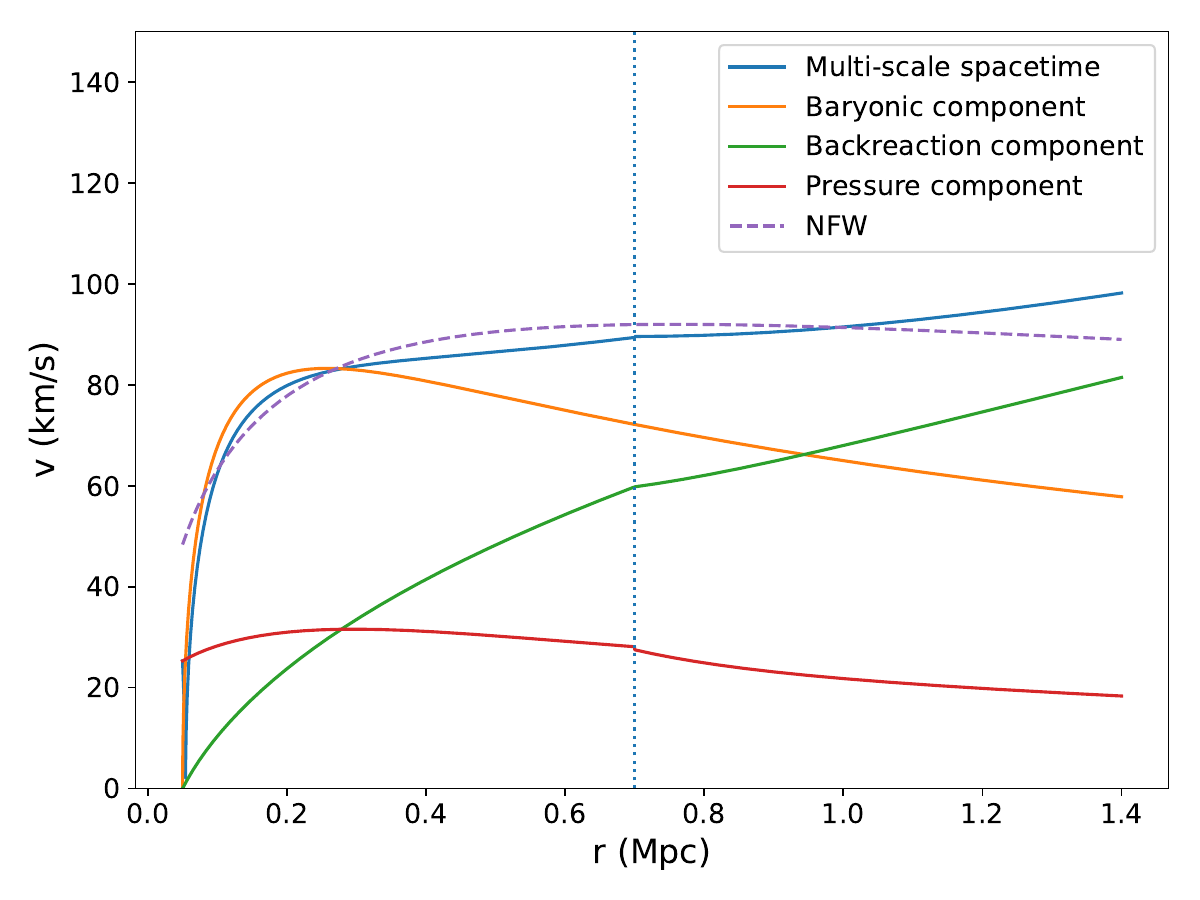}
\label{fig:rotation_curve}
\caption{Left panel: Galaxy rotation curve for a typical massive galaxy with  Hernquist density profile for the baryon density at $\tau_{\rm{gal}}$-hypersurface(The galaxy is not within a gravitationally bound system).  
Right panel: Galaxy rotation curve for a typical galaxy at $\tau_{{\rm{clus}}}$-hypersurface. That is a galaxy within a gravitationally bound cluster. 
The parameters of the  Hernquist density profile are  set to  $M = 1.5\times10^{12} M_{\otimes}$ and $a_{\pm} = 0.2$ Mpc. The velocity dispersion is set to  $\sigma_{1D\pm} =\sigma_{1TD \pm}  = 20$ [km/s] .
For the Navarro–Frenk–White (NFW)  rotation velocity, we used 
$ v_{\rm{NFW}}(r) = \sqrt{4\pi G \rho_s r_s^3 \frac{1}{r}\big[\ln\big(1 + \frac{r}{r_s}\big) - \frac{r/r_s}{1 + r/r_s}\big]}$,
where $\rho_s \sim 10^{25}-10^{-26} g/cm^{3}$ and we set $r_s \sim 0.2r_{{\star}} $Mpc~\cite{Navarro:1995iw}.
The rotation velocity becomes flat as soon as the contribution of the backreaction density field exceeds that of the baryon density field, and it then dominates in the outskirts. 
}

\end{figure}

 The log derivative of the density profiles is
 \begin{eqnarray}\label{eq:pressuregradient}
\frac{d \ln \rho_{m\pm}}{d \ln r_{\pm}} = -\frac{4r_{\pm} + a_{\pm}}{r_{\pm}+a_{\pm}}=-1 - \frac{3r_{\pm}}{r_{\pm}+a_{\pm}}\,,
\qquad \qquad 
\frac{d \ln \hat{\rho}_{\pm}}{d \ln r_{\pm}} = -2 + \frac{g_{\pm}(r_{\pm})}{Q_{\pm}(r_{\pm})}\,,
\end{eqnarray}
where $ {g(r_{\pm})} = {r_{\pm}^2} \frac{d\hat{\Phi}}{dr_{\pm}} $  is given in equation \eqref{eq:solng}  and the backreaction-induced density profile is given in equation \eqref{eq:backreactiondensityprofile}.
We show the galaxy rotation curves for a typical massive galaxy in Figure \ref{fig:rotation_curve}, illustrating the impact of the environment. 
The left panel shows a declining rotation curve after the matter horizon, while the right panel shows the other extreme, where the rotation curve is rising with radius instead of flattening. It is important to note that these features have been observed \cite{OBrien:2017ogc}. An exactly flat rotation curve can be readily accounted for by incorporating pressure effects into the Poisson equation or by modulating the baryon density. via $a_{\pm}$.

Modified Newtonian Dynamics (MOND) is an alternative theory of gravity that modifies Newton’s laws at extremely low accelerations~\cite{1983ApJ...270..365M}.  MOND does well to explain the fast rotation speeds of stars in galaxies without requiring invisible dark matter; however, it struggles to explain the behaviour of galaxy clusters and Cosmic Microwave Background (CMB) observations ~\cite{Famaey:2024uow}.
Although the exact functional form of the modification is usually given by an interpolating function, we can derive this result for a galaxy using the total gradient of gravitational potential, aka, gravitational Newtonian force ($\mathbf{a}_N$) (sum of the baryon component and the backreaction component)
       \begin{eqnarray}
  {a}_N & =& \frac{d\Phi_{\pm}}{dr_{\pm}} = \frac{G M_{bk\pm}}{r_{\pm}^{2}} \left[ 1 + \nu_{\pm}(r_{\pm}) \right]\,,
       \end{eqnarray}
  where   $\nu_{\pm}(r_{\pm}) = \left[ A_{\pm}\, g_{1\pm}(r_{\pm}) + g_{p\pm}(r_{\pm}) \right]/{G M_{bk\pm}} $. In Deep-MOND regime, it scales like $C/r_{\pm}$ largely independent of the particular solution for the galaxy in $\tau_{{\rm{clus}}}$ evolutionary phase. 

Finally, our analysis has focused on linear and spherically symmetric approximation; a more accurate representation of the physical reality will require going beyond this approximation. Most importantly, the behaviour of the backreaction energy density  $\tilde{\rho}$ could differ on large scales. 
This can be seen by solving equation \eqref{eq:shear_timehyper} on a perturbed FLRW spacetime. 
$
\tilde{\rho}_{\pm} 
=\< \bar{u}^a_{\pm}\bar{u}^{b}_{\pm} {\delta^{(2)}\tilde{\sigma}^{\pm}}_{ab} \>
+\< 2 v^{(a}_{\pm}\bar{u}^{b)}_{\pm}{\delta^{(1)}\tilde{\sigma}^{\pm}}_{ab}\>\,,
$
where $\tilde{\sigma}^{\pm}_{ab}  = {\delta^{(1)}\tilde{\sigma}^{\pm}}_{ab}  +{\delta^{(2)}\tilde{\sigma}^{\pm}}_{ab}$, 
${u}^a_{\pm}\approx \bar{u}^a_{\pm}+v^a_{\pm}$ and  $ \tilde\Theta = \bar{\tilde{\Theta}} + \delta \tilde{\Theta}$.
Beyond linear order, the perturbed shear tensor includes contribution of the square of the shear tensor
\begin{eqnarray}
\tilde\sigma_{\langle ab\rangle} &=& \frac{ \left[\tilde{\sigma}_{ab}\right]_{\rm{ini}}}{r^2} 
 + \frac{1}{r^2} \int_{r_{\rm{ini}}}^{r} \d r' {r'}^2\bigg[
   -\tfrac{2}{3}\,\left[  \delta \tilde{\Theta}\,\tilde\sigma_{ab}\right]
- \tilde\sigma_{ac}\tilde\sigma^c{}_{b}
-\mathcal{E}_{ab}
+ \tfrac{1}{2}\,\mathcal{R}_{ab}
\bigg]\,.
\label{eq:shearsecond}
\end{eqnarray}
We have already shown that on small scales, $\<\hat{\rho}\> \neq 0$, which indicates that it contributes to the background dynamics. 
A more detailed study is required to estimate $\<\tilde{\rho}\> $
while imposing suitable initial conditions on the gravitational potential.   This will show whether $\hat{\rho}_{\pm} $ can also act as dark energy on large scales because from equation \eqref{eq:shearsecond}, it is clear that $\<\tilde{\rho}\> \neq 0$.

\section{Conclusion}\label{sec:conc}

In this work, we have identified the  possible origin of the long-standing difficulty in modelling clustering of matter on nonlinear scales in general relativity: it stems from a breakdown of geodesic evolution at finite time or finite spatial extent. This breakdown is typically preceded by the formation of a matter horizon, driven by non-trivial local curvature for initial conditions set on an FLRW spacetime. This phenomenon is consistent with the cosmic censorship conjecture developed in the context of black hole event horizons~\cite{Penrose:1969pc}.

The failure of geodesic flow at finite time prevents a consistent, smooth extension of particle trajectories across the full dynamical range on a single sheet of spacetime. Importantly, this is not merely a limitation of the point-particle approximation, but rather a manifestation of the role of backreaction on particle trajectories.

To address this, we provide a consistent procedure for cutting the spacetime at the matter horizon and glueing it to another sheet of spacetime with opposite orientation. We explicitly demonstrate how to construct this cut-and-paste operation at the level of the action and show that it preserves smooth boundary conditions for the projected metric tensor at the shared boundary. These boundary conditions yield a more symmetric, smooth map between the two sheets of spacetime than the standard Darmois–Israel junction conditions, allowing us to join conformally related projected metric tensors. As a result, the extrinsic curvature does not vanish on either side; we interpret this non-zero extrinsic curvature as a backreaction contribution to the effective energy-momentum tensor.

Using variational calculus, we demonstrate how the boundary term arising from the variation of the Einstein-Hilbert action can be absorbed into the effective energy-momentum tensor. This term is interpreted as an effective fluid characterised by non-zero energy density, pressure, flux, and stress. As a practical application, we illustrated how these backreaction terms lead to flat rotation curves in the spherically symmetric limit for test particles orbiting a galaxy. We derived this explicitly by obtaining the evolution equation for the shear tensor on the timelike hypersurface; the contraction of this shear tensor with the matter four-velocity yields the backreaction energy density.

This construction introduces a hierarchical, multi-scale description of the universe, in which another sheet of spacetime with opposite orientation is glued through well-defined boundary conditions~\cite{Umeh:2026ajv}. It provides a covariant realisation of cosmological zoom-in simulations and establishes a consistent framework for resolving both temporal and spatial dynamical ranges. Crucially, the effects of finite extent enter through boundary contributions to the energy-momentum tensor, rather than through ad hoc regularisation of singular sources~\cite{Poisson:2003nc}.

As a concrete demonstration, we showed that the induced geometric backreaction reproduces flat galaxy rotation curves without invoking additional dark matter degrees of freedom, highlighting the physical relevance of the framework.

More broadly, this work suggests that hierarchical structure formation in the universe should be understood as a sequence of transitions between geodesic regimes separated by matter horizons. This opens a new avenue for connecting general relativity, cosmological simulations, and effective descriptions of matter on nonlinear scales.

  \section*{Acknowledgement}
  I benefited immensely from discussions with Sravan Kumar,   Mathew Hall,  David Wands and Robson Christie.  I appreciate the support of the CIC Foundation; without them, this work would not have seen the light of day.

  \section*{Data Availability}

The tensor algebraic computations in this paper were done with the open-source tensor algebra software, xPand, which can be found here(\url{https://github.com/Obinna/xPand})~\cite{Pitrou:2013hga}.  xPand is based on xPert(\url{http://www.xact.es/xPert/index.html})~\cite{Brizuela:2008ra}.
The numerical computations were done with Python \cite{van1995python} with extensive use of numpy~(\url{https://numpy.org/devdocs/user/building.html})~\cite{harris2020array} and Scipy~\cite{2020SciPy-NMeth}.
Plotting of results was done with matplotlib (\url{https://matplotlib.org/stable/users/project/citing.html}) \cite{Hunter:2007}.

\appendix

\section{Expansion in Fourier space}\label{sec:Fourier}

The last term in the square bracket in equation \eqref{eq:simpg00} can be simplified further by expanding in Fourier space 
\begin{eqnarray}
  { x^{i}_{F} x^{j}_{F}}\partial_{(i} \partial_{j)} \Phi  =  - \int \frac{\d^3 k}{(2\pi)^3}\frac{\partial}{\partial k^i} \frac{\partial}{\partial k^j}\left(k_i k_{j}\Phi(k)\right) e^{i{\k}\cdot {\x}}\,.
\end{eqnarray}
The position vector  and spatial derivative map into Fourier space according to $
x^i_{F} \rightarrow - i \partial_{k}^i $ and $ {\partial_x}^i \rightarrow i k^i $.
Focusing on the integrand alone gives to 
\begin{eqnarray}
\frac{\partial}{\partial k^i} \frac{\partial}{\partial k^j}\left(k_i k_{j}\Phi(k)\right) &=&\left[12+ 8 \frac{k^i}{\Phi }  \frac{\partial \Phi (k)}{\partial k^{i}} +\frac{k^i k^j}{\Phi }  \frac{\partial^2 \Phi(k)}{\partial k^{i} \partial k^{j}} \right]\Phi(k)
\end{eqnarray}
At the leading order, modes that contribute to $\Phi$ evolve independently, hence can be decomposed into spatial and temporal part  $\Phi(\eta,{\bf{k}}) = T(\eta,k) \varphi_{{\bf{k}}} =g(\eta)T(k) \varphi_{{\bf{k}}}$, where $g(\eta)$ is the growth function,  $T(k)$ is the transfer function and  $\varphi_{{\bf{k}}}$ is related to the the primordinal curvature perturbation. The assumption of independent evolution of modes is an approximation, however, it remains until shell crossing in a matter-dominated universe~\cite{Ma:1995ey,Jelic-Cizmek:2018gdp}.
\begin{eqnarray}
\frac{k^i}{\Phi }  \frac{\partial \Phi (k)}{\partial k^{i}}  &=&\frac{k^i}{\varphi_{\bf{k} } }  \frac{\partial \varphi_{{\bf{k}}} (k)}{\partial k^{i}}   +\frac{k^i}{ T(k) }  \frac{\partial T(k)}{\partial k^{i}}  \,,
\\
\frac{k^i k^j}{\Phi }  \frac{\partial^2 \Phi(k)}{\partial k^{i} \partial k^{j}} & =& \frac{k^ik^j}{\varphi_{\bf{k} } }  \frac{\partial^2 \varphi_{{\bf{k}}} (k)}{\partial k^{i} \partial k^j}   +\frac{k^i k^j}{ T(k) }  \frac{\partial^2 T(k)}{\partial k^{i}\partial k^{j}} 
+2\frac{k^{(i} }{\varphi_{{\bf{k}}} }  \frac{\partial \varphi_{{\bf{k}}} (k)}{\partial k^{i}}   \frac{k^{j)}}{ T(k) }  \frac{\partial T(k)}{\partial k^{j}} \,.
\end{eqnarray}
Using the dictionary given in \cite{Cabass:2016cgp}, we can change the directional derivative in Fourier space wrt to log derivative of the wavenumber. Furthermore, we can convert the log derivative of the wavenumber to the power law of the primordial curvature perturbation~\cite{Umeh:2019qyd,Umeh:2019jqg}
\begin{eqnarray}
\frac{1}{\varphi_{{\bf{k}}}} \frac{\partial  \varphi_{{\bf{k}}}}{\partial k^i} &=& \frac{k^i}{k^2} \bigg[ \frac{d\log (k^3\varphi_{{\bf{k}}})}{d\log k} -3\bigg] ],,
\\
\frac{1}{\varphi_{{\bf{k}}}} \frac{\partial^2 \varphi_{{\bf{k}}}}{\partial k^i \partial k^j} &=& \frac{\delta^{ij}}{k^2}\bigg[ \frac{d\log (k^3\varphi_{{\bf{k}}})}{d\log k} -3\bigg]
+ \frac{k^i k^j}{k^4} \bigg[ \frac{d^2\log (k^3\varphi_{{\bf{k}}})}{d(\log k)^2} +\left[\frac{d\log (k^3 \varphi_{{\bf{k}}})}{d\log k} \right]^2 -8 \frac{d\log (k^3 \varphi_{{\bf{k}}})}{d\log k}  +15\bigg] \,.
\end{eqnarray}
Putting all these together leads 
\begin{eqnarray}
  { x^{i}_{F} x^{j}_{F}}\partial_{(i} \partial_{j)} \Phi & =&    - \int \frac{\d^3 k}{(2\pi)^3}\frac{\partial}{\partial k^i} \frac{\partial}{\partial k^j}\left(k_i k_{j}\Phi(k)\right) e^{i{\k}\cdot {\x}}
=
    \int \frac{\d^3 k}{(2\pi)^3} \mathcal{Z}(k)\Phi(k) e^{i{\k}\cdot {\x}} = \mathcal{F}^{-1} \left[  \mathcal{Z}(k)\Phi(k) \right]
\end{eqnarray}
In general, $ \mathcal{Z}(k) $ is a scale-dependent weighting function that depends on the quantum mechanics of the early and the physical process of gravitational structure formation
\begin{eqnarray}
 \mathcal{Z}(k) = - \bigg[  \mathcal{Y}  +
 \left(\mathcal{X}  -3\right) \frac{\d \log T(k)}{\d \log k} 
+\frac{1}{ T(k) } \frac{d^2T(k)}{d\log k^2} + \frac{7}{ T(k) } \frac{d T(k)}{d\log k}   \bigg]\,.
\end{eqnarray}
$\mathcal{X} $ and $\mathcal{Y} $ are sourced by the quantum mechanics of the early universe
\begin{eqnarray}
\mathcal{X} &=& \frac{d\log (k^3\varphi_{{\bf{k}}})}{d\log k} \,,
\\
\mathcal{Y} &=&\frac{d^2\log (k^3 \varphi_{{s{\k}}})}{d(\log k)^2} +\left[\frac{d\log (k^3 \varphi_{{s{\k}}})}{d\log k} \right]^2+ \frac{d\log (k^3 \varphi_{{s{\k}}})}{d\log k} \,.
\end{eqnarray}
The modulation of the gravitational potential by the physical process of gravitational structure formation is contained in the log derivatives of the transfer function.  
Within the slow-roll single inflaton model, $\mathcal{X}$ is related to the spectral index  of the power spectrum of the initial curvature perturbation  \cite{dePutter:2015vga}
\begin{eqnarray}
\frac{d\log (k^3\varphi_{{\bm{k}}})}{d\log k}  \equiv\frac{d\log \sqrt{(k^3P_{\varphi}({\k}))}}{d\log k} = \frac{1}{2}\frac{d\log (k^3P_{\varphi}({\k}))}{d\log k}  = \frac{1}{2}(n_s-1)\,.
\end{eqnarray}
The spectral index $n_s$ arises directly from the quantum mechanics of the inflationary perturbations. The precise value of $n_s$
  depends on the inflationary potential  $V(\phi)$.  $\mathcal{Y}$ is related to the running of the spectral index. 
At the leading order in slow-roll approximation, the running of the spectral index satisfies~\cite{Zarei:2014bta}
\begin{eqnarray}\label{eq:runningspectralindex}
\frac{d^2\log (k^3 \varphi_{{{\k}}})}{d(\log k)^2} &=&0=\frac{ d n_s}{d \log k}\, .
\end{eqnarray}
 Furthermore, the spectral index is related to the amplitude of the primordial non-Gaussianity according to $\fnl =  {5}(1-n_s)/{12}\,. $
This is the well-known  Maldacena relation~\cite{Maldacena:2002vr}.  
The transfer function may be approximated with a fitting function ~\cite{Bardeen:1985tr,Eisenstein:1997ik}


\providecommand{\href}[2]{#2}\begingroup\raggedright\endgroup

\end{document}